\documentclass[fleqn,usenatbib]{mnras}

\usepackage[figuresright]{rotating}
\usepackage{longtable}

\usepackage[T1]{fontenc}

\DeclareRobustCommand{\VAN}[3]{#2}
\let\VANthebibliography\thebibliography
\def\thebibliography{\DeclareRobustCommand{\VAN}[3]{##3}\VANthebibliography}

\usepackage{graphicx}	
\usepackage{amsmath}	
\usepackage{amssymb}	
\newcommand{\mJyb}{mJy beam$^{-1}$}
\newcommand{\angstrom}{\text{\normalfont\AA}}

\title[VLBI Observations of PG  quasars]{VLBI Observations of a sample of Palomar-Green quasars I: parsec-scale morphology}

\author[Wang et al.]{
Ailing Wang,$^{1,2}$
Tao An,$^{1,2}$\thanks{E-mail: antao@shao.ac.cn}
Xiaopeng Cheng$^{3}$,
Luis C. Ho$^{4,5}$,
Kenneth I. Kellermann$^{6}$, \newauthor
Willem A. Baan$^{7,8}$,
Jun Yang$^{9}$,
Yingkang Zhang$^{1}$
\\
$^{1}$ Shanghai Astronomical Observatory, CAS, 80 Nandan Road, Shanghai 200030, China \\
$^{2}$ School of Astronomy and Space Sciences, University of Chinese Academy of Sciences, No. 19A Yuquan Road, Beijing 100049, China  \\
$^{3}$ Korea Astronomy and Space Science Institute, 776 Daedeok-daero, Yuseong-gu, Daejeon 34055, Korea \\
$^{4}$ Kavli Institute for Astronomy and Astrophysics, Peking University, Beijing 100871, China \\
$^{5}$ Department of Astronomy, School of Physics, Peking University, Beijing 100871, China \\
$^{6}$ National Radio Astronomy Observatory, 520 Edgemont Rd., Charlottesville, VA 22903, USA \\
$^{7}$ Xinjiang Astronomical Observatory, Chinese Academy of Sciences, 150 Science 1-Street, 830011 Urumqi, P.R. China \\
$^{8}$ Netherlands Institute for Radio Astronomy ASTRON, NL-7991 PD Dwingeloo, the Netherlands \\
$^{9}$ Department of Space, Earth and Environment, Chalmers University of Technology, Onsala Space Observatory, SE-439 92 Onsala, Sweden 
}

\date{Accepted XXX. Received YYY; in original form ZZZ}

\pubyear{2015}

\begin{document}
\label{firstpage}
\pagerange{\pageref{firstpage}--\pageref{lastpage}}
\maketitle

\begin{abstract}
We observed 20 Palomar-Green (PG) quasars at low redshift ($z<0.5$) with total flux density > 1 mJy, including 4 radio-loud quasars (RLQs) and 16 radio-quiet quasars (RQQs), using the Very Long Baseline Array (VLBA) at 5 GHz. 
Ten RQQs are clearly detected in the VLBA images, and a compact radio core is identified in eight of them, indicating the prevalence of active galactic nucleus (AGN)-related radio emission in this flux-density-limited RQQ sample. 
The RQQs and RLQs in our sample have a division at $\sim$30~mJy.
The radio emission from RQQs appears to be the result of a combination of star formation and AGN-associated activities.
All RQQs in our sample have a 5 GHz flux density ratio of Very Large Array (VLA) A-array to  D-array $f_{\rm c} = S_{\rm A}^{\rm VLA}/S_{\rm D}^{\rm VLA}$  above 0.2. 
The RQQs with $f_a$ (VLBA and VLA flux density ratio $S^{\rm VLBA}/S_{\rm A}^{\rm VLA}) > 0.2$ versus $f_a < 0.2$ show significant differences in morphology, compactness and total flux density. 
$f_{\rm a}$ of RQQs is systematically lower than that of RLQs, probably due to the extended jets or relic jets of RQQs on 10s to 100s parsecs which are resolved out in VLBA images. 
Future larger samples, especially with the addition of milli-arcsec resolution radio images of RQQs with total flux densities below 1 mJy, can test the conclusions of this paper and contributes to the understanding of the radio emission mechanism of RQQs, and the dichotomy and physical connection between RQQs and RLQs.
\end{abstract}




\section{Introduction}



Usually, quasars are divided into two populations according to the radio-to-optical flux density ratio \citep{1989AJ.....98.1195K,2016ApJ...831..168K}, the so-called radio-loudness parameter $\rm R=S_{5 GHz}/S_{4000\angstrom}$: radio-loud quasars (RLQs) and radio-quiet quasars (RQQs). 
It is now recognized that the radio emission from RLQs is dominated by relativistic jets \citep[e.g.][]{2015MNRAS.452.1263P,2019ARA&A..57..467B}. In contrast to RLQs, the nature of radio emission from RQQs has remained an open question for decades. 
In this paper, we adopt the same nomenclature as \citet{2019NatAs...3..387P}, defining collimated outflows as jets and extended large-aperture outflows as winds. It has been suggested that RQQs may just be scaled-down versions of RLQs with lower jet powers \citep[e.g.][]{2005ApJ...621..123U,2015MNRAS.448.2665W,2021MNRAS.506.5888M}, or that the radio emission from RQQs may come from star formation \citep[e.g.][]{1993ApJ...419..553B,2011ApJ...740...20P,2013ApJ...768...37C,2016ApJ...831..168K}, or a mixture of both \citep[][]{2021MNRAS.506.5888M}.
Due to the complex characteristics and feedback of the active galactic nucleus (AGN), as well as the diverse galactic environment, the radio emission of RQQs may come from star formation, accretion disk winds, coronal disk emission, low-power jets, or a combination of some of these components  \citep[reviewed by][]{2019NatAs...3..387P}.

In RQQs with clear signs of AGN activity observed in optical and X-rays, large-scale extended jets or lobes are rarely observed and have only been observed on a few hundred parsecs in a few cases \citep{1993MNRAS.263..471P,1998AJ....115..928C,2021MNRAS.507..991S}. 
A challenging issue in low-resolution (e.g., arcsec scale) observations of RQQs is that the total radio flux density obtained is a mixture of multiple emission sources discussed above
and cannot be directly linked to the optical and hard X-ray emission from the AGN. Parsec-scale radio emission is closely associated with the central engine, and is therefore crucial for studying AGN physics (accretion and outflowing) and the interactions between radio sources and host galaxies (co-evolution of AGN and host galaxies, jet-mode AGN feedback). 

Very long baseline interferometry (VLBI), with pc or even sub-pc resolution imaging capacity, enables a clear separation of AGN and non-AGN radio emission, thus establishing a broad-band coverage of AGN emission from radio to hard X-rays, which helps to constrain the AGN spectral energy distribution (SED) more accurately.
In particular, the VLBI data may help to decompose the SEDs into AGN, torus and host components. Thus, the true AGN luminosity $(L_{\rm bol}$) and accretion rate can be estimated.
So far, only a few RQQs have been observed by VLBI \citep{1998MNRAS.299..165B,2004A&A...417..925M,2005ApJ...621..123U,2009ApJ...706L.260G,2009MNRAS.398..176K,2013MNRAS.432.1138P,2016A&A...589L...2H}. Inconsistencies in the observational samples, source selection methods and observation conditions in these studies render a systematic VLBI imaging survey of RQQs lacking to date, and the radio emission from RQQs closely related to the central AGN on the parsec scales is not yet fully understood.  

In this study, we select a sample of RQQs and observe them with the Very Long Baseline Array (VLBA) at 5 GHz to obtain radio images with parsec resolution.\footnote{Throughout, we adopt the following cosmological parameters: H$_0$ = 71 km s$^{-1}$ Mpc$^{-1}$, $\Omega_\Lambda = 0.73$ and $\Omega_m = 0.27$. At a redshift of 0.1, an angular size of 1 mas corresponds to a projected linear size of 1.8 pc. } The radio morphology and brightness temperature derived from VLBI data can directly distinguish whether the radio emission is from star formation activity or from AGN, and determine what percentage of the radio emission comes from the AGN contribution.

\section{Observations and Data Reduction}

\subsection{Sample Selection}

The proper selection of a representative sample of the RQQ population is crucial to reveal the physical nature of the RQQs. Such a sample needs to have multiple indicators of the AGN emission characteristics. Another important factor to  consider is that both the optical and X-ray emission of an AGN contains the emission from the host galaxy, and with a fixed angular resolution, the decoupling between the AGN and its host galaxy is only possible in nearby galaxies.
Based on these two considerations, low-redshift ($z < 0.5$) Palomar-Green (PG) quasars \citep{1983ApJ...269..352S,1992ApJS...80..109B} are the most suitable sample for studying the multi-band astrophysical properties of RQQs. The PG  sample has rich observational data ranging from radio to hard X-rays and is the best studied bright low-$z$ quasar sample. Moreover, PG  quasars have a wealth of astrophysical information, including accurate BH masses either directly from reverberation mapping or from H$_\beta$ viral masses \citep[e.g.][]{2000ApJ...533..631K,2006ApJ...641..689V}, bolometric luminosity measurements and multi-band SEDs \citep[e.g.][]{2011ApJS..196....2S}, morphology and structure of the host galaxies obtained from the high-resolution Hubble Space Telescope images \citep[e.g.][]{2017ApJS..232...21K,2021ApJ...911...94Z},
properties of the host galaxy's gas environment \citep[e.g.][]{2015ApJS..219...22P,2020ApJS..247...15S,2018ApJ...854..158S,2006AJ....132.2398E}, star formation rate measurements \citep[e.g. ][]{2014ApJS..214...23S,2021arXiv210202695X}. Therefore, correlations of the AGN radio luminosity with the black hole mass and accretion rate can be established.

Among the 87 PG  quasars at $z < 0.5$, there are 16 RLQs and 71 RQQs. 
Since most of these samples have no previous VLBI observations, in our first  experiment, we selected 16 RQQs with  VLA core flux densities greater than 1 mJy at 5 GHz. Future observational studies can be extended to the weaker radio sources in the optically-selected quasar sample.
We included four RLQs for which no VLBI data were available, and  also collected  archive VLBI data for the remaining 12 RL PG  quasars, so that we have a complete sample of RL PG  quasars at $z<0.5$ as a control sample for a comparative study of the radio emission properties of RLQs and RQQs.  Details of the present sample are given in Table \ref{tab:image} as follows: Column (1): source name; Columns (2): redshift; (3): right ascension and declination coordinates of the peak radio emission; Column (4): offset between the optical \textit{Gaia} nucleus position and the radio peak position; Columns (5): observation date; Columns (6): major axis and minor axis of the restoring beam, and the position angle of the major axis, measured from north to east; Column (7): peak flux density; Column (8): root-mean-square noise of the image, measured in off-source regions; Column (9): contour levels; Column (10): phase-reference calibrator name; Column (11): the radio loudness; Column (12): the total radio flux density observed by VLA in the D configuration at 5 GHz; Column (13): the unresolved radio flux density observed by VLA in the A configuration at 5 GHz.  Note: Columns (11)-(13) are obtained from \citet{1989AJ.....98.1195K,1994AJ....108.1163K}. The VLBI archive data for RLQs were adopted as the closest in time to our VLBA observations.

\subsection{VLBA Observations}

The observations were carried out in August 2015 using the Very Long Baseline Array (VLBA) of the National Radio Astronomy Observatory of the US using all ten telescopes. To facilitate scheduling and to obtain an optimized (\textit{u,v}) coverage, we divided the 20 sources into three groups according to their right ascension (RA) distribution. The observations were made at 5 GHz in a dual-circular polarization mode with a data recording rate of 2 gigabits per second. The data were recorded in 8 intermediate frequency (IF) channels, each  32 MHz wide. Due to the weakness of radio-quiet PG  quasars, we used  phase-referencing  with "calibrator (1.5 min) - target (3.5 min)" cycles. The four radio-loud PG  quasars themselves are bright enough for fringe searching and self-calibration. Fourteen snapshot scans were performed for each RQQ, with an effective time of $\sim$50 min for each source.
The typical baseline sensitivity is 1.7 mJy for a 30-s integration time allowing for a signal-to-noise ratio greater than 50 in the fringe search of the calibrators. The thermal noise of an image for 50 min on-source time is expected to be $\sim0.025$  mJy beam$^{-1}$.
\footnote{\url{https://science.nrao.edu/facilities/vlba/docs/manuals/oss}}
The actual image noise is a factor of 0.8--1.4 of this theoretical noise estimate (see Table \ref{tab:image}) due to various factors, such as coherence loss in the phase-referenced observations, insufficient \textit{(u,v)} coverage in snapshot mode degrading the image quality, and a part of the source emission being resolved and contributing to the apparent noise.
The observational data were correlated using the DiFX software correlator \citep{2011PASP..123..275D} at Socorro with an averaging time of 2s, 128 frequency channels per IF, and uniform weighting.

\subsection{Data Reduction}
\label{sec:obs-data}

The correlated data were downloaded to the China Square Kilometre Array Regional Centre \citep{2019NatAs...3.1030A} via the internet for further calibration and imaging. 
The data calibration was done using a pipeline that we developed using Python and Parseltongue \citep{,2022arXiv220613022A}, which calls programs integrated in the Astronomical Image Processing System \citep[\textsc{AIPS}:][]{2003ASSL..285..109G} and executed in a scripted manner.  

We first applied ionospheric corrections to the visibility data using the VLBATECR task, which applied the Global Positioning System (GPS) models of the electron content in the ionosphere to correct for the dispersive delays. Then, we corrected the sampler voltage offsets with the auto-correlation data using the task ACCOR. Amplitude calibration was conducted using the information extracted from the gain curve (GC), system temperature (TY), and weather (WX) tables for all antennas to correct the atmospheric opacity. A phase correction for parallactic angle effects was carried out before any other phase corrections were made using task VLBAPANG. The instrumental single-band delay and phase offsets were corrected using 2-min observational data of the calibrators 3C~84, 4C~39.25, and 3C~345. Throughout the calibration procedure, we used the Pie Town (PT) telescope as the reference antenna. Global frequency- and time-dependent phase error solutions were derived from the phase-referencing calibrators using the global fringe fitting with a 2-min solution interval and a point-source model \citep{1995ASPC...82..189C} by averaging over all the IFs. We then applied these solutions derived from the calibrators and interpolated them into their corresponding target sources. Next, we calibrated the bandpass shape of each telescope. Finally, we applied all calibration solutions to the data and exported the single source files by averaging all the channels in each IF and averaging the data over 2 min.

\begin{figure*}
\centering
  \begin{tabular}{cc}
  \includegraphics[width=0.24\textwidth]{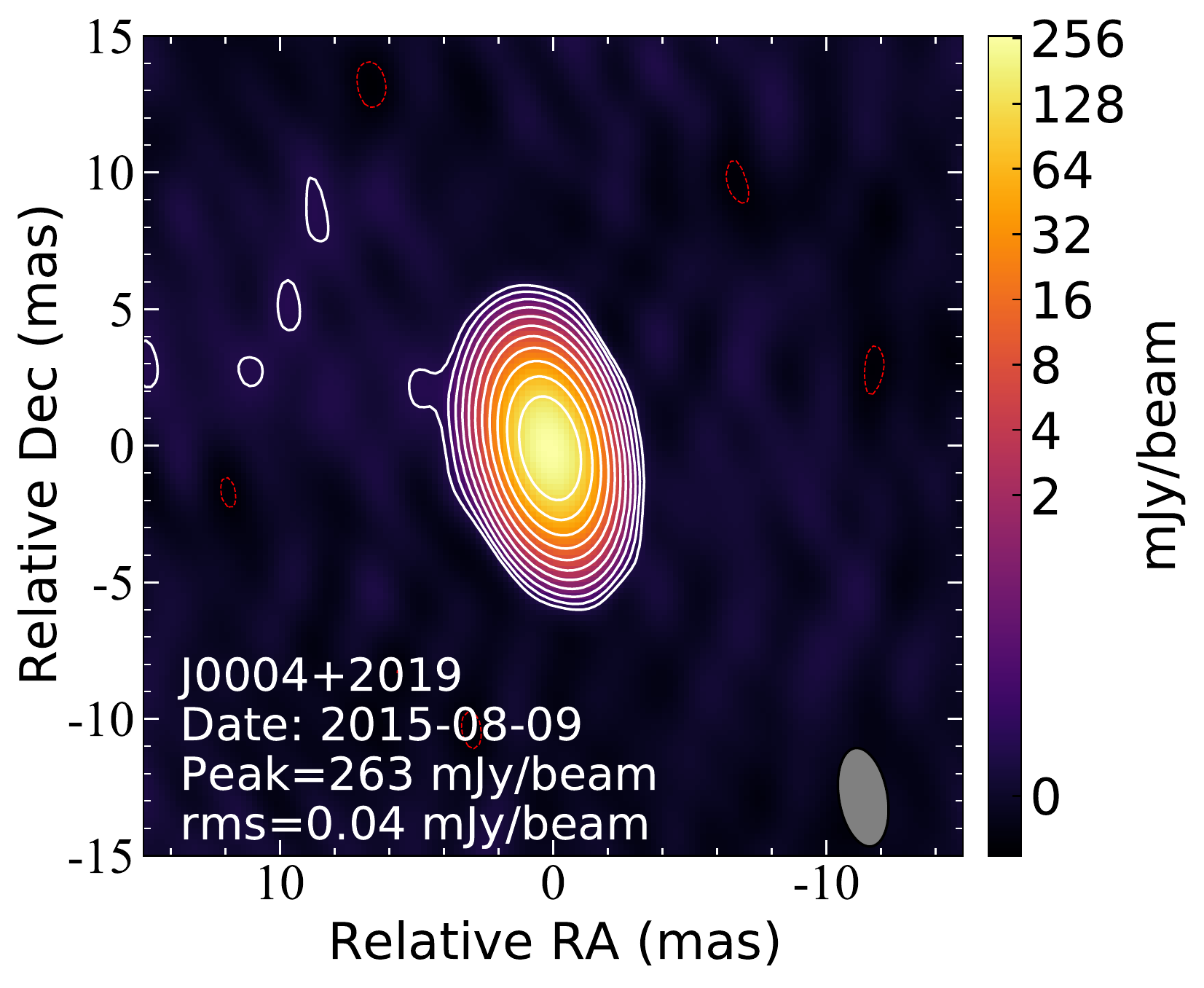}    
  \includegraphics[width=0.24\textwidth]{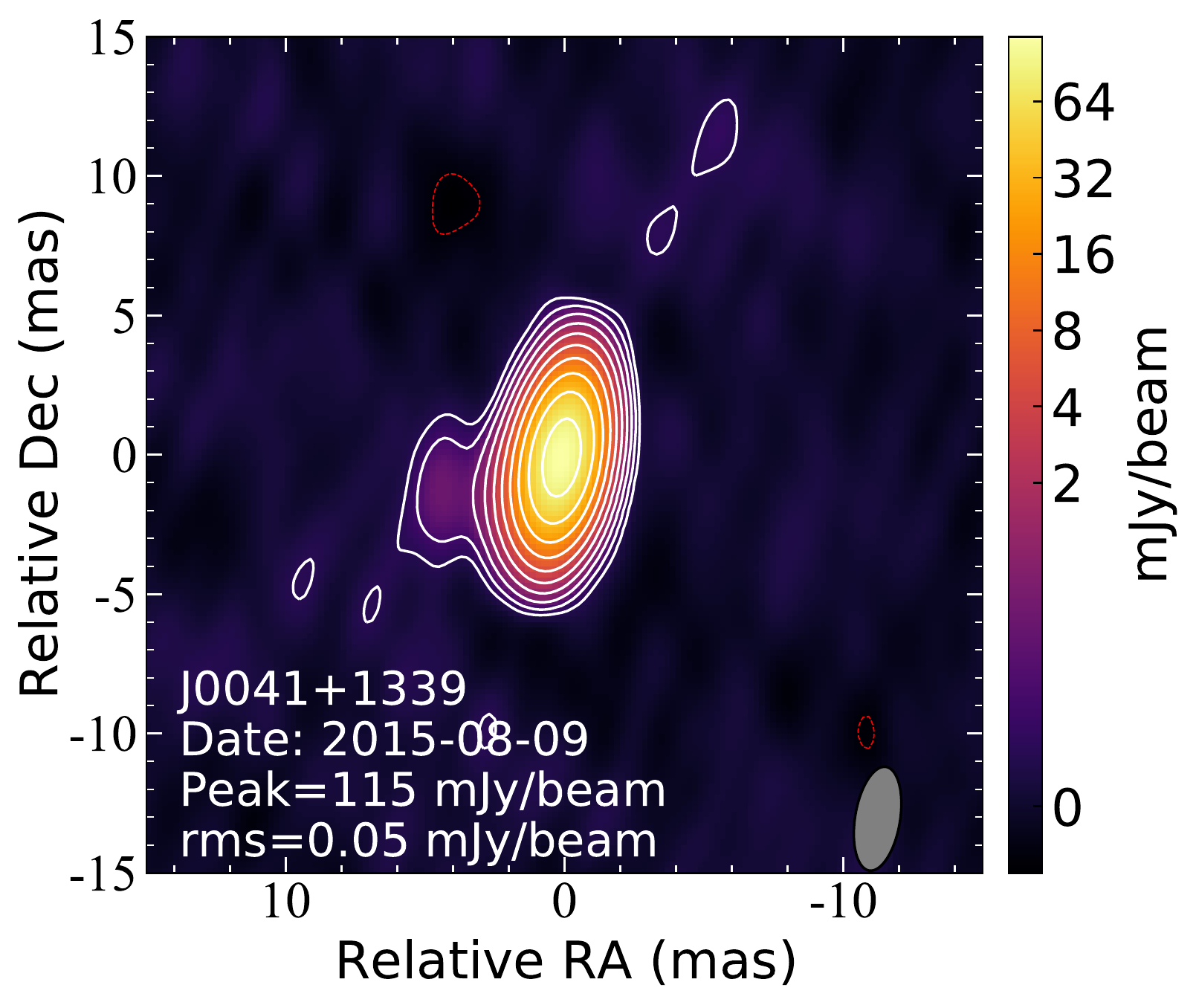}  
  \includegraphics[width=0.24\textwidth]{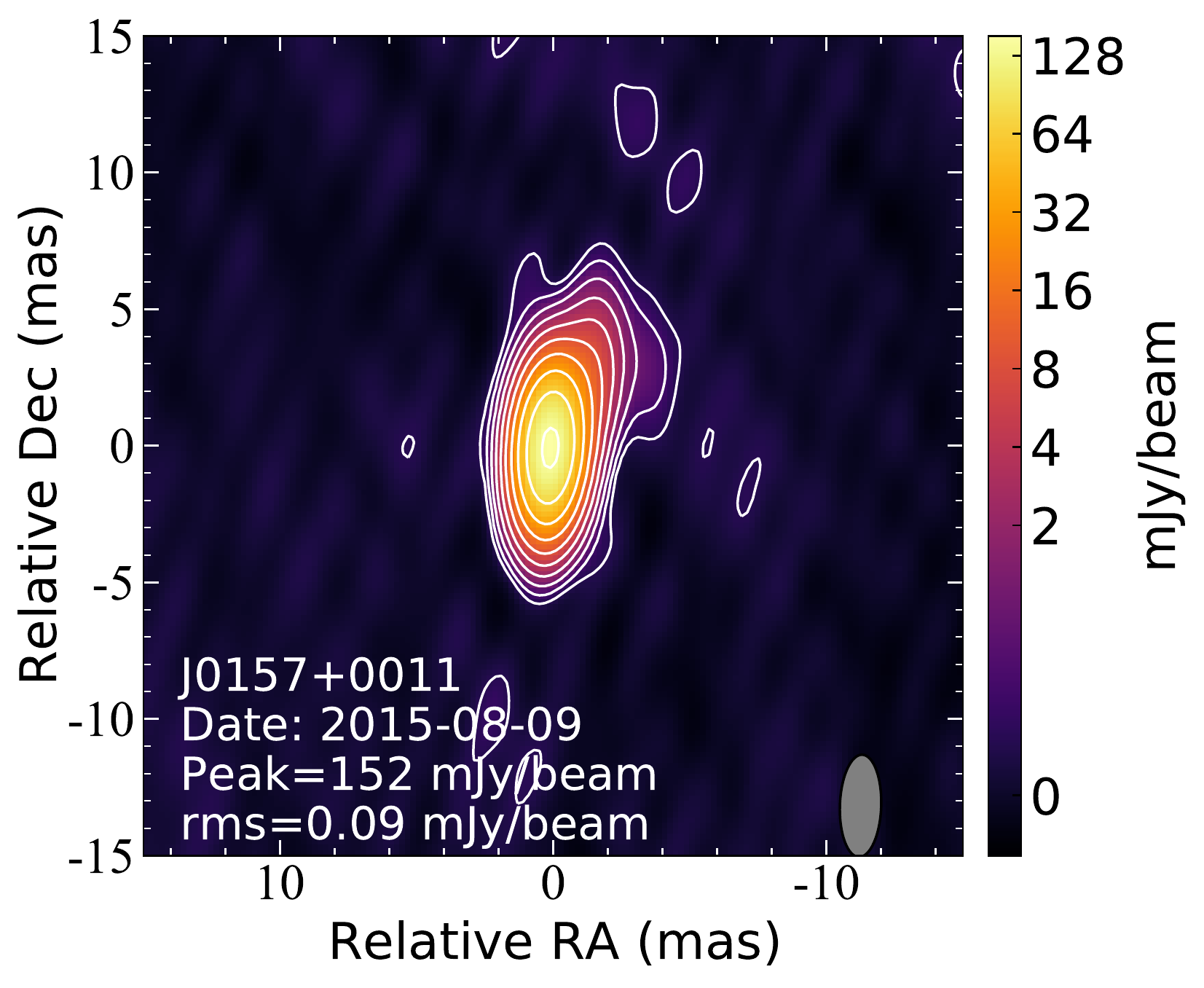} 
  \includegraphics[width=0.24\textwidth]{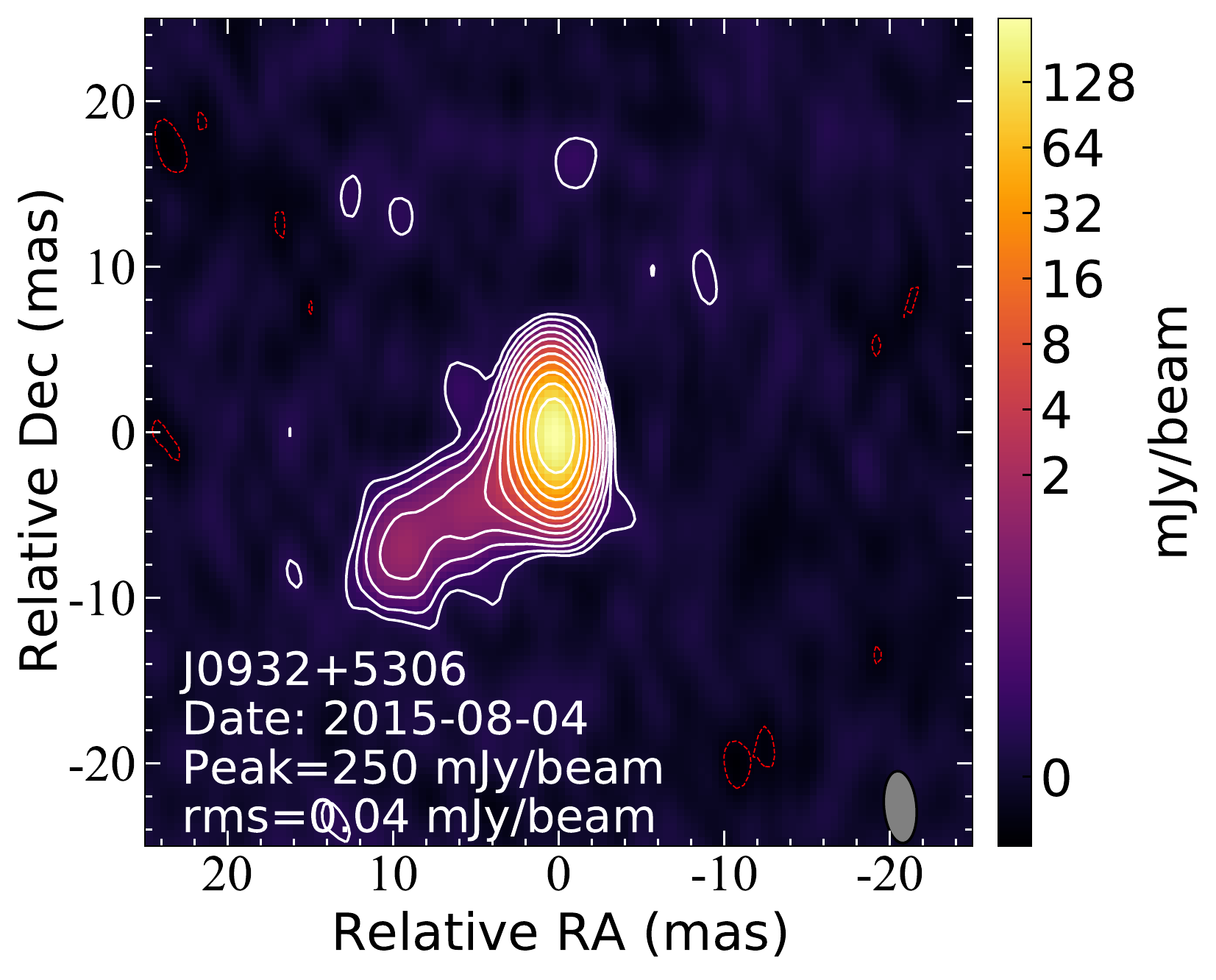} \\  
  \includegraphics[width=0.24\textwidth]{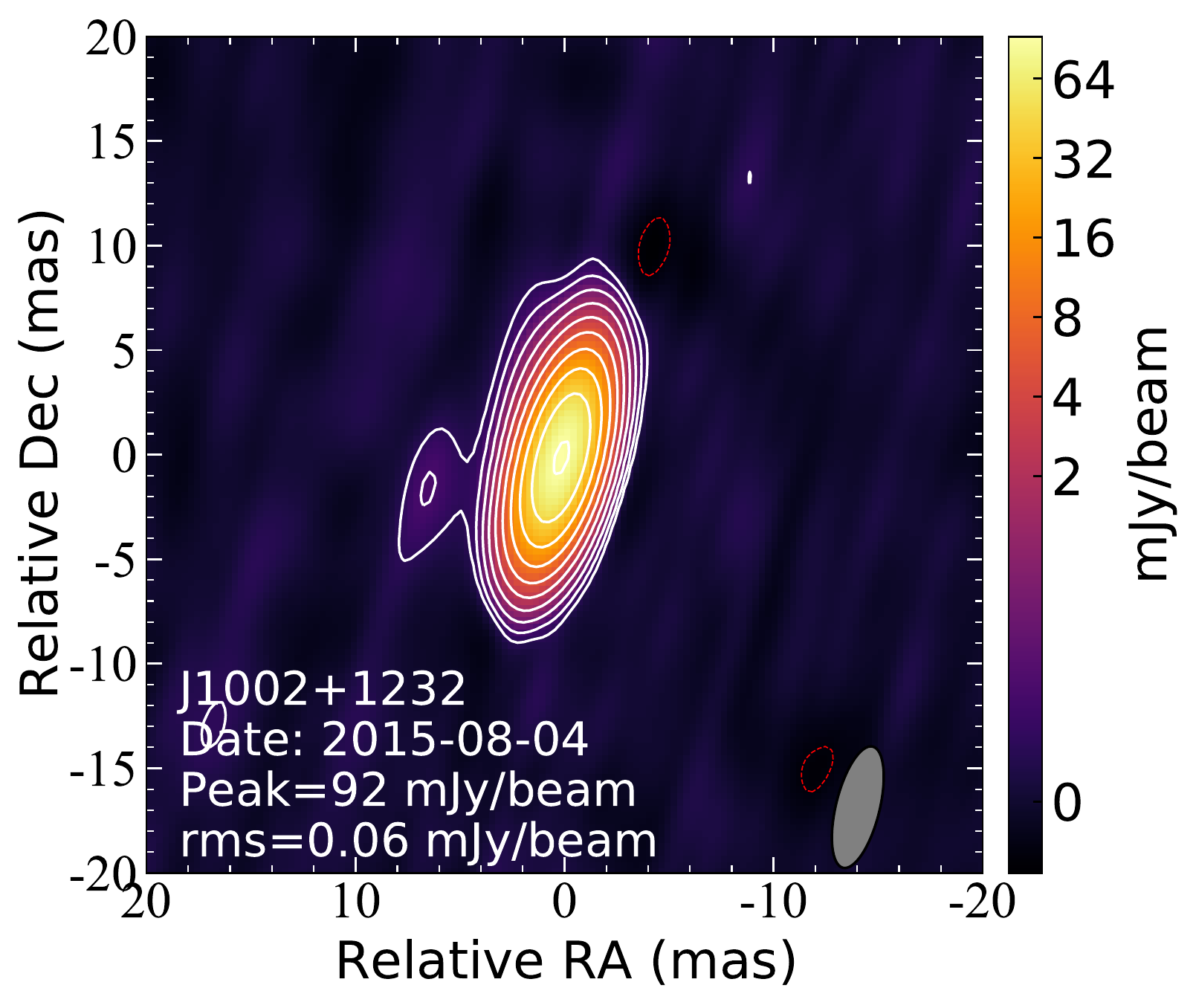} 
  \includegraphics[width=0.24\textwidth]{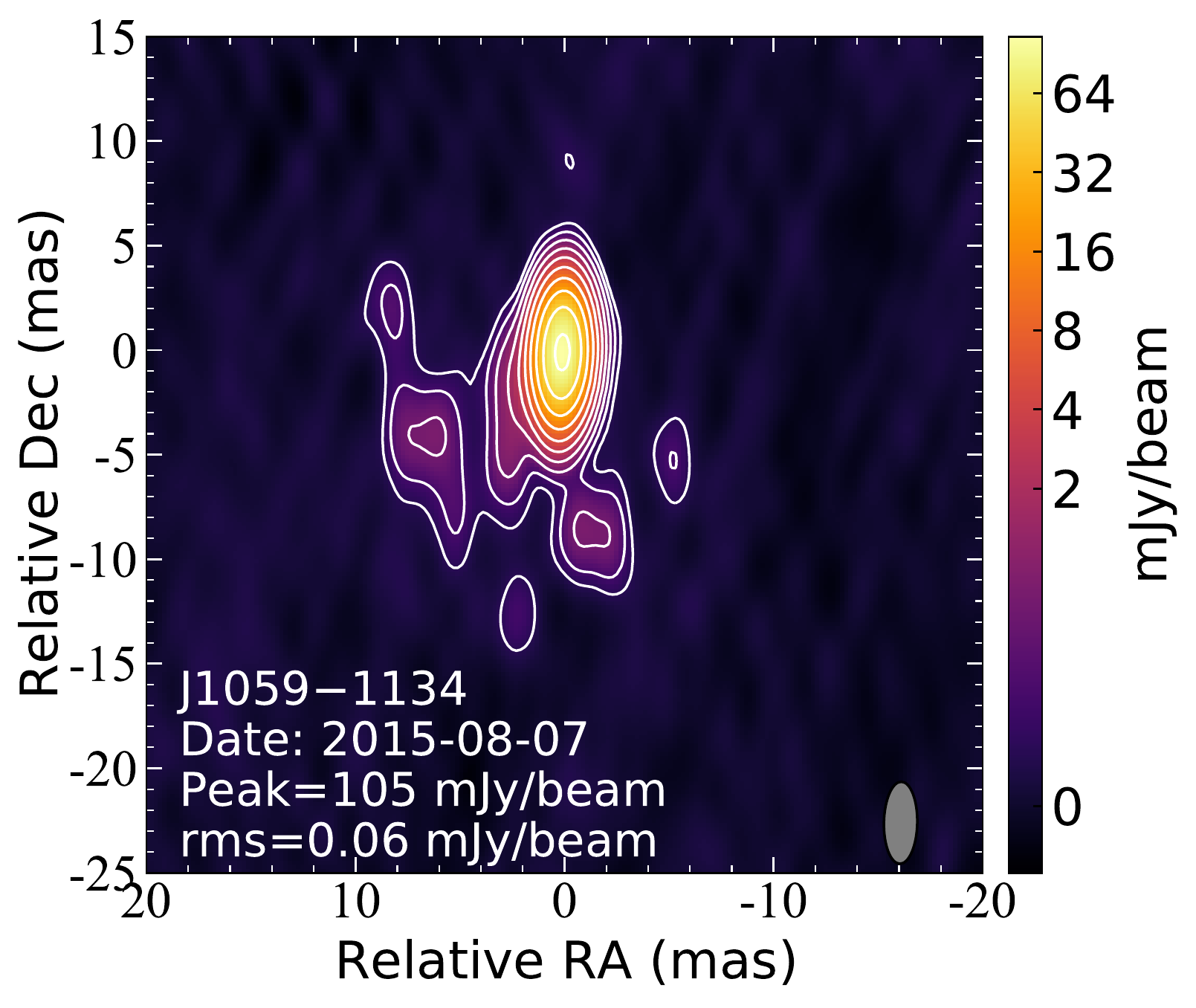}  
  \includegraphics[width=0.24\textwidth]{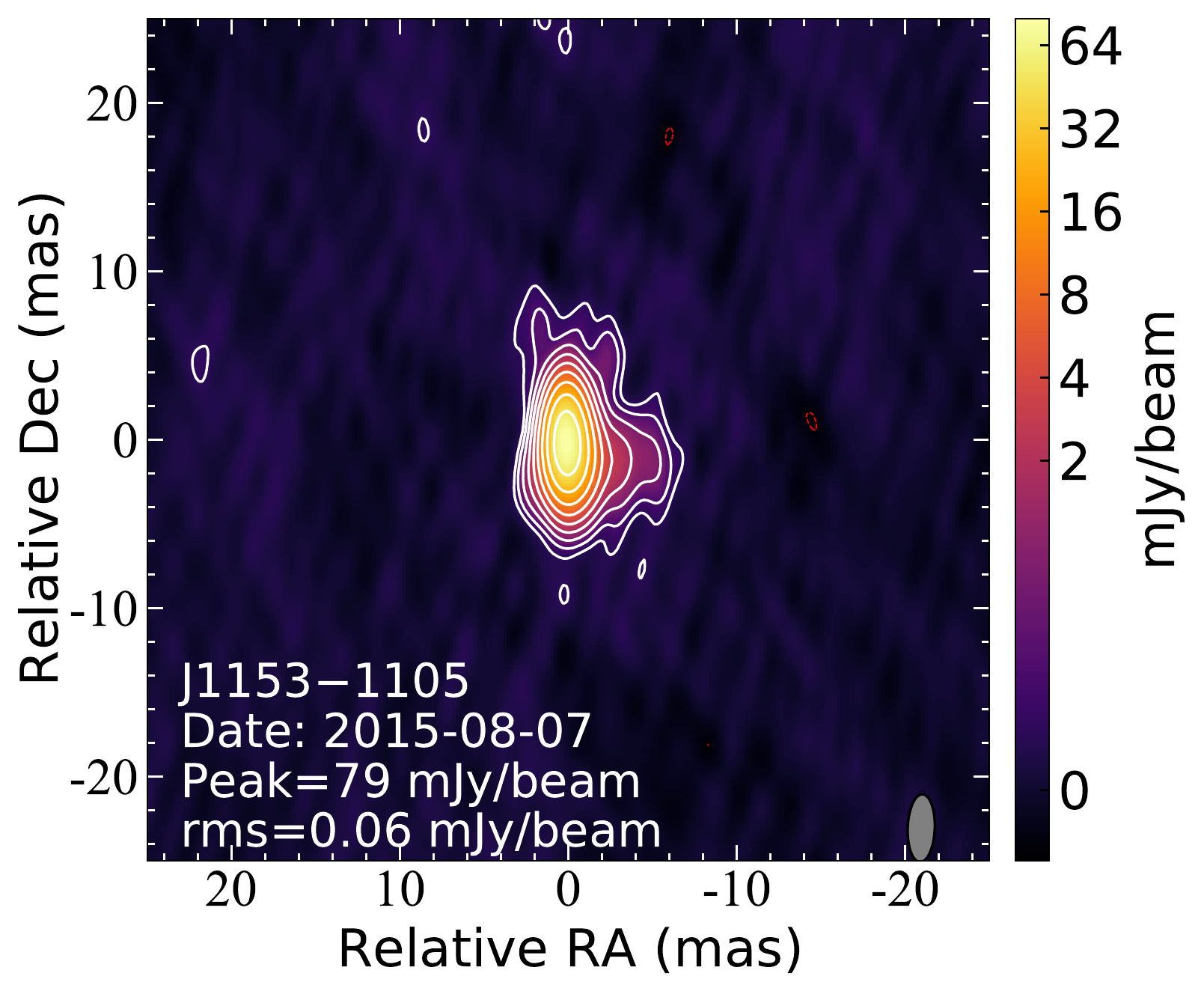}
  \includegraphics[width=0.24\textwidth]{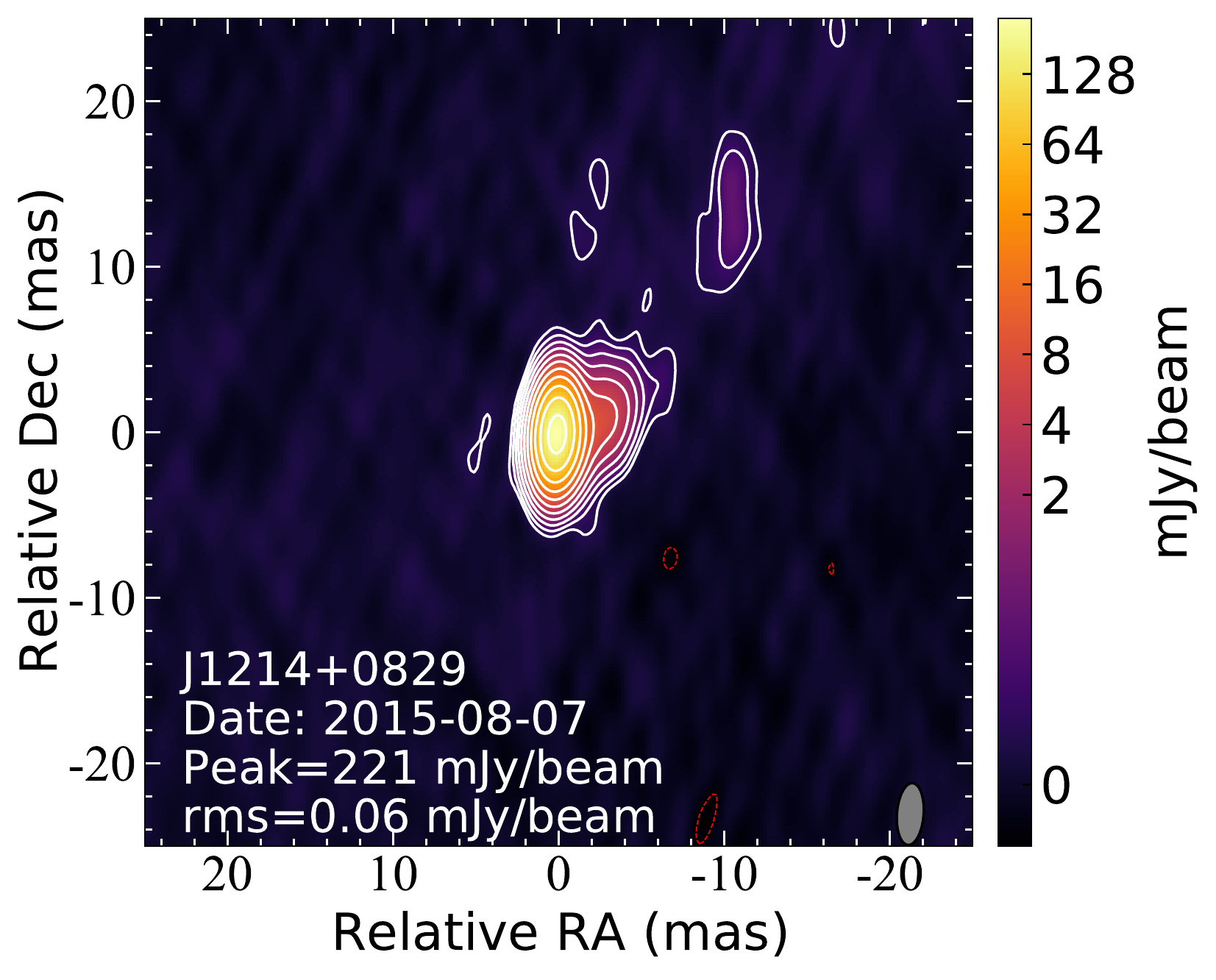} \\ 
  \includegraphics[width=0.24\textwidth]{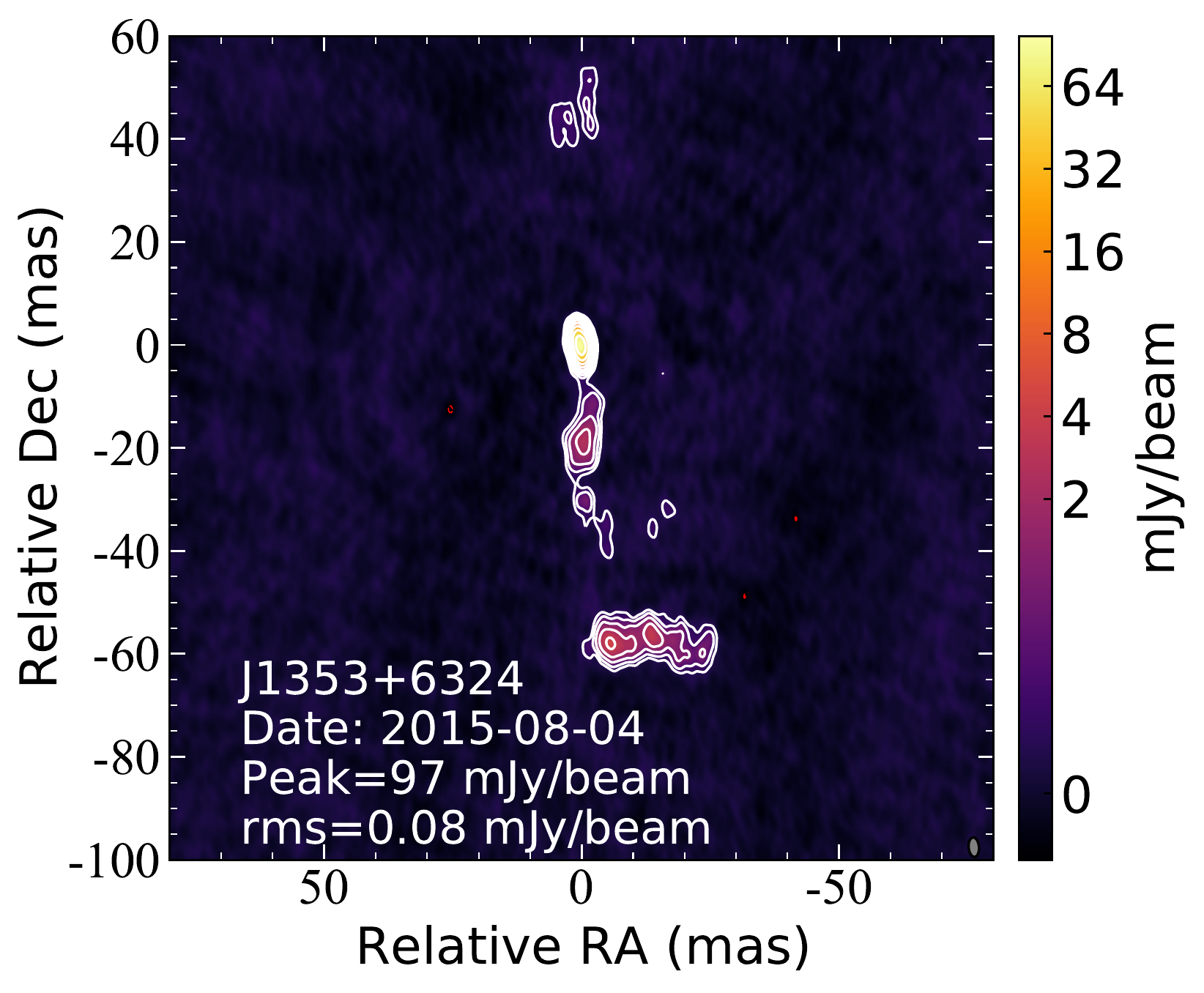} 
  \includegraphics[width=0.24\textwidth]{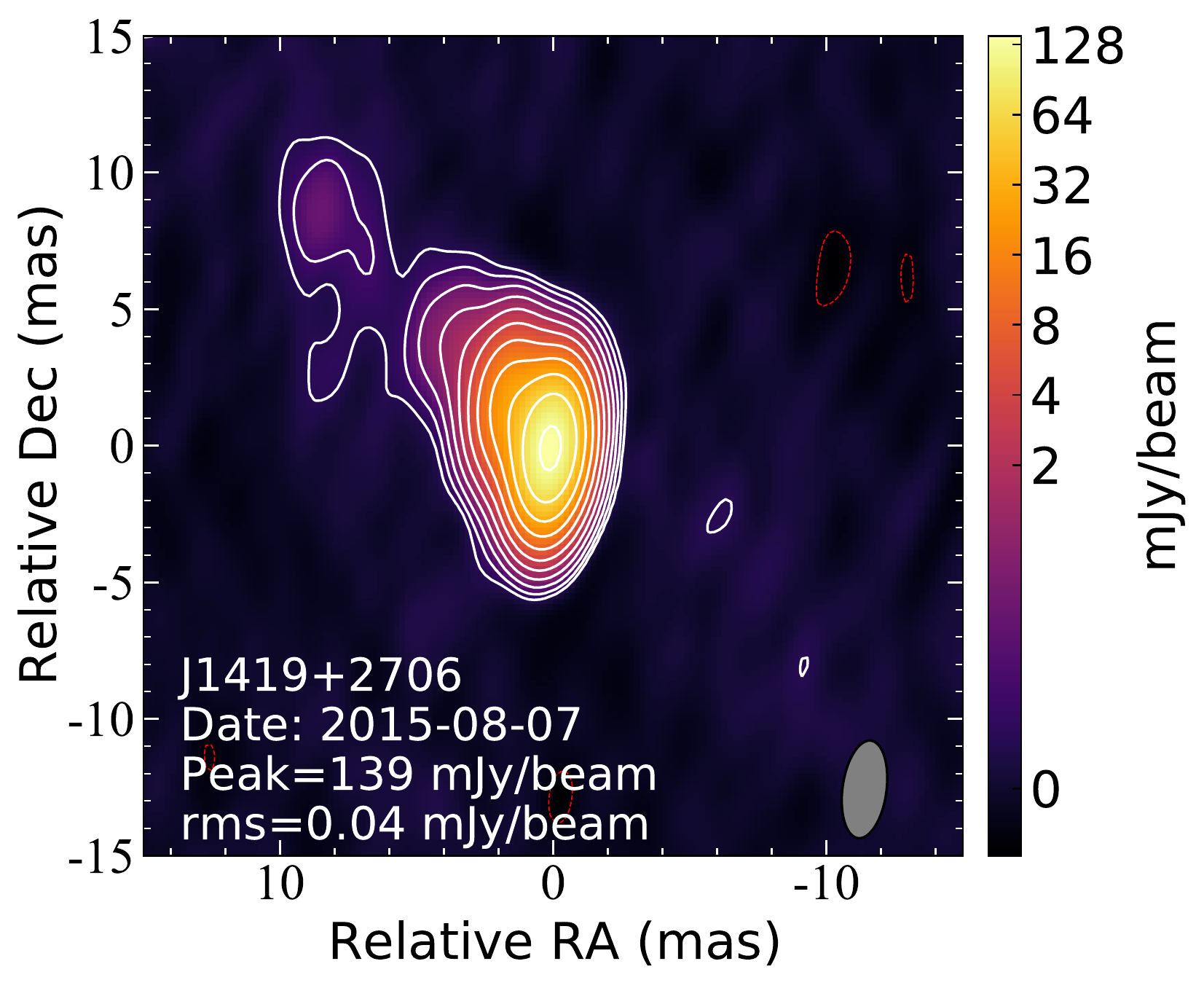}  
  \includegraphics[width=0.24\textwidth]{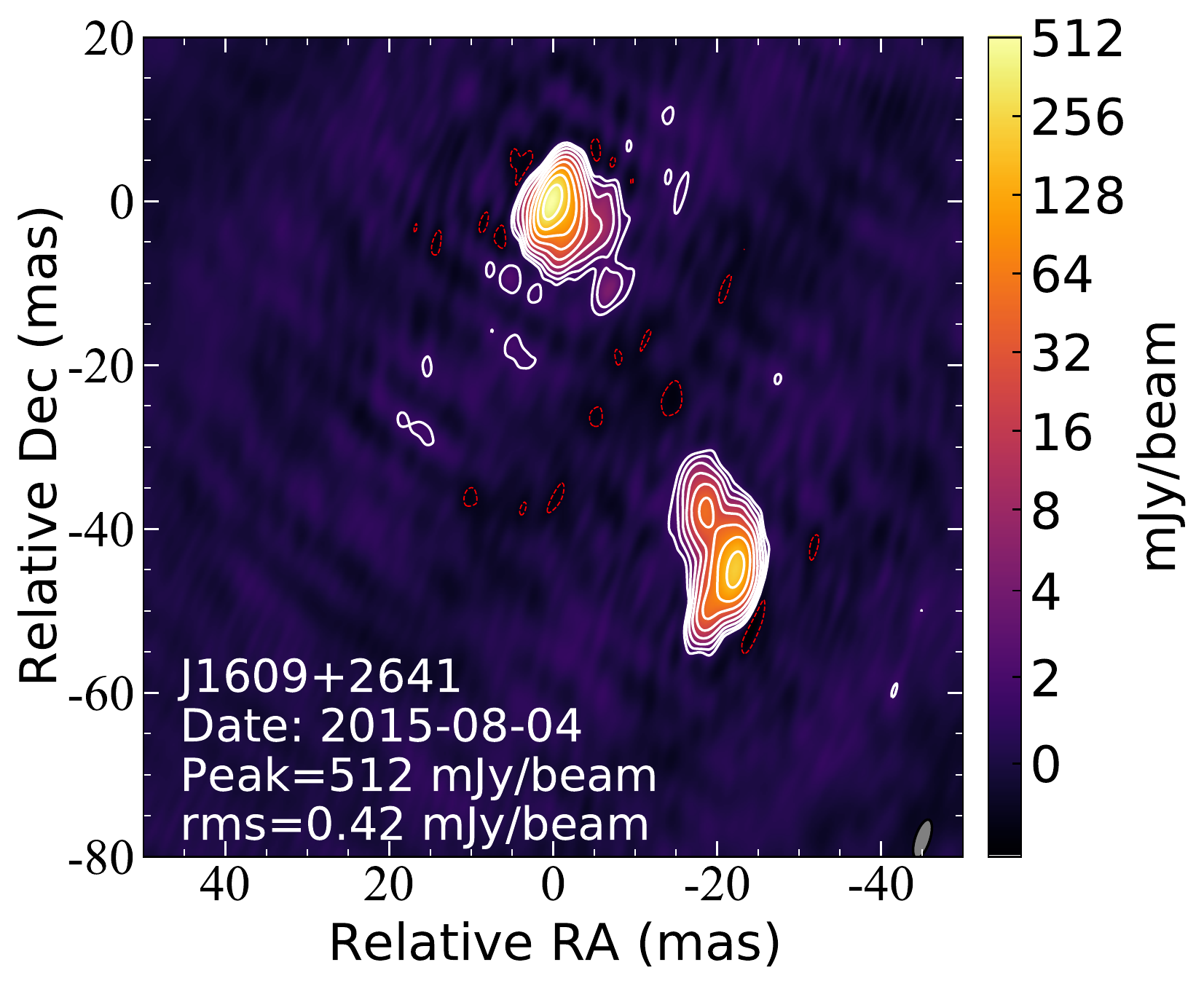}
  \includegraphics[width=0.24\textwidth]{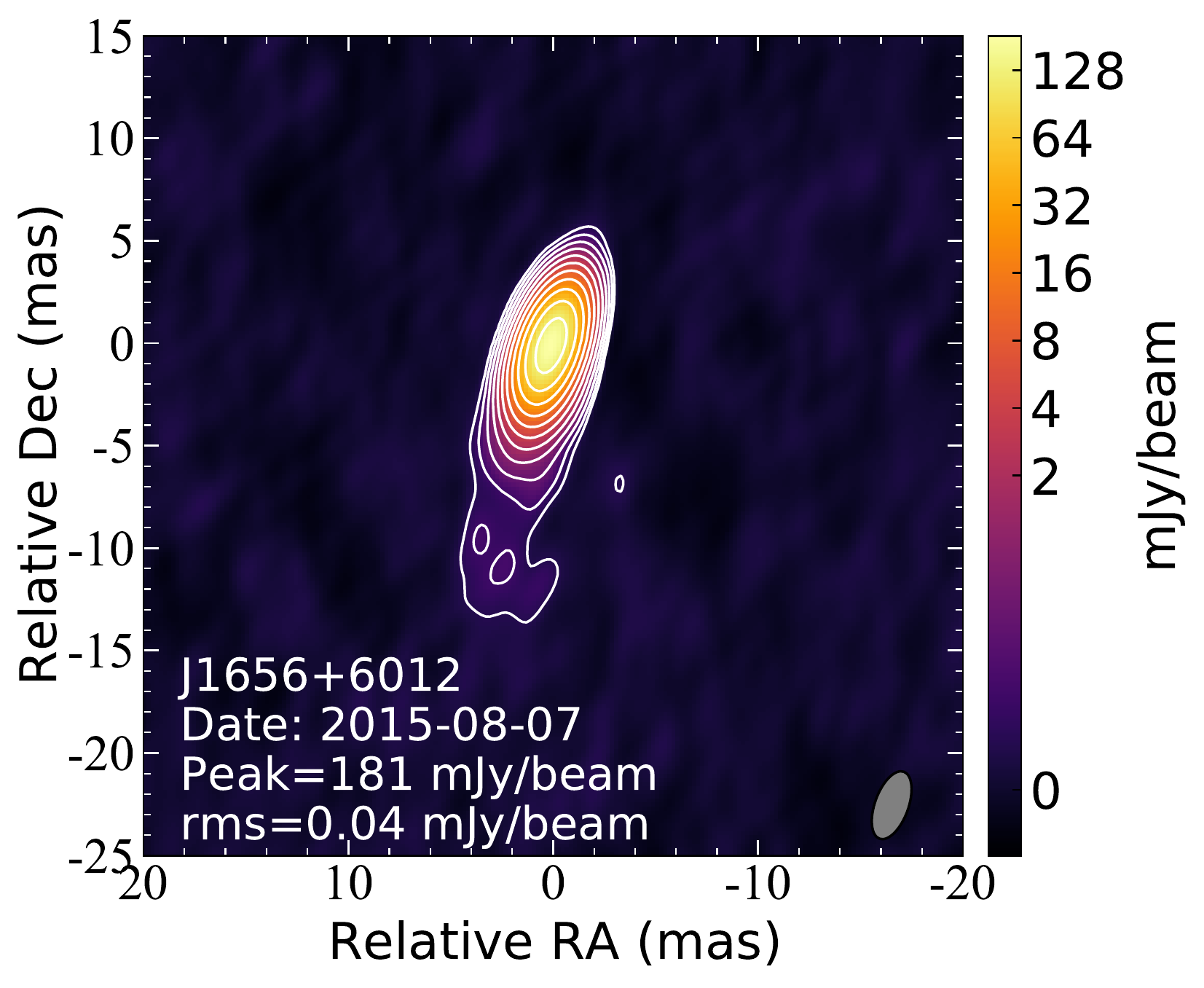} \\ 
  \includegraphics[width=0.24\textwidth]{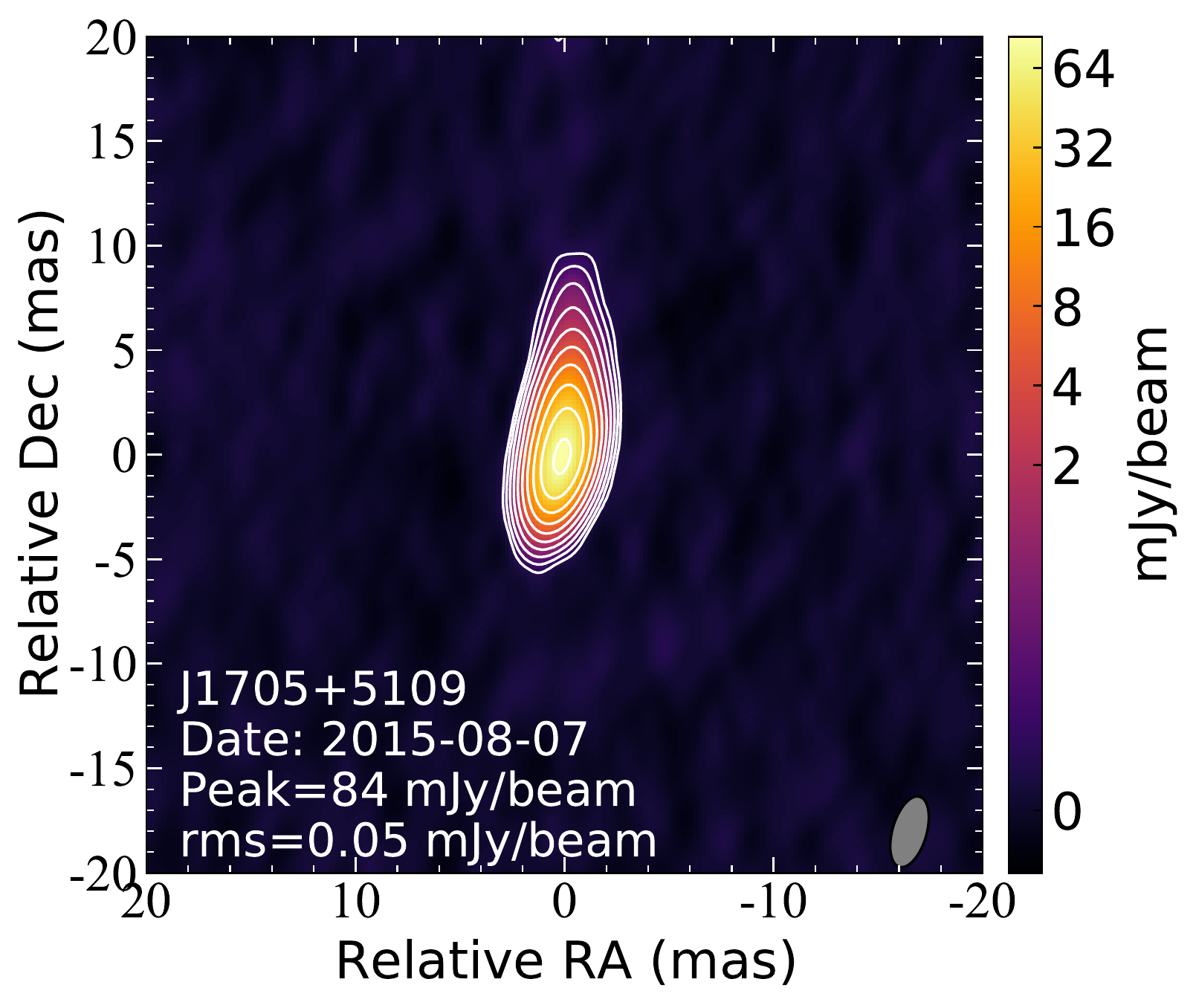} 
  \includegraphics[width=0.24\textwidth]{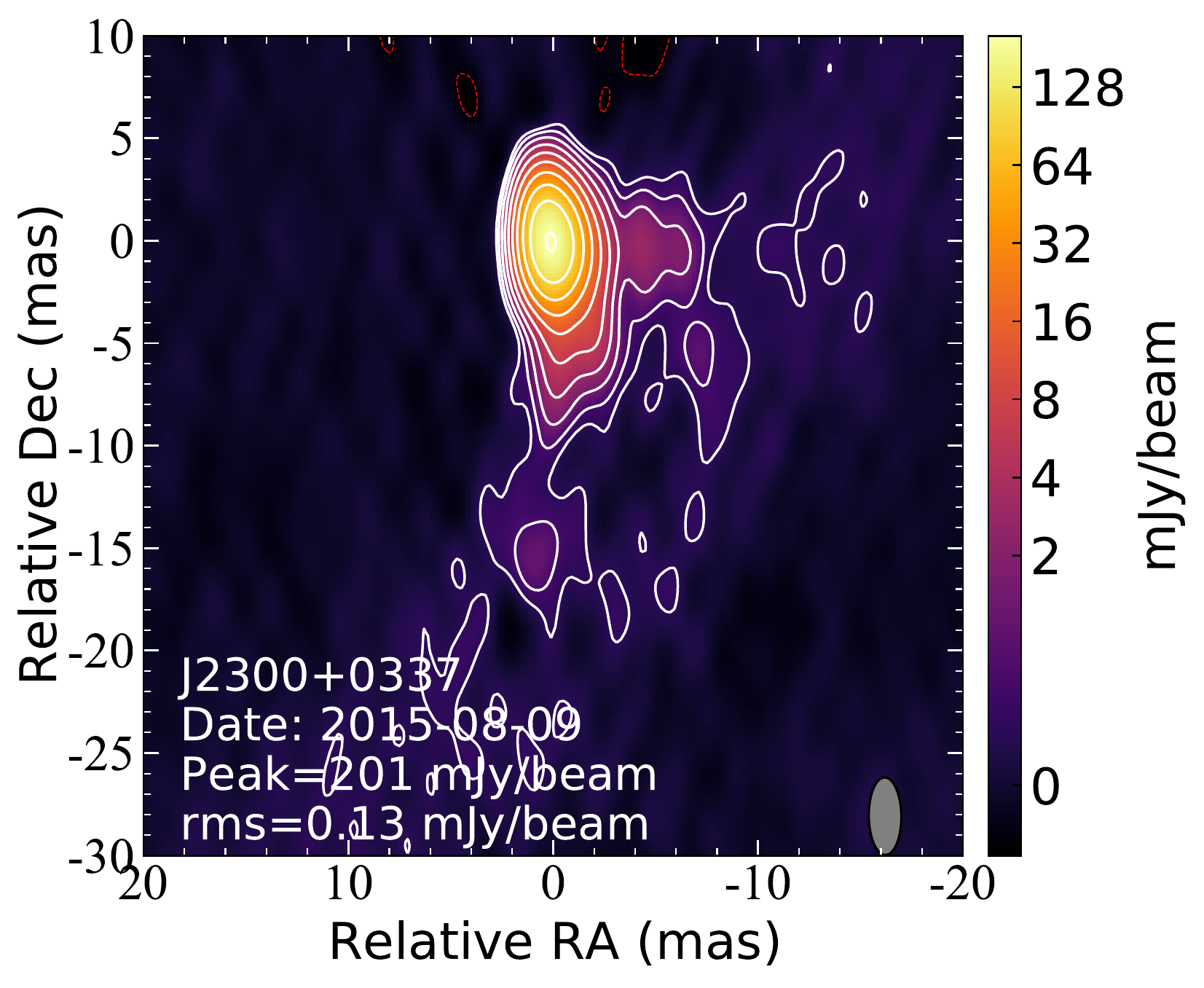} \\
    \end{tabular}
  \centering
  \caption{Images of calibrators in our VLBA observations.}
  \label{fig:CalImage}
\end{figure*}

The visibility data of calibrators were imported into \textsc{Difmap} software package \citep{1997ASPC..125...77S} for self-calibration and imaging.
In \textsc{Difmap}, we followed the hybrid mapping method \citep{1984ARA&A..22...97P} by running a loop of Fourier transform imaging and deconvolution (so-called CLEAN), phase and amplitude self-calibration on the phase calibrators. The time interval of the amplitude self-calibration was gradually reduced from 90~min to 1~min. The final \textsc{CLEAN} images of phase-reference calibrators were obtained as shown in Figure \ref{fig:CalImage}.

The well-calibrated FITS files of the phase calibrators were then loaded back to AIPS and were applied again in a global fringe fitting process. This step may solve the extra phase errors introduced by the core-jet structure of phase calibrators themselves (see Figure \ref{fig:CalImage}).  Except for J1353+6324 and J1609+2641, the rest of the sources all show either an unresolved core or a 'core + one-sided' jet morphology. Their radio structures are consistent with the radio continuum properties of the general flat-spectrum radio-loud quasars or blazars. J1353+6324 displays a complex core and two-sided jet morphology. The southernmost and northernmost jet knots are more than 100 mas apart. The southern jet is much longer and brighter. The southernmost clumps are aligned along a direction perpendicular to the southern jet. The 2.3-GHz VLBI archive image in the astrogeo database\footnote{Astrogeo database maintained by L. Petrov,  \url{http://astrogeo.org}.} reveals a jet extending continuously from the core to the southernmost tip, with a sharp bend  of more than 90\degr\ at the end of the jet. J1609+2641 shows two compact components with an angular separation of about 50 mas between them. In the 5-GHz VLBI image, both components show an edge-brightening brightness distribution and compact morphology, consistent with the definition of Compact Symmetric Object (CSO) source \citep{1982A&A...106...21P,1988ApJ...328..114P,2005ApJ...622..136G,2012ApJS..198....5A}. Due to the complex radio structures of J1353+6324 and J1609+2641, we carefully mapped them and used their CLEAN models as the input model for fringe fitting, which eventually calibrated the phase error well and successfully resulted in the detection of the corresponding target sources (i.e., PG 1351+640 and PG 1612+261). Moreover, the calibration factors for the visibility magnitude were obtained from the calibrator data and applied to the target source data. These amplitude calibration factors are in the range of 0.9--1.1, indicating that the typical uncertainty in the visibility magnitude calibration is about 10\%.

\begin{table*}
\setlength{\tabcolsep}{5pt}
\scriptsize
\caption{Image parameters. }

\begin{sideways}
	\begin{tabular}{cclcclcclcccc}
	\hline \hline
Name         &$z$    &  (R.A., Dec)             &$\Delta_{\rm p}$  &obs date         &Beam                  & S$_{\rm peak}$    &$ \sigma$ & Contours                           &Calibrator         &R      &$S^{\rm VLA}_{\rm D}$  &$S^{\rm VLA}_{\rm A}$ \\
             &       & (J2000)                  & (mas)            &yy-mm-dd         &(mas$\times$mas,deg)  &(\mJyb)            & (\mJyb)  &                                    &                   &       &mJy            &mJy           \\
 (1)         &(2)   & (3)                       & (4)              & (5)             &(6)                   &(7)                & (8)      &(9)                                 &(10)               &(11)   &(12)           &(13)          \\

\hline                                                    
\multicolumn{12}{c}{radio-quiet quasars} \\                
PG 0003+199   &0.026     &00:06:19.5373, 20:12:10.617     &0.73     &2015-08-09       &4.88$\times$2.77, $-$7.35   &0.21     &0.025 & $3\sigma \times (1, \sqrt{2}, 2, 2\sqrt{2})$      &J0004+2019    &0.27 &3.92          &3.03          \\
PG 0050+124   &0.061     &00:53:34.9340, 12:41:35.927     &11.41    &2015-08-09       &3.89$\times$1.86, $-$7.16   &0.12     &0.021 & $1\sigma \times (3, 4, 5)$                        &J0041+1339    &0.33 &2.60          &1.80          \\
PG 0157+001   &0.163     &01:59:50.2542, 00:23:40.869     &24.08    &2015-08-09       &3.61$\times$1.45, $-$1.88   &0.17     &0.023 & $1\sigma \times (3, 4, ..., 7, 8)$                &J0157+0011    &2.12 &8.00          &5.58          \\
PG 0921+525   &0.035     &09:25:12.8478, 52:17:10.386     &0.86     &2015-08-04       &4.72$\times$2.01, 8.71      &0.83     &0.027 & $3\sigma \times (1, \sqrt{2}, ..., 4\sqrt{2}, 8)$ &J0932+5306    &1.49 &3.80          &1.87          \\
PG 0923+129   &0.029     &09:26:03.2695, 12:44:03.731     &...      &2015-08-07       &4.83$\times$2.37, $-$16     &...      &0.027 &...                                                &J0921+1350    &2.07 &10            &...           \\    
PG 1116+215   &0.177     &11:19:08.6786, 21:19:17.988     &...      &2015-08-04       &4.44$\times$1.86, $-$31     &...      &0.026 &...                                                &J1119+2226    &0.72 &2.80          &1.94          \\
PG 1149$-$110 &0.049     &11:52:03.5504, $-$11:22:24.090  &3.80     &2015-08-07       &4.51$\times$1.84, $-$8.39   &0.29     &0.033 & $3\sigma \times (1, \sqrt{2}, 2, 2\sqrt{2}) $     &J1153$-$1105  &0.88 &2.60          &1.00          \\
PG 1211+143   &0.081     &12:14:17.6739, 14:03:13.183     &...      &2015-08-04       &4.99$\times$2.45, $-$21     &...      &0.025 &...                                                &J1213+1307    &0.13 &0.80          &...           \\
PG 1216+069   &0.331     &12:19:20.9318, 06:38:38.468     &0.69     &2015-08-07       &3.72$\times$1.58, $-$3.45   &0.94     &0.028 & $3\sigma \times (1, \sqrt{2}, ..., 8, 8\sqrt{2})$ &J1214+0829    &1.65 &4.00          &4.95          \\
PG 1351+640   &0.088     &13:53:15.8313, 63:45:45.684     &1.48     &2015-08-04       &3.75$\times$1.95, 6.21      &2.50     &0.029 & $3\sigma \times (1,  2,  4,  8, 16) $             &J1353+6324    &4.32 &13            &20            \\
PG 1448+273   &0.065     &14:51:08.7648, 27:09:26.962     &...      &2015-08-04       &4.58$\times$2.01, $-$37     &...      &0.027 &...                                                &J1453+2648    &0.25 &1.01          &1.09          \\
PG 1534+580   &0.031     &15:35:52.4030, 57:54:09.514     &...      &2015-08-07       &3.98$\times$1.76, $-$51     &...      &0.025 &...                                                &J1551+5806    &0.70 &1.92          &1.80          \\
PG 1612+261   &0.131     &16:14:13.2050, 26:04:16.225     &2.15     &2015-08-04       &5.02$\times$1.73, $-$18     &0.17     &0.023 & $3\sigma \times (1, \sqrt{2}, 2) $                &J1609+2641    &2.81 &5.07          &1.66          \\
PG 1700+518   &0.292     &17:01:24.8264, 51:49:20.451     &3.46     &2015-08-07       &3.75$\times$1.85, $-$1.85   &1.06     &0.026 & $3\sigma \times (1, \sqrt{2}, ..., 8, 8\sqrt{2})$ &J1705+5109    &2.36 &7.20          &2.05          \\
PG 2130+099   &0.063     &21:32:27.8163, 10:08:19.252     &...      &2015-08-09       &6.16$\times$2.25, $-$13     &...      &0.028 &...                                                &J2130+0843    &0.32 &2.05          &1.30          \\
PG 2304+042   &0.042     &23:07:02.9147, 04:32:57.102     &0.49     &2015-08-09       &3.83$\times$1.68, $-$3.96   &0.46     &0.028 & $3\sigma \times (1, \sqrt{2}, ... 4, 4\sqrt{2}) $ &J2300+0337    &0.25 &0.77          &1.10          \\
\hline                                                    
\multicolumn{12}{c}{radio-loud quasars from this paper} \\                 
PG 1004+130   &0.241     &10:07:26.0966, 12:48:56.185     &0.80     &2015-08-04       &6.17$\times$2.57, $-$12.8   &27       &0.028 & $3\sigma \times (1, 2, ...,  128, 256)$    &J1002+1232           &228  &440          &27             \\
PG 1048$-$090 &0.345     &10:51:29.9161, $-$09:18:10.192  &1.34     &2015-08-07       &3.75$\times$1.63, $-$2.58   &40       &0.029 & $3\sigma \times (1, 2, ...,  128, 256)$    & J1059$-$1134        &377  &680          &53             \\
PG 1425+267   &0.364     &14:27:35.6075, 26:32:14.541     &0.12     &2015-08-07       &3.53$\times$1.55, $-$6.20   &22       &0.023 & $3\sigma \times (1, 2, ..., 128, 256)$     & J1419+2706          &53   &131          &41             \\
PG 1704+608   &0.372     &17:04:41.3744, 60:44:30.520     &0.57     &2015-08-07       &3.70$\times$1.92, $-$27.03  &6.41     &0.025 & $3\sigma \times (1, \sqrt{2}, ..., 8, 8\sqrt{2})$   & J1656+6012 &645  &1230         &8.00           \\ 
\hline  
\multicolumn{12}{c}{radio-loud quasars from archive} \\ 
PG 0003+158	  &0.450     &00:05:59.2378, 16:09:49.022 	 &0.48      &2017-03-27       &1.93$\times$0.88, 90        &117       &0.20 & $3\sigma \times (1, 2, ..., 32, 64)$   &...                      &180  &329	        &130              \\
PG 0007+107   &0.089     &00:10:31.0059, 10:58:29.504 	 &0.58      &2015-08-31       &4.59$\times$1.93, 93        &143       &0.12 & $3\sigma \times (1, 2, ..., 32, 64)$&...                         &200  &321	        &151               \\
PG 1100+772	  &0.313     &11:04:13.8611, 76:58:58.195 	 &0.57      &2016-09-14       &1.49$\times$1.05, 63        &72        &0.29 & $3\sigma \times (1, 2, ..., 32, 64)$&...                         &320  &660	        &76                \\
PG 1103$-$006 &0.425     &11:06:31.7743, $-$00:52:52.381 &0.75      &2013-04-21       &5.70$\times$2.49, 80        &96        &0.28 & $3\sigma \times (1, 2, ..., 32, 64)$ &...                        &270  &482	        &76                \\	
PG 1226+023	  &0.158     &12:29:06.6997, 02:03:08.598 	 &1.94      &2007-04-30       &3.75$\times$1.54, 89        &6358      &28   & $3\sigma \times (1, 2, ... 4096)$ &...                           &1100 &37000	    &26400             \\
PG 1302$-$102 &0.286     &13:05:33.0150, $-$10:33:19.428 &0.31      &2019-08-20       &3.90$\times$2.10, 72        &448       &0.45 & $3\sigma \times (1, 2, ..., 256, 512)$  &...                     &190  &780	        &780               \\
PG 1309+355   &0.184     &13:12:17.7527, 35:15:21.086    &0.69      &2017-05-27       &1.80$\times$1.01, 82        &45        &0.14 & $3\sigma \times (1, 2, ...,16, 32)$  &...                        &18   &54	        &51                \\ 
PG 1512+370	  &0.371     &15:14:43.0683, 36:50:50.357 	 &0.93      &2017-09-06       &2.27$\times$1.03, 112       &42        &0.36 & $3\sigma \times (1, 2, 4, 8, 16)$    &...                        &190  &352	        &62                \\
PG 1545+210	  &0.266     &15:47:43.5378, 20:52:16.614 	 &0.86 	    &2018-12-01       &4.55$\times$1.69, 82        &29        &0.23 & $3\sigma \times (1, 2, 4, 8, 16)$    &...                        &420  &720	        &32                \\ 
PG 2209+184	  &0.070     &22:11:53.8889, 18:41:49.862 	 &0.14      &2015-01-23       &1.90$\times$0.99, 90        &195       &0.20 & $3\sigma \times (1, 2, ..., 64, 128)$ &...                       &54   &290	        &280               \\
PG 2251+113	  &0.320     &22:54:10.4214, 11:36:38.748 	 &0.84      &2017-01-31       &2.80$\times$1.05, 110       &38        &0.31 & $3\sigma \times (1, 2, ..., 32, 64)$   &...                      &370  &523	        &2                 \\
PG 2308+099	  &0.432     &23:11:17.7545, 10:08:15.757 	 &0.45      &2012-05-27       &1.64$\times$0.81, 91        &85        &0.19 & $3\sigma \times (1, 2, ..., 32, 64)$   &...                      &190  &303	        &87                \\
\hline                      
	\end{tabular}
 \label{tab:image} 
\end{sideways} 
\end{table*}


\begin{table*}
\centering \setlength{\tabcolsep}{4pt}
\caption{Radio properties of the PG  quasar sample in this paper. Column (1) source name,  (2) total flux density observed with VLBA at 5 GHz, (3) the flux density ratio $f_{\rm a}={S^{\rm VLBA}}/{S_{A}^{\rm VLA}} $, $S_{A}^{\rm VLA}$ measured by VLA A-array from \citet{1989AJ.....98.1195K},  (4) the flux density of the core component, (5) component size (full width at half-maximum of the fitted Gaussian component), (6) brightness temperature of the core component, (7) monochromatic radio luminosity of VLBA at 5 GHz.} 
	\begin{tabular}{lcccccc}
	\hline 
Name           &$S^{\rm VLBA}$      &$f_{a}$      & $S_{\rm core}^{\rm VLBA}$ & $\theta_{\rm core}^{\rm VLBA}$  &log($T_{\rm b}$)  &log(P$^{\rm VLBA}$)    \\         
              &(mJy)                          &             & (mJy)           & (mas)                 & log(K)                     &log(W Hz$^{-1}$)  \\
(1)           &(2)                            &(3)          &(4)              &(5)                    &(6)                         &(7)                \\    
\hline 
\multicolumn{6}{c}{VLBI-detected radio-quiet quasars} \\     
PG 0003+199            &1.06                           &0.35         &$0.23\pm0.02$    &$1.11\pm0.33$          &$6.97\pm0.60$       &21.20               \\ 
PG 0050+124$^{*}$      &0.55                           &0.31         &...              &...                    &...                 &21.67               \\ 
PG 0157+001$^{*}$      &0.99                           &0.18         &...              &...                    &...                 &22.82                 \\ 
PG 0921+525            &1.05                           &0.56         &$1.04\pm0.10$    &$0.93\pm0.06$          &$7.78\pm0.16$       &21.45               \\ 
PG 1149$-$110          &0.33                           &0.33         &$0.32\pm0.03$    &$<0.47$                &$>7.87$             &21.25               \\ 
PG 1216+069            &1.25                           &0.25         &$1.20\pm0.12$    &$0.77\pm0.05$          &$8.12\pm0.16$       &23.59               \\
PG 1351+640            &5.32                           &0.27         &$1.98\pm0.20$    &$0.28\pm0.02$          &$9.13\pm0.19$       &22.98               \\
PG 1612+261            &0.19                           &0.11         &$0.19\pm0.02$    &$<0.56$                &$>7.51$             &21.90               \\
PG 1700+518            &1.35                           &0.66         &$1.17\pm0.12$    &$0.64\pm0.04$          &$8.25\pm0.15$       &23.50               \\
PG 2304+042            &0.58                           &0.53         &$0.54\pm0.05$    &$0.58\pm0.07$          &$7.91\pm0.27$       &21.36               \\ 

\hline 
\multicolumn{6}{c}{VBI-non-detected radio-quiet quasars} \\
PG 0923+129            &$<0.081$                      &$<0.01$       &...              &...                    &...                 &...              \\
PG 1116+215            &$<0.078$                      &$<0.04$       &...              &...                    &...                 &...              \\
PG 1211+143            &$<0.075$                      &$<0.09$       &...              &...                    &...                 &...              \\
PG 1448+273            &$<0.081$                      &$<0.06$       &...              &...                    &...                 &...              \\
PG 1534+580            &$<0.075$                      &$<0.04$       &...              &...                    &...                 &...              \\
PG 2130+099            &$<0.084$                      &$<0.06$       &...              &...                    &...                 &...              \\
\hline 
\multicolumn{6}{c}{radio-loud quasars from this paper} \\   
PG 1004+130            &29.46                         &1.09         &$27.80\pm2.78$         &$0.12\pm0.01$          &$11.08\pm0.11$      &24.66       \\     
PG 1048$-$090          &52.30                         &0.99         &$29.54\pm2.95$         &$0.22\pm0.03$          &$10.62\pm0.29$      &25.25       \\ 
PG 1425+267            &33.75                         &0.82         &$23.01\pm2.30$         &$0.69\pm0.02$          &$ 9.51\pm0.11$      &25.11       \\ 
PG 1704+608            &7.21                          &0.90         &$7.11\pm0.71$          &$1.58\pm0.06$          &$ 8.28\pm0.13$      &24.46       \\

\hline
\multicolumn{6}{c}{radio-loud quasars from archive} \\   

PG 0003+158$^{x}$ 	  &152.31                         &1.17         &116.06$\pm$11.61        &0.27$\pm$0.02          &10.58$\pm$0.17      &25.97               \\
PG 0007+107           &144.43                         &0.96         &144.43$\pm$14.44        &0.13$\pm$0.01          &11.76$\pm$0.20      &24.43               \\
PG 1100+772$^{x}$	  &64.82                          &0.85         &72.66$\pm$7.27          &0.24$\pm$0.02          &10.42$\pm$0.19      &25.25               \\
PG 1103$-$006   	  &125.36                         &1.66         &106.58$\pm$10.66        &1.04$\pm$0.12          &9.97$\pm$0.25       &25.83               \\
PG 1226+023	          &23701.53                       &0.90         &4694.79$\pm$469.48      &$<0.08$                &$>13.75$            &27.16               \\
PG 1302$-$102         &446.98                         &0.57         &445.98$\pm$44.60        &$<0.04$                &$>13.32$            &26.00               \\
PG 1309+355$^{x}$     &47.47                          &0.93         &46.32$\pm$4.63          &0.14$\pm$0.02          &10.64$\pm$0.27      &24.61               \\
PG 1512+370$^{x}$     &54.14                          &0.88         &47.91$\pm$4.79          &0.51$\pm$0.09          &9.62$\pm$0.36       &25.33               \\
PG 1545+210	          &36.96                          &1.16         &26.78$\pm$2.68          &0.14$\pm$0.02          &11.03$\pm$0.33      &24.85               \\
PG 2209+184$^{x}$	  &175.01                         &0.63         &189.29$\pm$18.93        &0.19$\pm$0.02          &10.96$\pm$0.23      &24.29               \\
PG 2251+113$^{x}$	  &45.27                          &22.64        &40.94$\pm$4.09          &0.38$\pm$0.06          &9.79$\pm$0.34       &25.11               \\
PG 2308+099$^{x}$     &84.76                          &0.97         &87.33$\pm$8.73          &0.26$\pm$0.02          &10.49$\pm$0.20      &25.68               \\
\hline
	\end{tabular} \\
 \label{tab:radio}
Note:  The core is identified as the VLBI component close to the optical nucleus position. For those without 5 GHz VLBA data, we use 8.7-GHz VLBA images instead to obtain their radio structure; they are marked with $x$. 
Their total flux densities at 5 GHz (Column 2) were estimated from the total flux densities in S- and X-bands, and the values in Columns 4-6 come from model fitting of the X-band VLBA data. 
The two RQQs (PG 0050+124 and PG 0157+001) marked with an asterisk lack an identifiable core, so their parameters are not showed here. For the PG quasars (PG 1149$-$110, PG 1612+261, PG 1226+023 and PG 1302$-$102), their fitted Guassian size were consistent with to a point source, so upper limit of the component sizes were used in place of them \citep{2005astro.ph..3225L}. Considering our limited resolution capability, the core size of RQQs in column 5 is actually an upper limit, and the real value is smaller than the one obtained by the model fitting; therefore, the brightness temperature in column 6 can be regarded as a lower limit.
\end{table*}

\section{Parsec-scale morphology}
 \label{sec:result}

\subsection{Radio quiet quasars}

\begin{figure*}
\centering
  \begin{tabular}{cc}
  \includegraphics[width=0.35\textwidth]{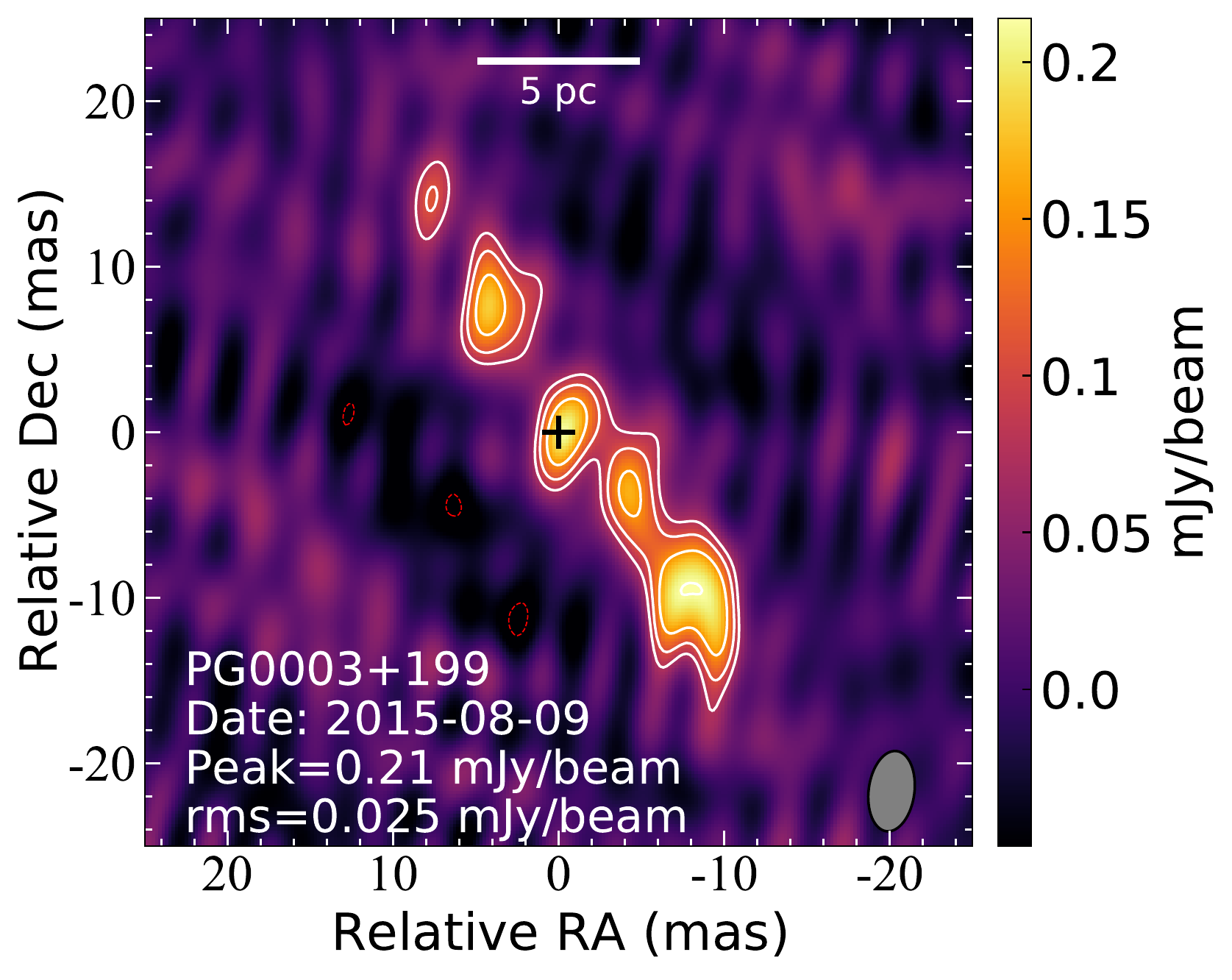}    
  \includegraphics[width=0.35\textwidth]{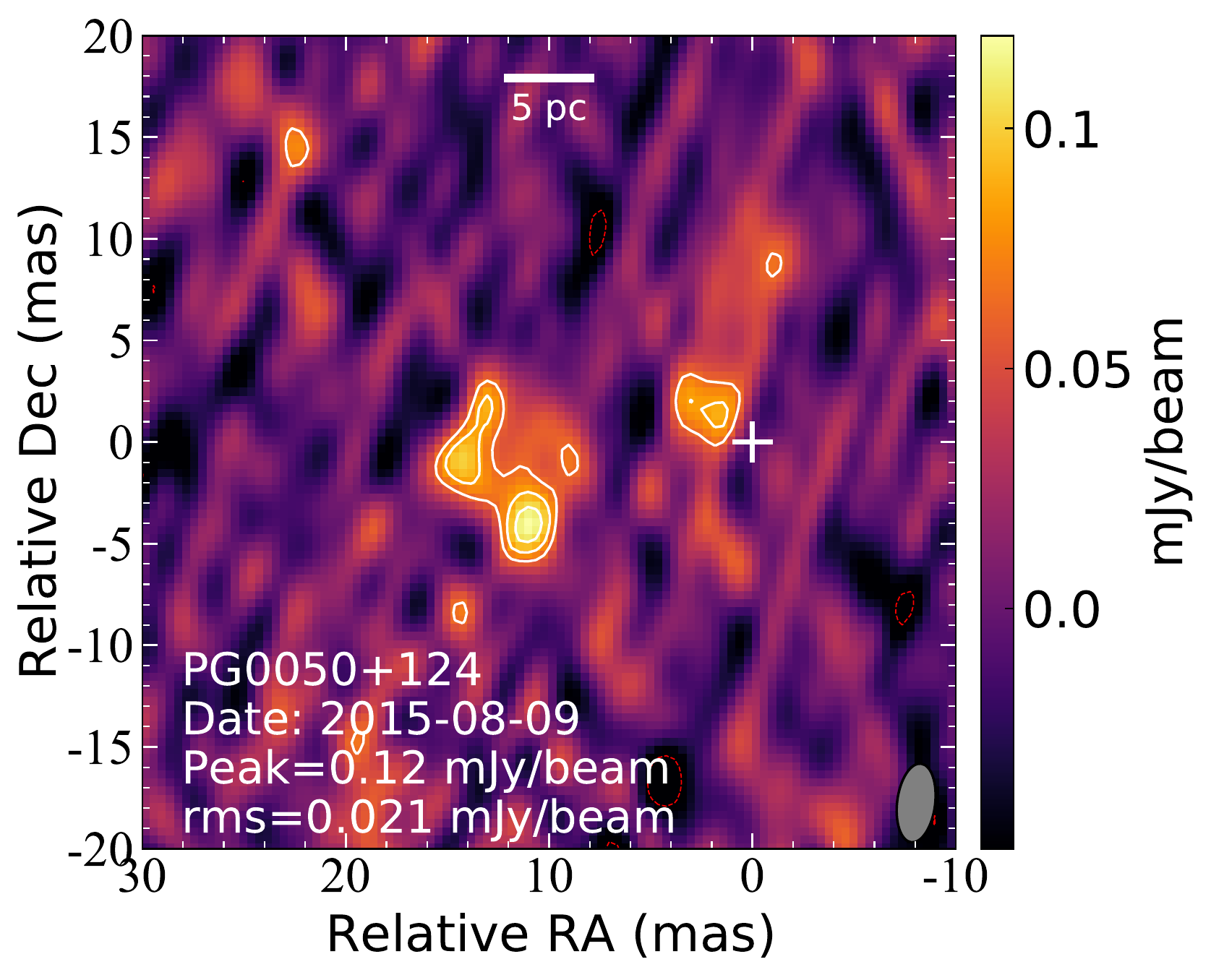}  
  \includegraphics[width=0.35\textwidth]{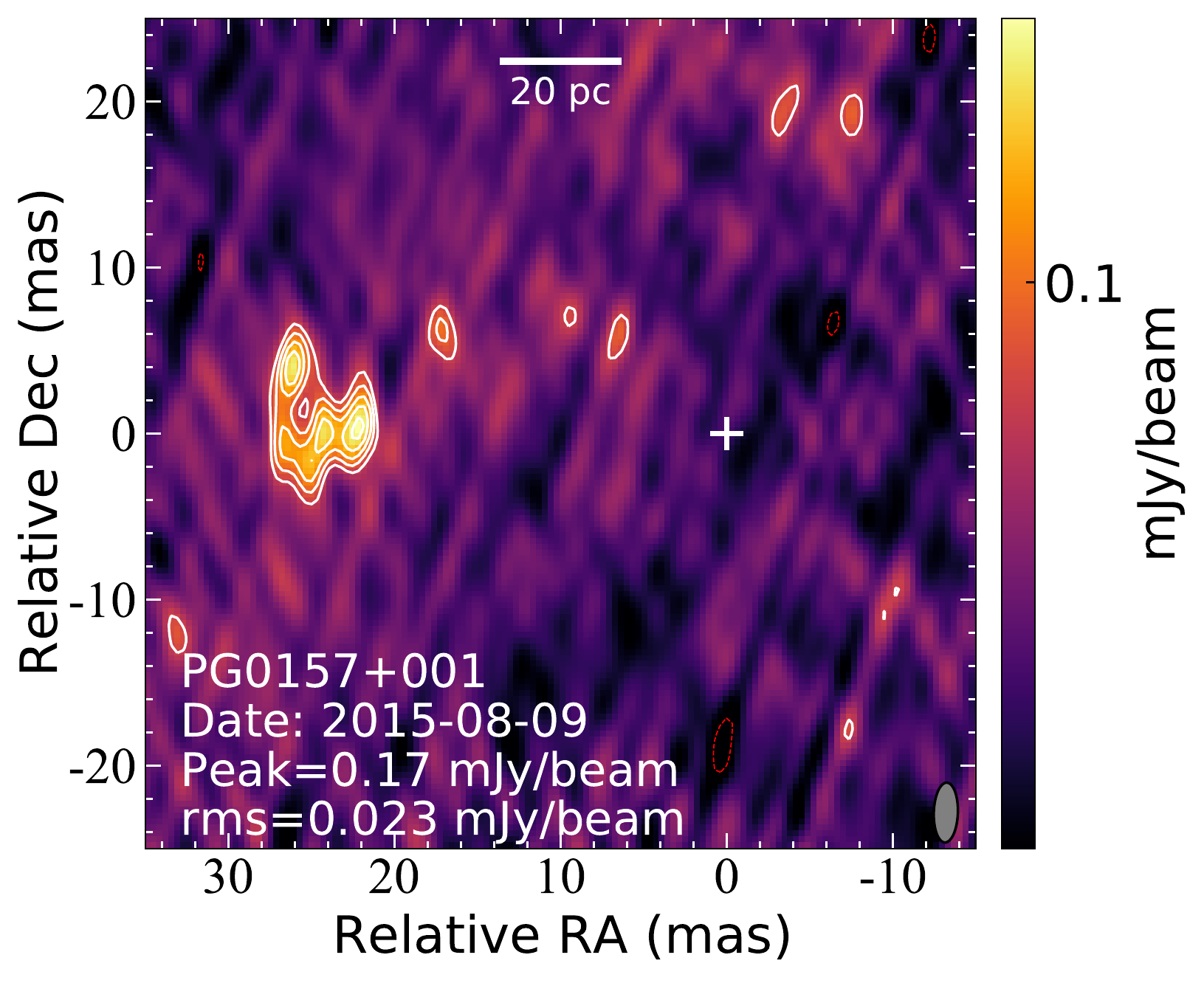} \\
  \includegraphics[width=0.35\textwidth]{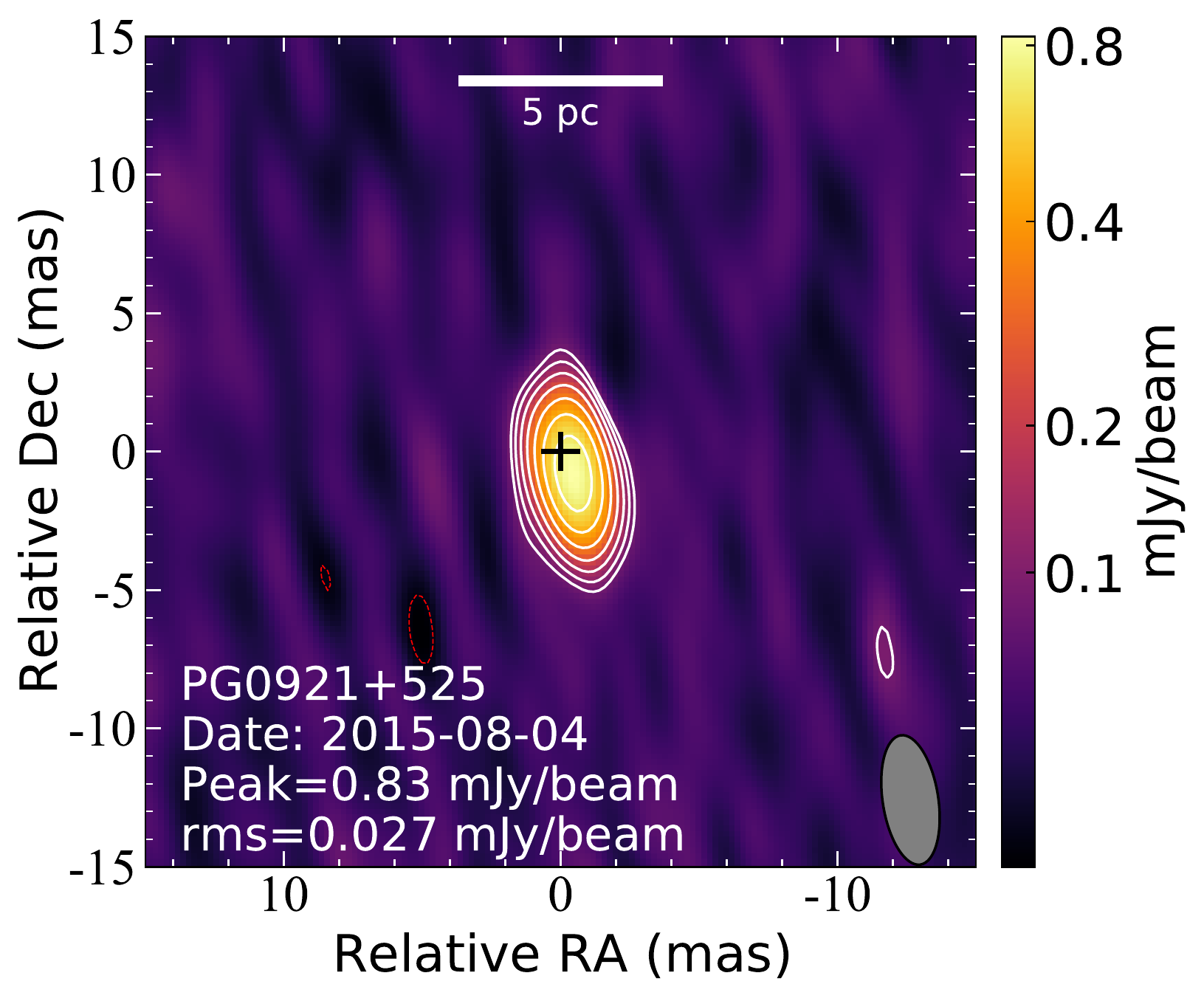}  
  \includegraphics[width=0.35\textwidth]{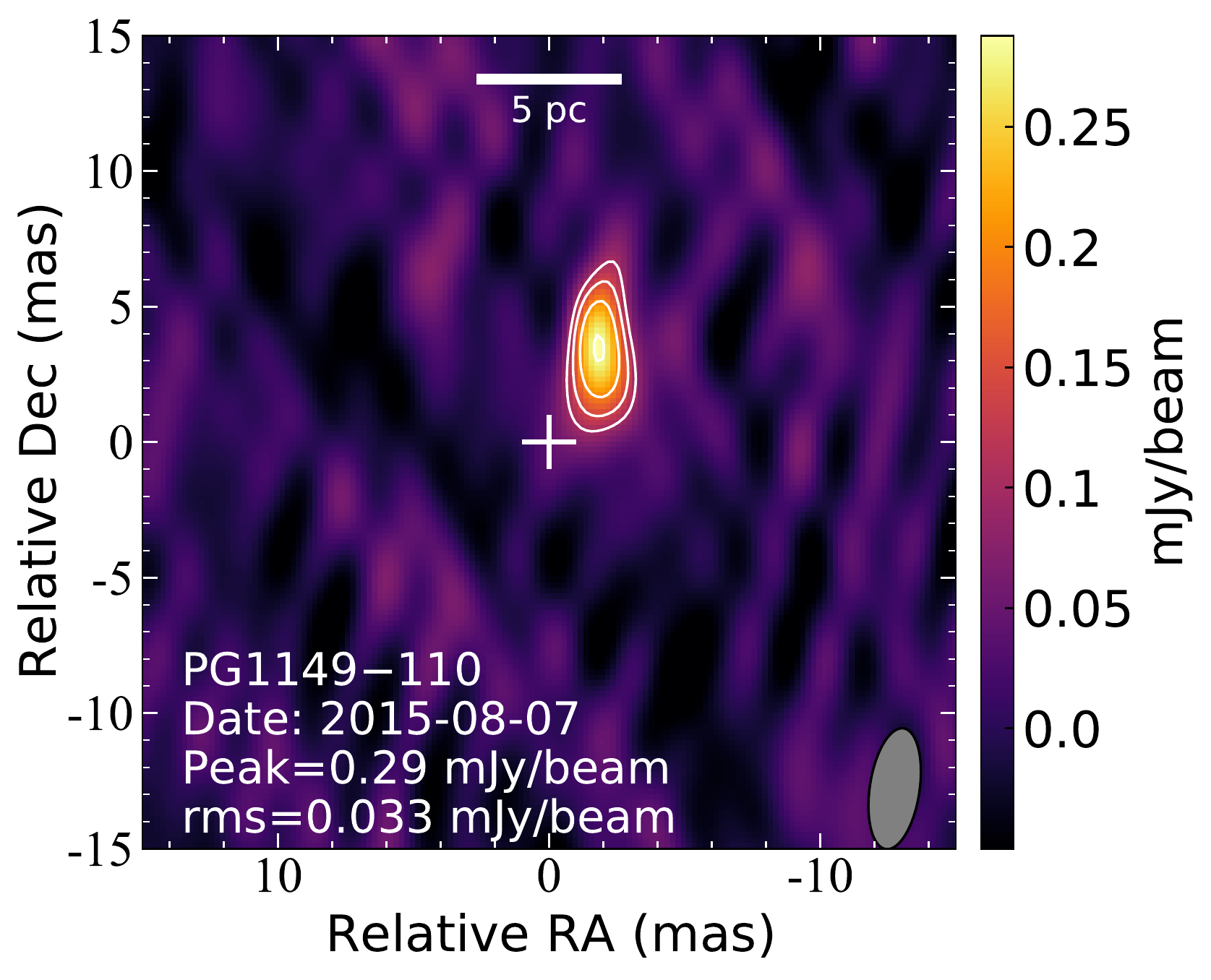}
  \includegraphics[width=0.35\textwidth]{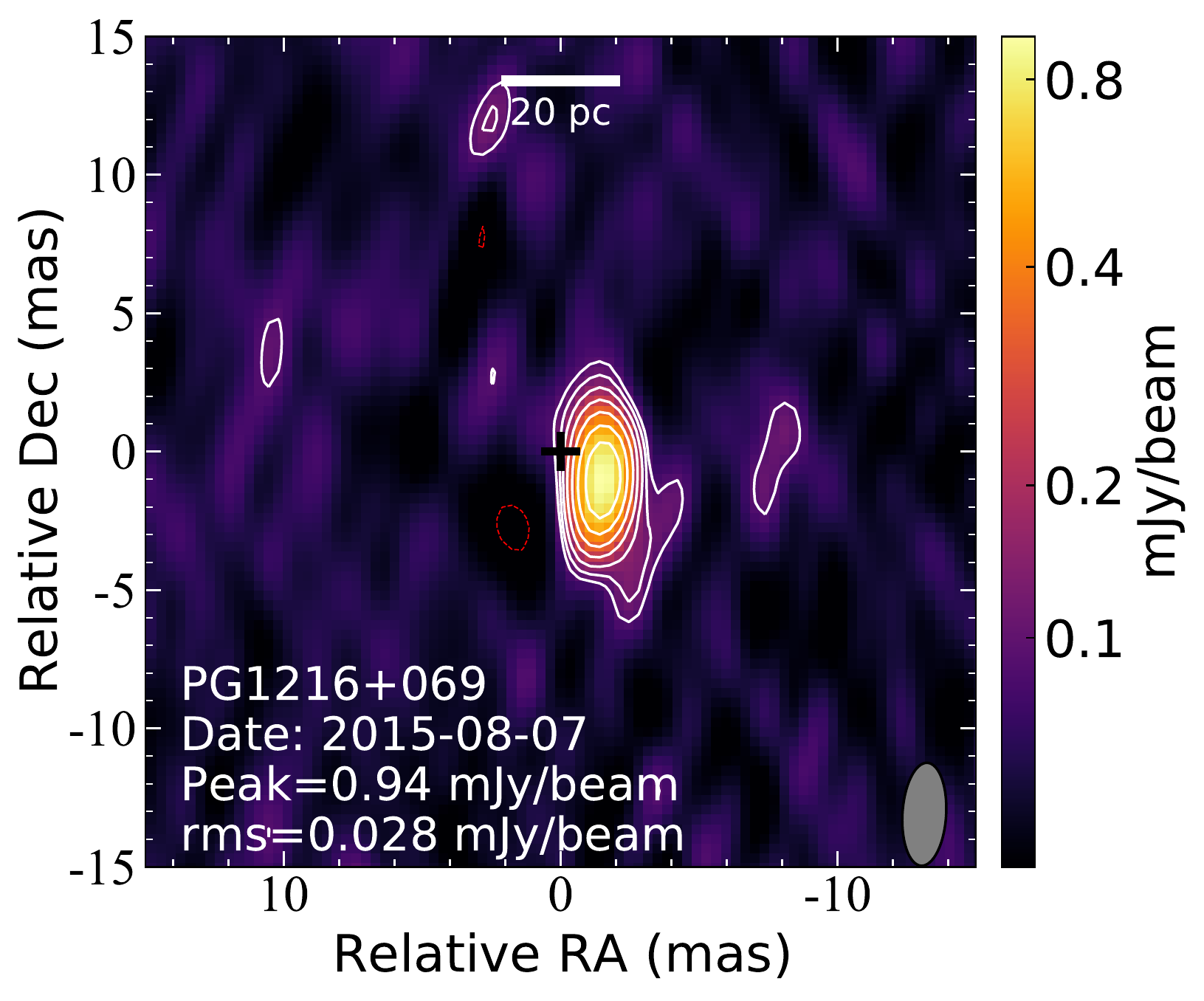} \\
  \includegraphics[width=0.35\textwidth]{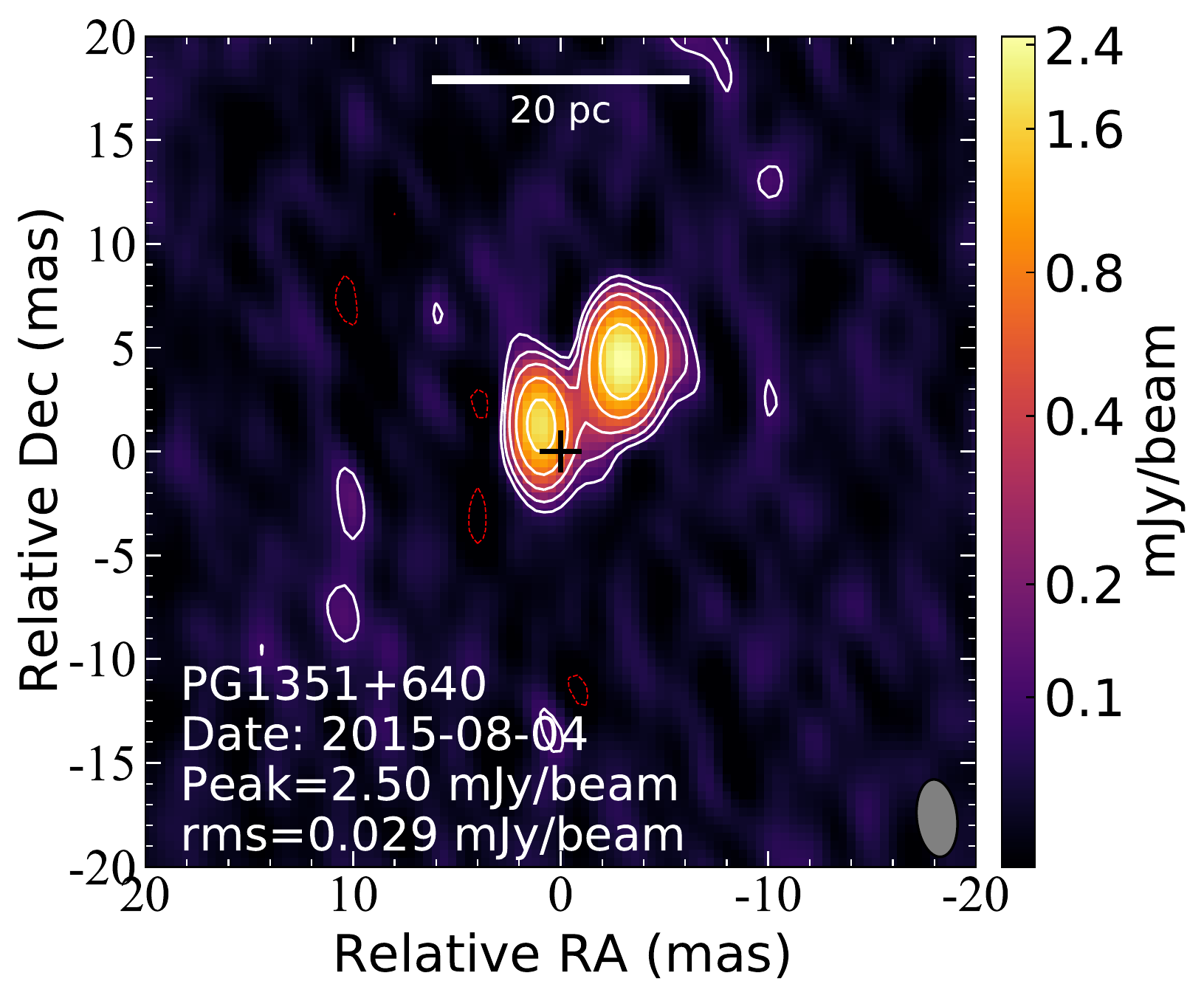} 
  \includegraphics[width=0.35\textwidth]{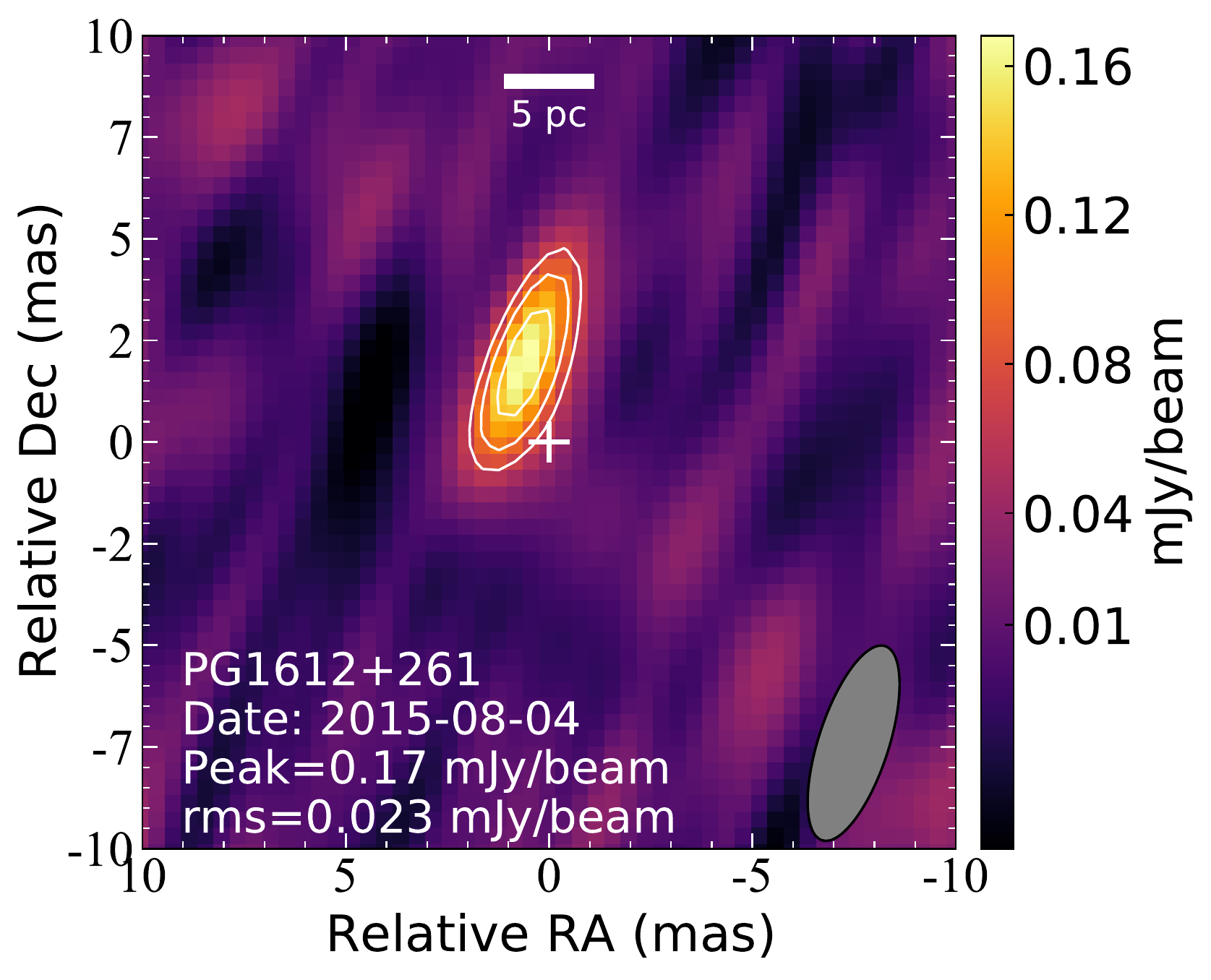} 
  \includegraphics[width=0.35\textwidth]{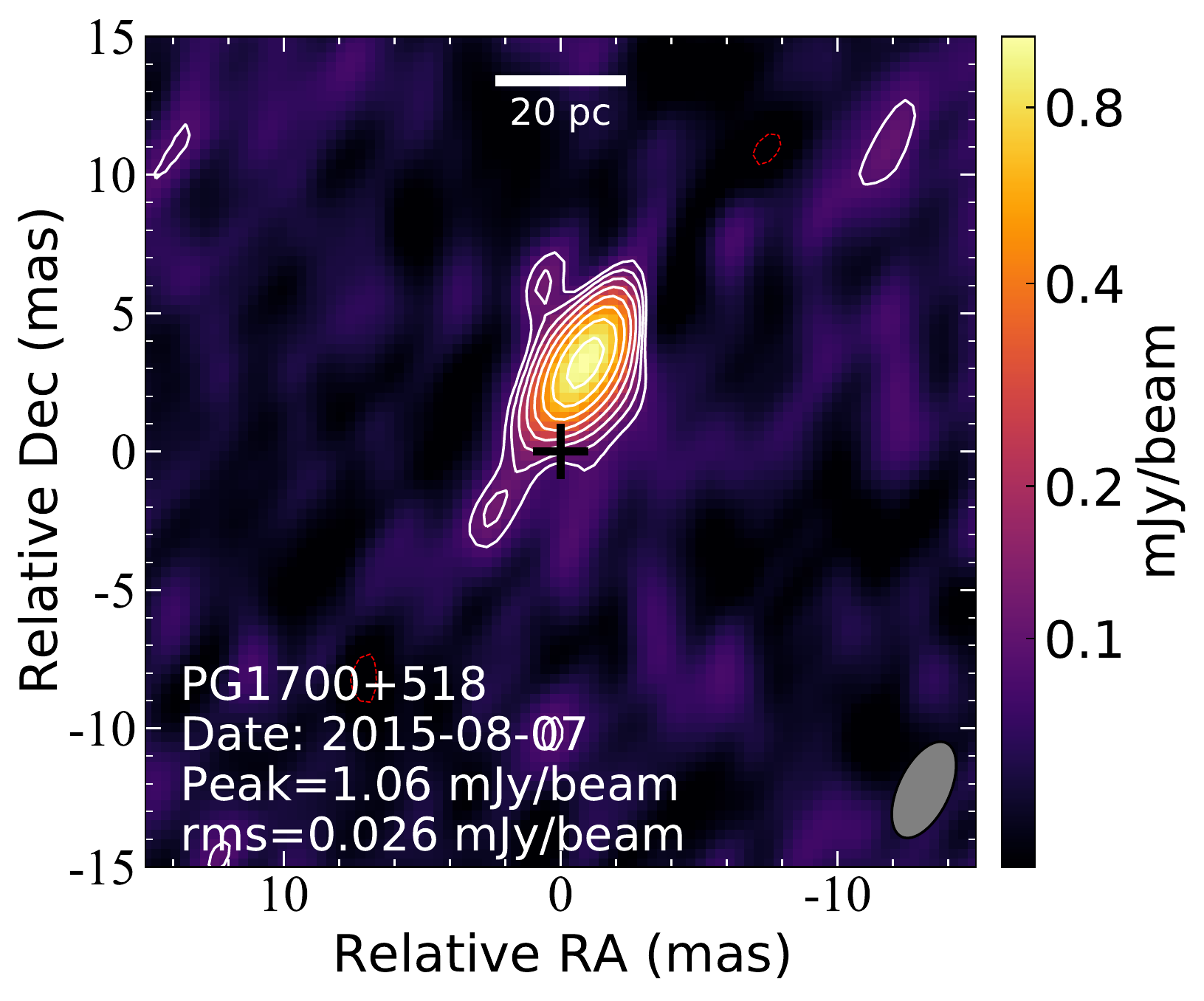} \\
  \includegraphics[width=0.35\textwidth]{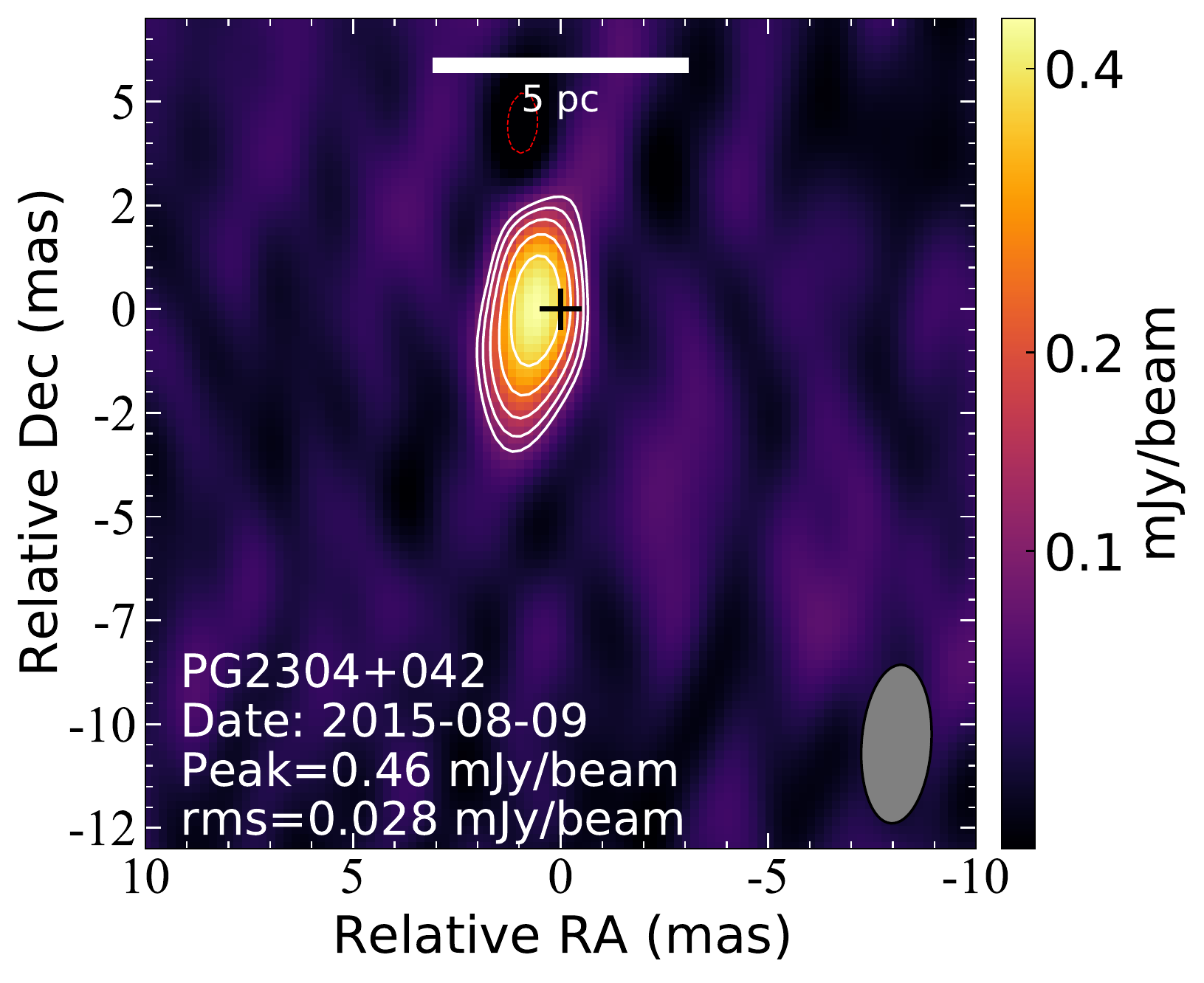} 
    \end{tabular}
  \centering
  \caption{VLBA images of radio quiet quasars at 5 GHz. Images of undetected sources are not shown here. The observation date, peak intensity, and root mean square noise level for each source are shown at the bottom-left corner of each map. The image parameters are referred to Table \ref{tab:image}. The cross in each panel center marks the position of the \textit{Gaia} DR3 optical nucleus \citep{2022yCat.1355....0G}. The grey ellipse in the bottom-right corner represents the shape of the restoring beam.}
  \label{fig:RQAGN}
\end{figure*}

Of the 20 target sources, all four RLQs are detected, and 10 of the 16 RQQs are successfully detected. Images of the ten RQQs are shown in Figure~\ref{fig:RQAGN}, and image parameters are tabulated in Table \ref{tab:image}. 
According to the 5-GHz VLBA images and the position of the optical nucleus in the radio structure, the morphology of these RQQs in our sample can be divided into four classes: single core, one-sided jet, two-sided jet, and complex structure.  

Six RQQs (PG 0921+525, PG 1149-110, PG 1216+069, PG 1612+261, PG 1700+518, and PG 2304+042) show a single compact component at or close to the optical nucleus. The minimum brightness temperature of these components ranges from $3.23\times 10^7$~K to $1.79\times10^8$~K, typical of non-thermal origin. 
The compact structure in the nuclear region (0.5 -- 1.1 pc) and high brightness temperature allow them to be identified as the radio cores. 

PG 0003+199 (Mrk~335) displays a stripe structure which is resolved into several clumps along the northeast--southwest direction with an extent of about 32 mas (17 pc). 
The optical nucleus is associated with the central component,  which is the most compact and brightest component in the VLBA image. It is classified as 'a core + two-sided jet' morphology. Our image is consistent with the 1.5-GHz VLBA image obtained by \citet{2021MNRAS.508.1305Y}. The difference is that our image has a higher resolution and focuses more on the compact jet knots; Yao et al.'s image reveals a more extended structure.

PG 1351+640 shows two compact components separated by $\sim$6 mas ($\sim$10 pc), 
resembling a CSO \citep{1982A&A...106...21P,2000ApJ...534...90P,2005ApJ...622..136G,2012ApJ...760...77A}. However, the proximity of the optical nucleus to the southeast component leads us to identify it as a 'core + one-sided jet' structure.

The two remaining sources (PG 0050+124 and PG 0157+001) show a complex radio emission structure with multiple discrete weak components lacking a direct connection between them in the present images. Three components are on the same side of the \textit{Gaia} position of PG 0050+124 (I Zw 1), one of which is close to the optical nucleus. 
\citet{2022arXiv220801488A} has recently published 1.4 and 5 GHz VLBA images of eight RQQs, among which the 5-GHz image of PG 0050+124 is very similar to our results, and their 1.4-GHz VLBA image shows a more extended emission structure elongated along the east-west direction.
The brightest component in PG 0157+001 is clearly away from the optical peak and obviously cannot be associated with the radio core. Deeper VLBA images of PG 0157+001 at lower frequencies are required to reveal its radio structure. Possible explanations for these knotty morphologies of PG 0050+124 and PG 0157+001 are a resolved jet \citep[e.g., 3C~216: ][]{2013MNRAS.433.1161A}, disrupted jet  (e.g. 3C~48 in \citealt{2010MNRAS.402...87A}) or hotspots due to jet collisions with the interstellar medium (e.g. NGC~3079 in \citealt{2007MNRAS.377..731M}).

Of the ten RQQs detected in our observations, nine RQQs have previously reported VLBI observations (see details also in the notes of individual sources in Section~\ref{sec:individual}): 
PG 0003+199 (Mrk 335, \citealt{2021MNRAS.508.1305Y} at 1.5 GHz), 
PG 0050+124 (\citealt{2022arXiv220801488A} at 1.4 and 5 GHz),
PG 0921+525 (Mrk 110, \citealt{2013ApJ...765...69D} at 1.7 GHz, \citealt{2022MNRAS.510..718P} at 5 GHz), 
PG 1149$-$110  (\citealt{2022arXiv220801488A} at 1.4 and 5 GHz), 
PG 1216+069 (\citealt{1998MNRAS.299..165B} at 8.4 GHz, \citealt{2005ApJ...621..123U} at 1.4 and 5 GHz), 
PG 1351+640 (\citealt{2005ApJ...621..123U} at 1.4 and 5 GHz), 
PG 1612+261 (\citealt{2022arXiv220801488A} at 1.4 GHz),
PG 1700+518 (\citealt{2012MNRAS.419L..74Y} at 1.7 GHz), 
and PG 2304+042 (\citealt{2022arXiv220801488A} at 1.4 and 5 GHz). The VLBI image of PG 0157+001 was obtained for the first time.  

High-resolution VLBI imaging is an effective way to determine the physical nature of compact radio components.
Using the VLBA observables (flux density and source size), we calculated the brightness temperatures ($T_{\rm b}$) of the radio core, all $\gtrsim 10^{7}$ K (Table~\ref{tab:radio}). Since the measured component size of RQQ from the VLBA image is an upper limit, the resulting brightness temperature represents a lower limit. The high values of $T_{\rm b}$  rule out a thermal origin of the radio components \citep{1992ARA&A..30..575C, 1998MNRAS.299..165B}.
As a comparison, we also calculated the radio core brightness temperatures of 16 RL PG  quasars with $z<0.5$. All RLQs have $T_{\rm b} > 10^9$~K, and some even exceed the Compton catastrophe brightness  temperature limit \citep{1969ApJ...155L..71K,1994ApJ...426...51R}, indicating a very strong relativistic beaming effect of these jets.

Of the six undetected RQQs, only PG 0923+129 has reported  1.7-GHz VLBI observations \citep{2013ApJ...765...69D}, but that detection remains doubtful (see discussion in Section \ref{sec:undetected}).

The 5 GHz VLA-A flux density ($S_{\rm A}^{\rm VLA}$) of these six VLBA-undetected sources is below 2 mJy (Table \ref{tab:image}). Our VLBA observations were made in snapshot mode, and the average noise level of the images is $\sim$0.026 \mJyb. A firm VLBA detection would require a compact component with at least 0.125 mJy ($\geq 5 \sigma$). The non-detection suggests that the percentage of compact radio core/jet in these RQQs is less than (6-10)\%. Future VLBA observations with higher sensitivity are needed to explore their weak pc-scale jets.
Similarly, the ratio of the VLA flux densities obtained in different array configurations ($f_{\rm c} = S_{\rm A}^{\rm VLA} / S_{\rm D}^{\rm VLA}$: \citealt{1989AJ.....98.1195K,1994AJ....108.1163K}) at the same frequency of 5 GHz does not show a systematic low $f_{\rm c}$ for four undetected sources by our VLBA observations  (PG 1116+215, PG 1448+273, PG 1534+580 and PG 2130+099: Figure~\ref{fig:fc}).
Except for a very few sources (accounting for 12.5\%) with $f_{\rm c}$ smaller than 0.5, the vast majority of RQQs in our sample have $f_{\rm c}$ larger than 0.5, indicating that, in these brightest RQQs selected by flux density limit of $>$1 mJy, the radio flux is mainly from within about 1 kpc (corresponding to the VLA A-array resolution). Therefore, it is possible that intrinsic differences in the radio emission structure on parsec scales and sub-kpc scales lead to differences in the VLBA images of RLQs and RQQs.
One possibility is that extended jets or relic jets of RQQs on tens to hundreds of pc scales are resolved out in the VLBA images. 
Another possible source of the resolved emission in RQQs is the accretion disk wind that could extend over several parsec distances. The farthest outer boundary that the wind can reach is still an interesting topic worth exploring.

\subsection{Radio loud quasars}

\begin{figure*}
\centering
  \includegraphics[width=0.24\textwidth]{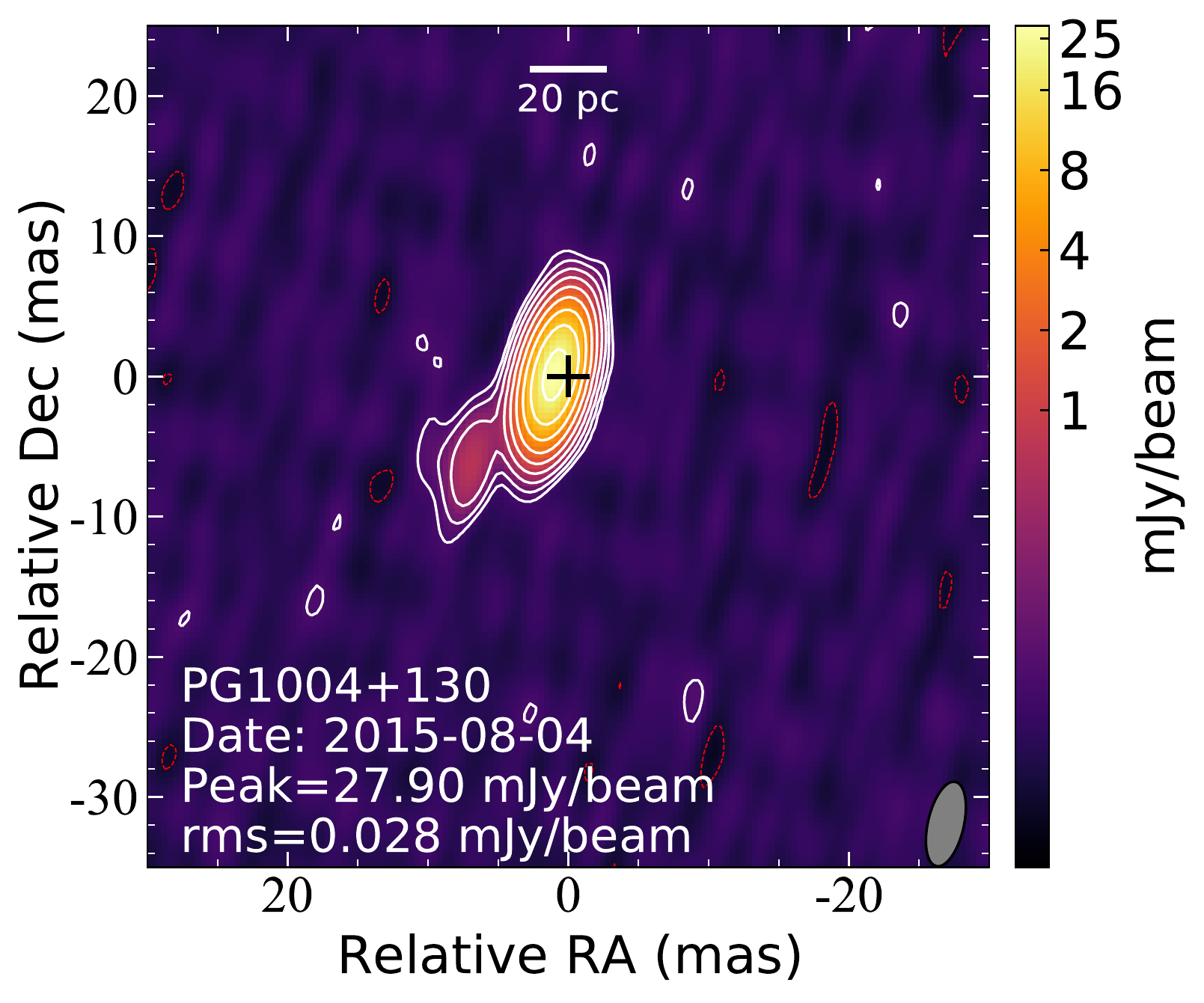} 
  \includegraphics[width=0.24\textwidth]{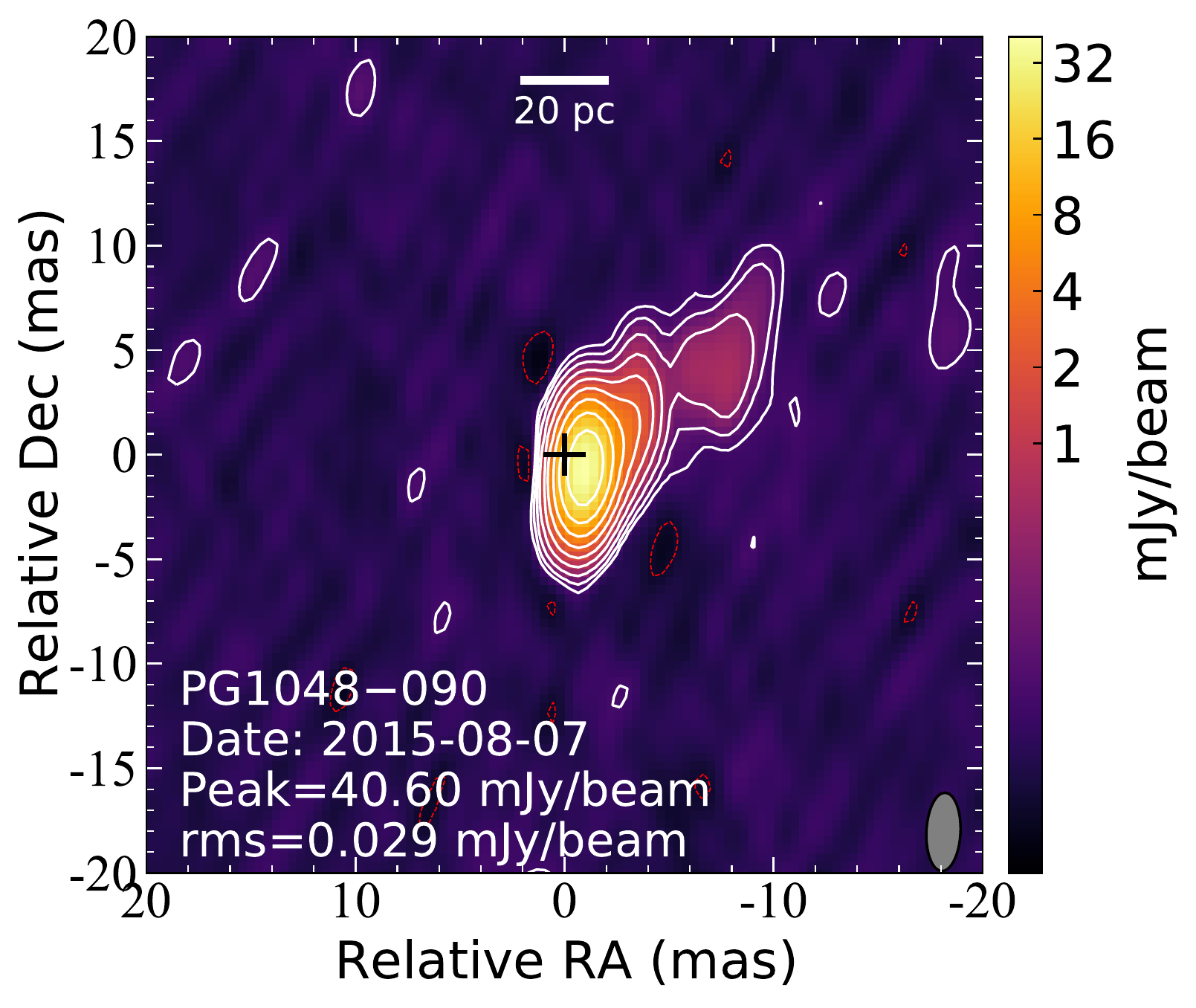}  
  \includegraphics[width=0.24\textwidth]{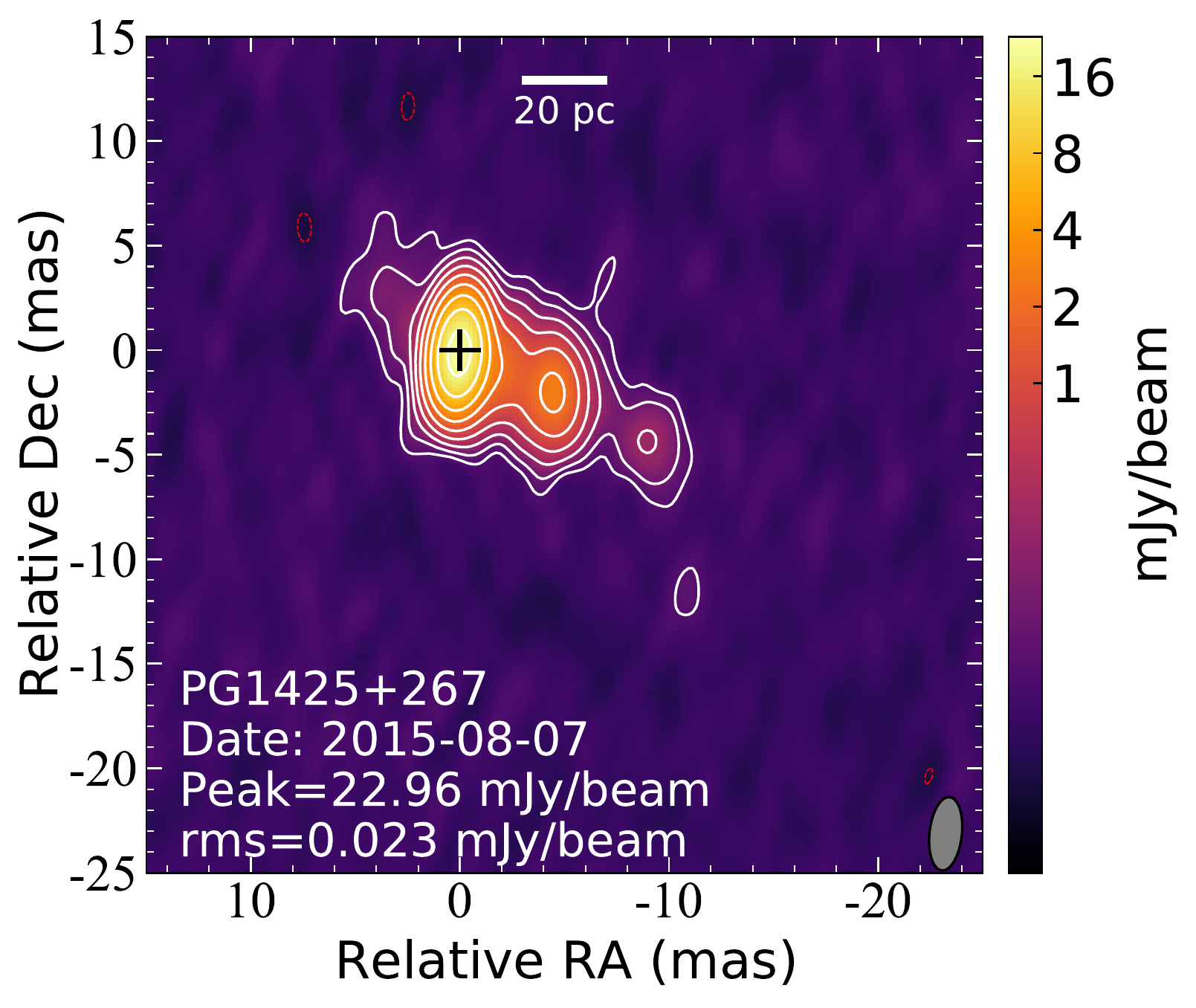}  
  \includegraphics[width=0.24\textwidth]{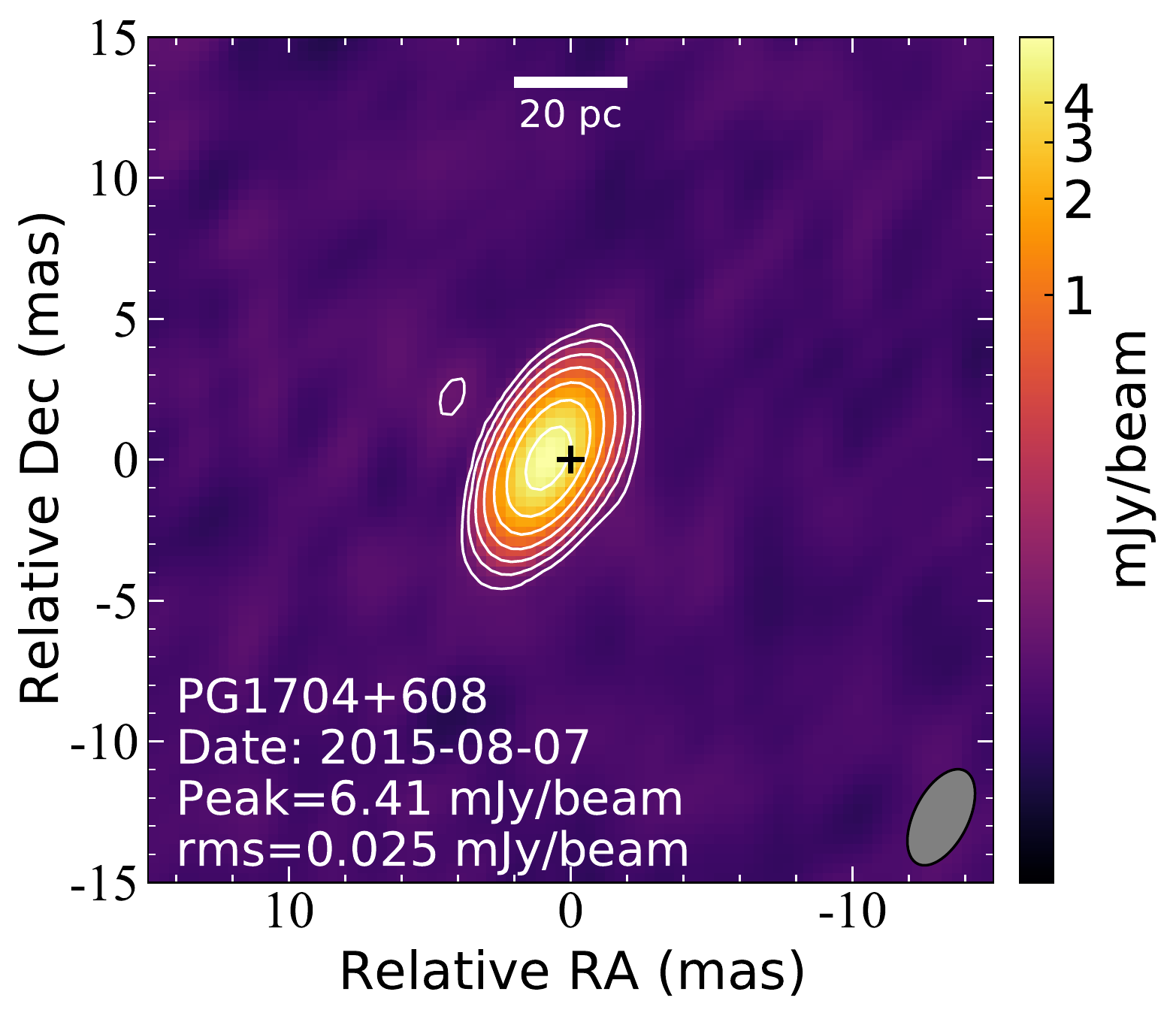}  \\
  \caption{5 GHz VLBA images of radio loud quasars in our sample. The image parameters are referred to Table \ref{tab:image}. The cross in each image center indicates the position of the \textit{Gaia} optical nucleus. The grey ellipse in the bottom-right corner represents the shape of the restoring beam.}
  \label{fig:RLAGN-4}
\end{figure*}

\begin{figure*}
\centering
  \begin{tabular}{cccc}
  \includegraphics[width=0.24\textwidth]{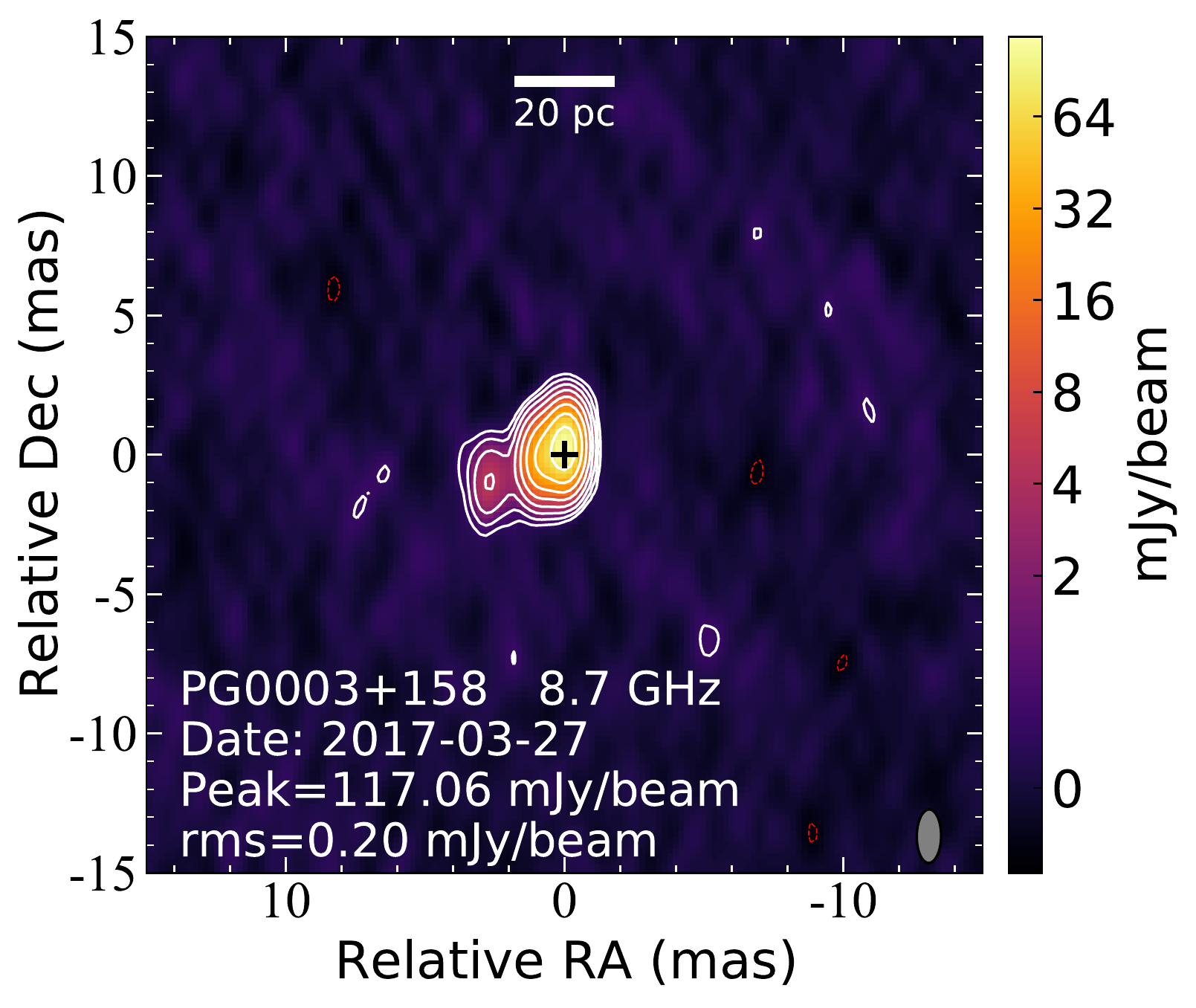} 
  \includegraphics[width=0.24\textwidth]{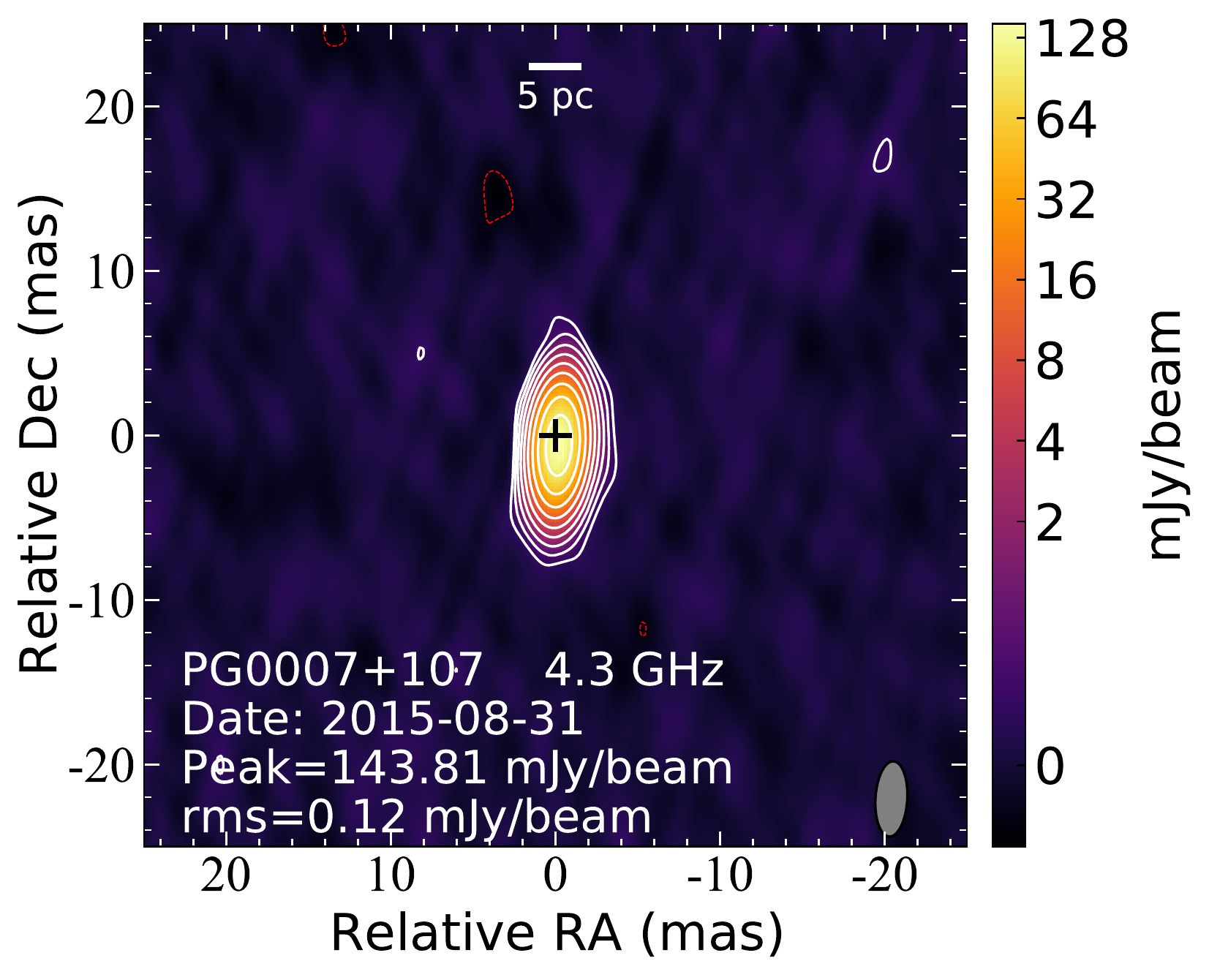}
  \includegraphics[width=0.24\textwidth]{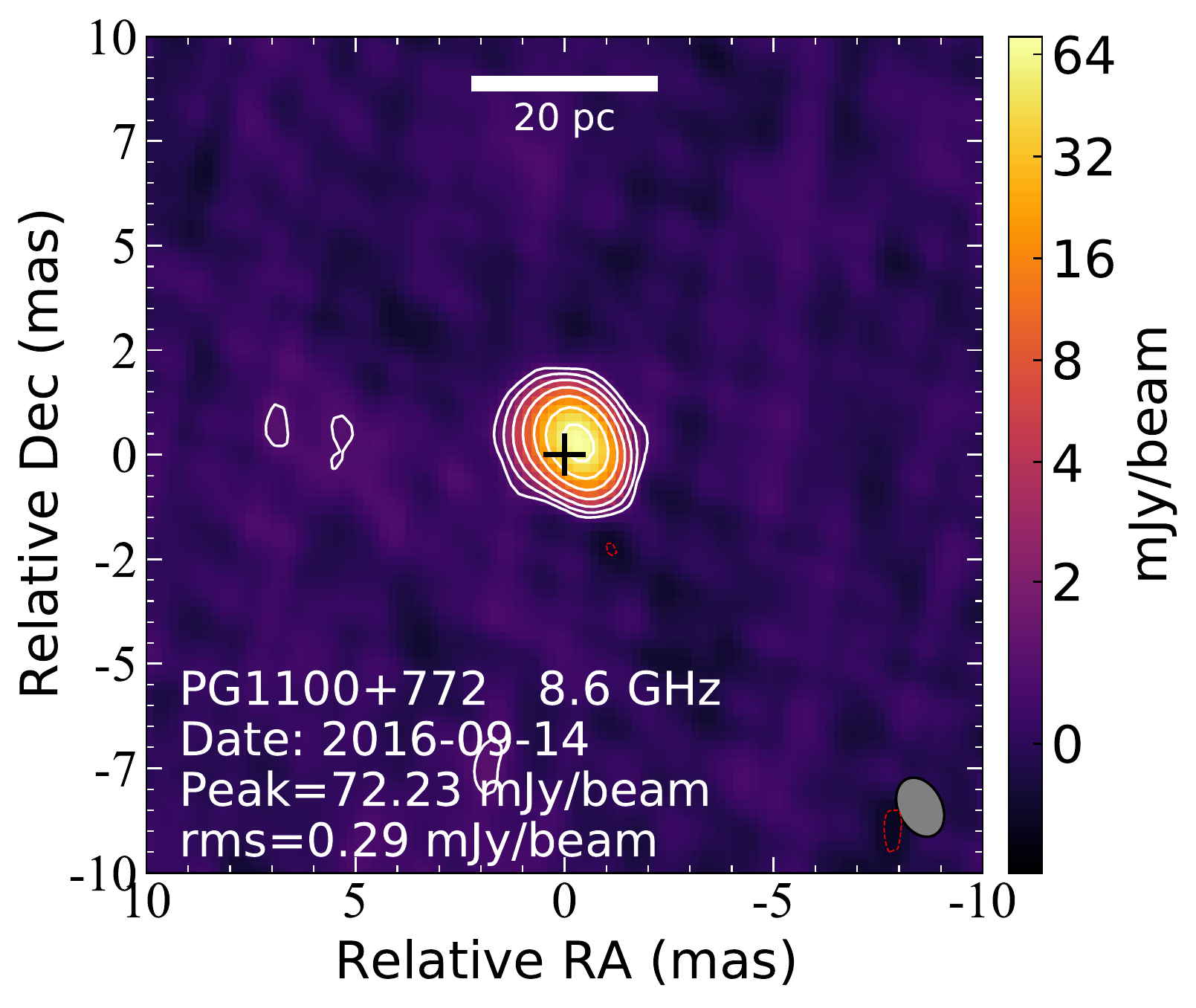}
  \includegraphics[width=0.24\textwidth]{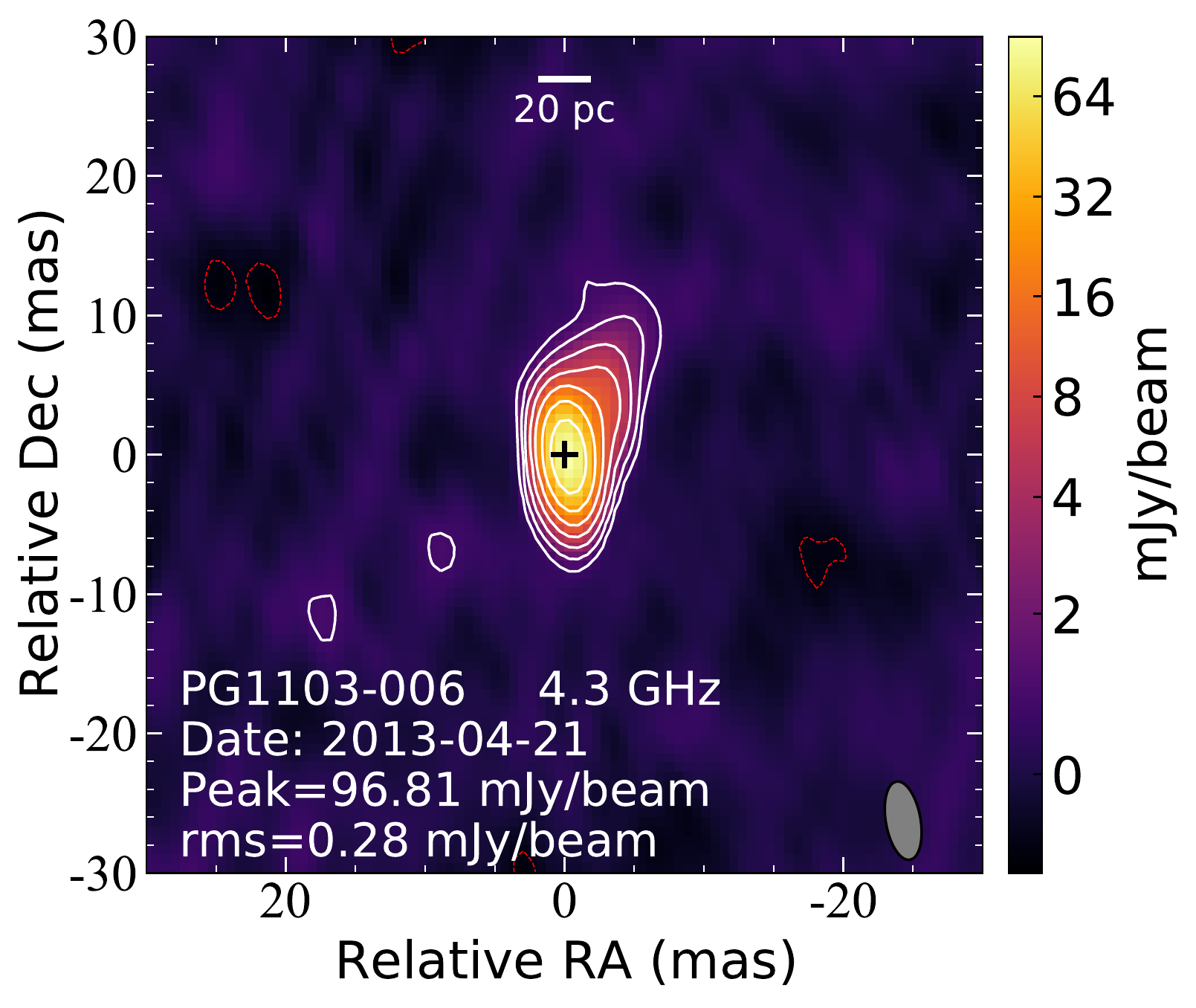}\\
  \includegraphics[width=0.24\textwidth]{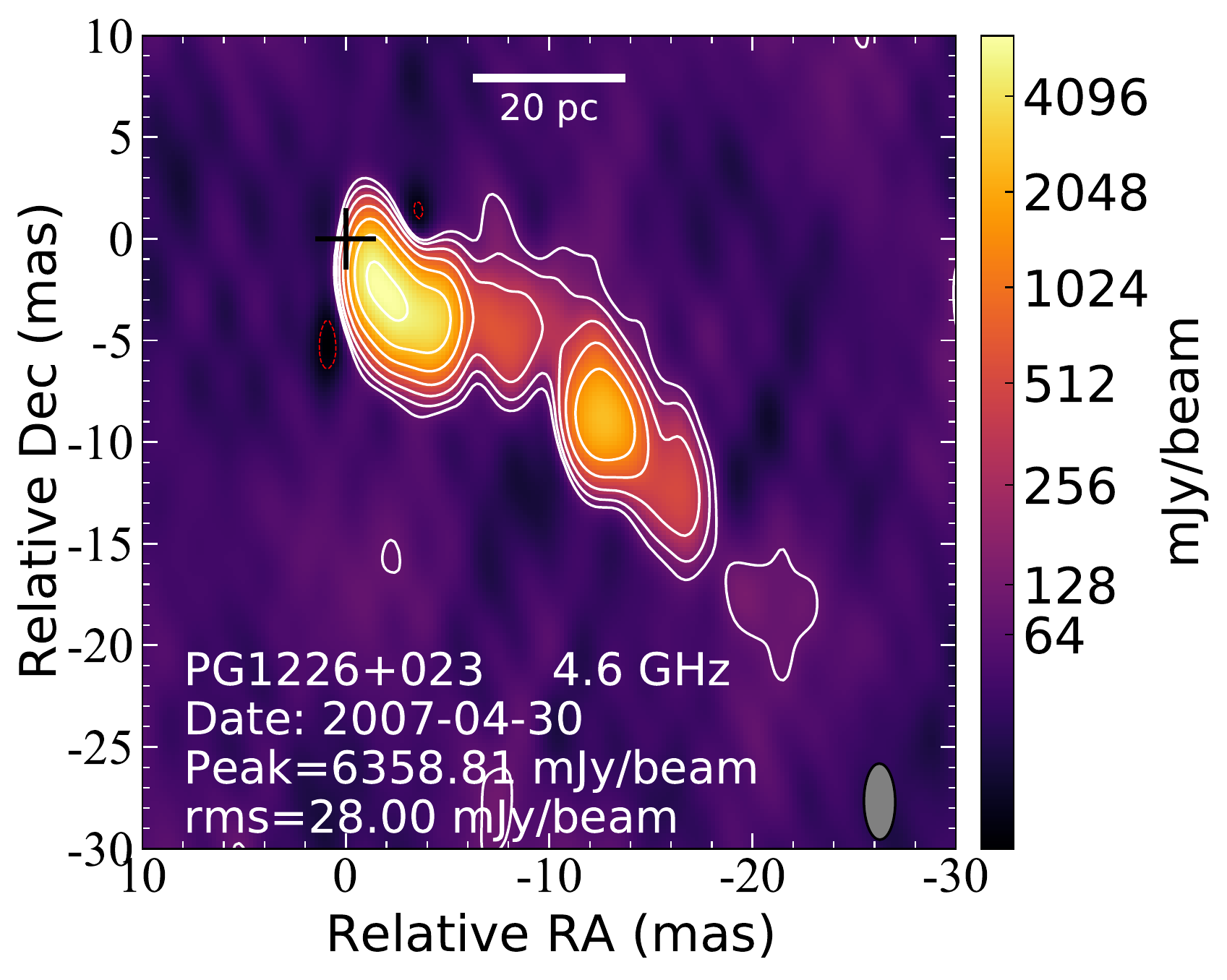} 
  \includegraphics[width=0.24\textwidth]{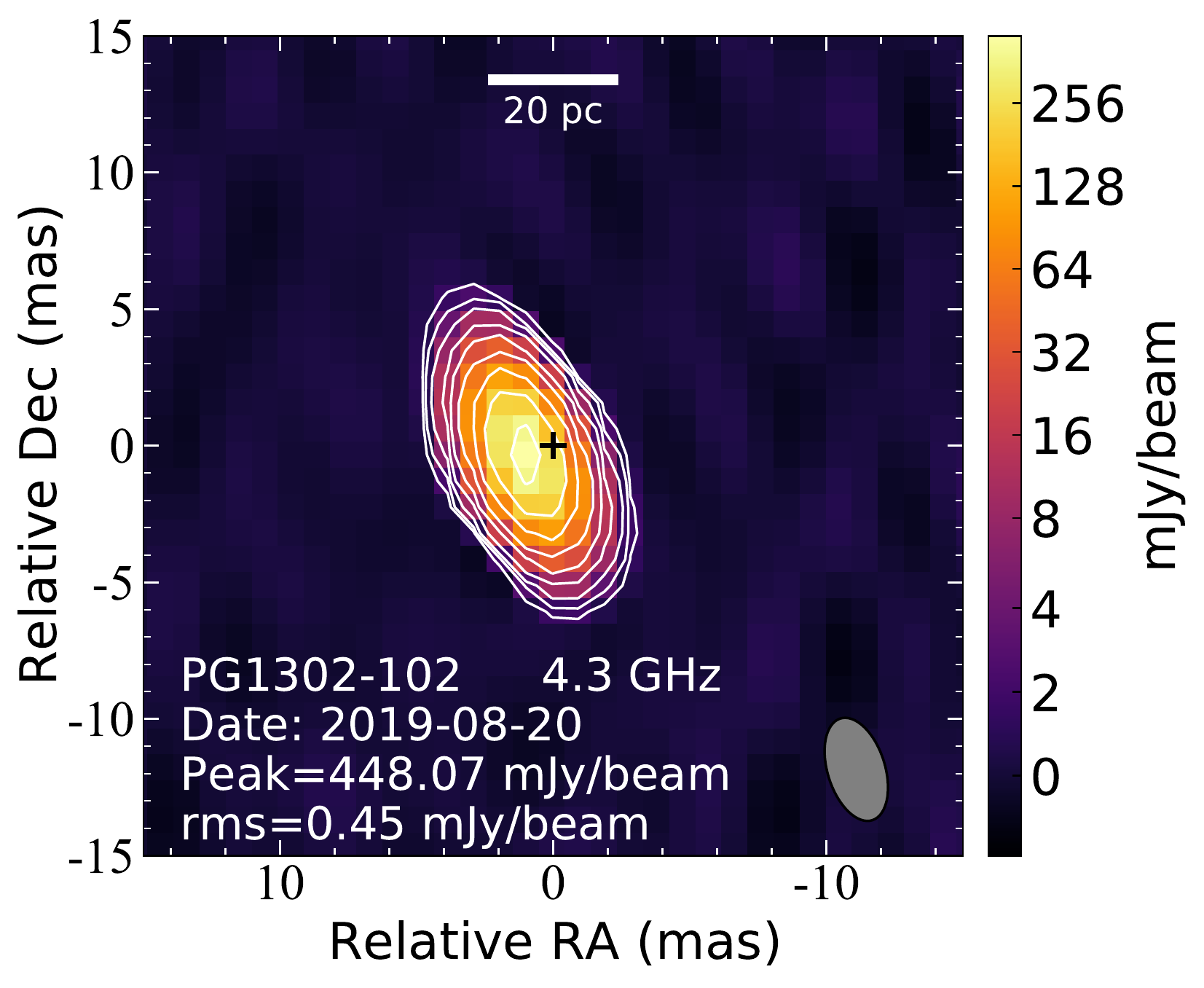}
  \includegraphics[width=0.24\textwidth]{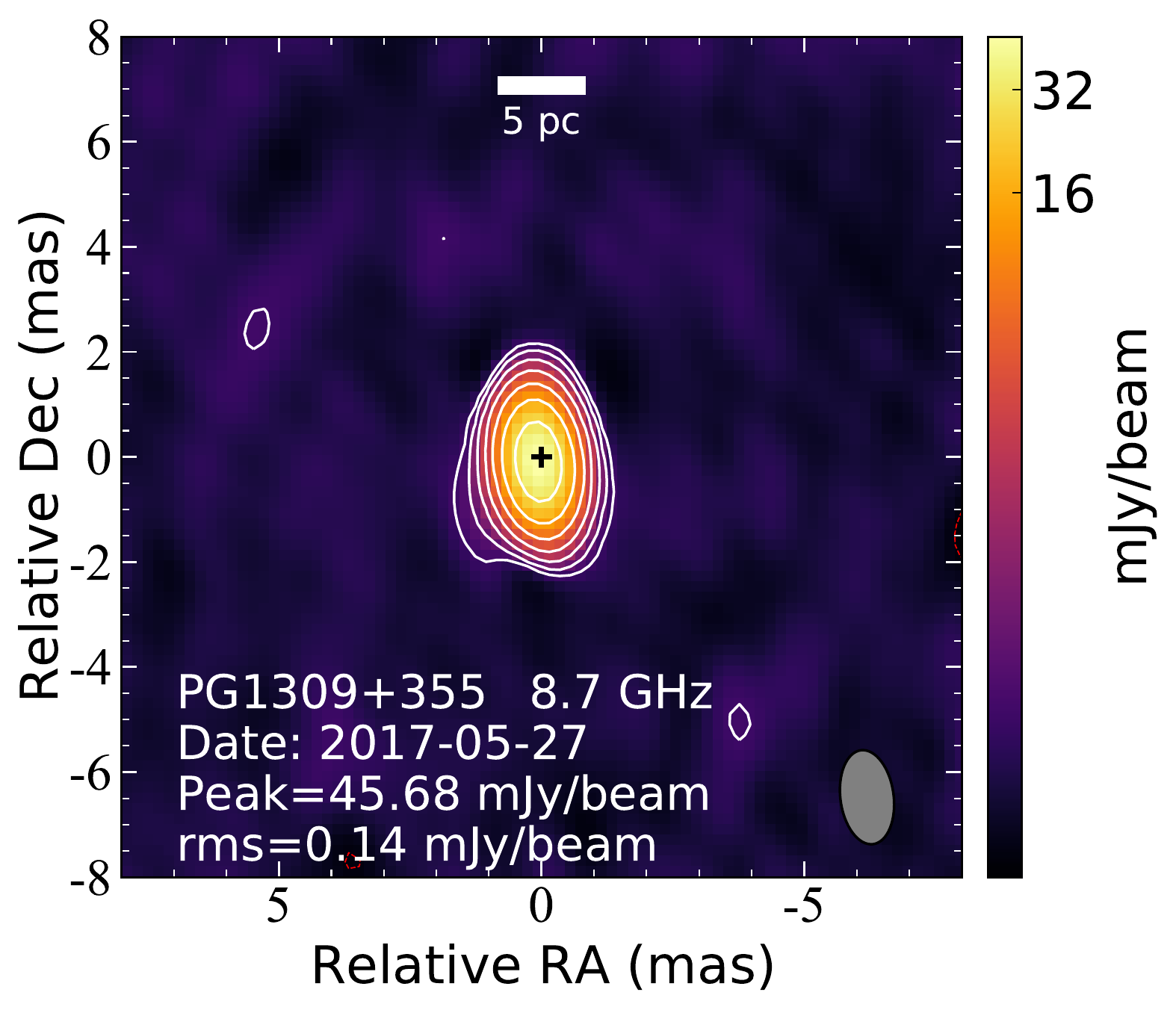}
  \includegraphics[width=0.24\textwidth]{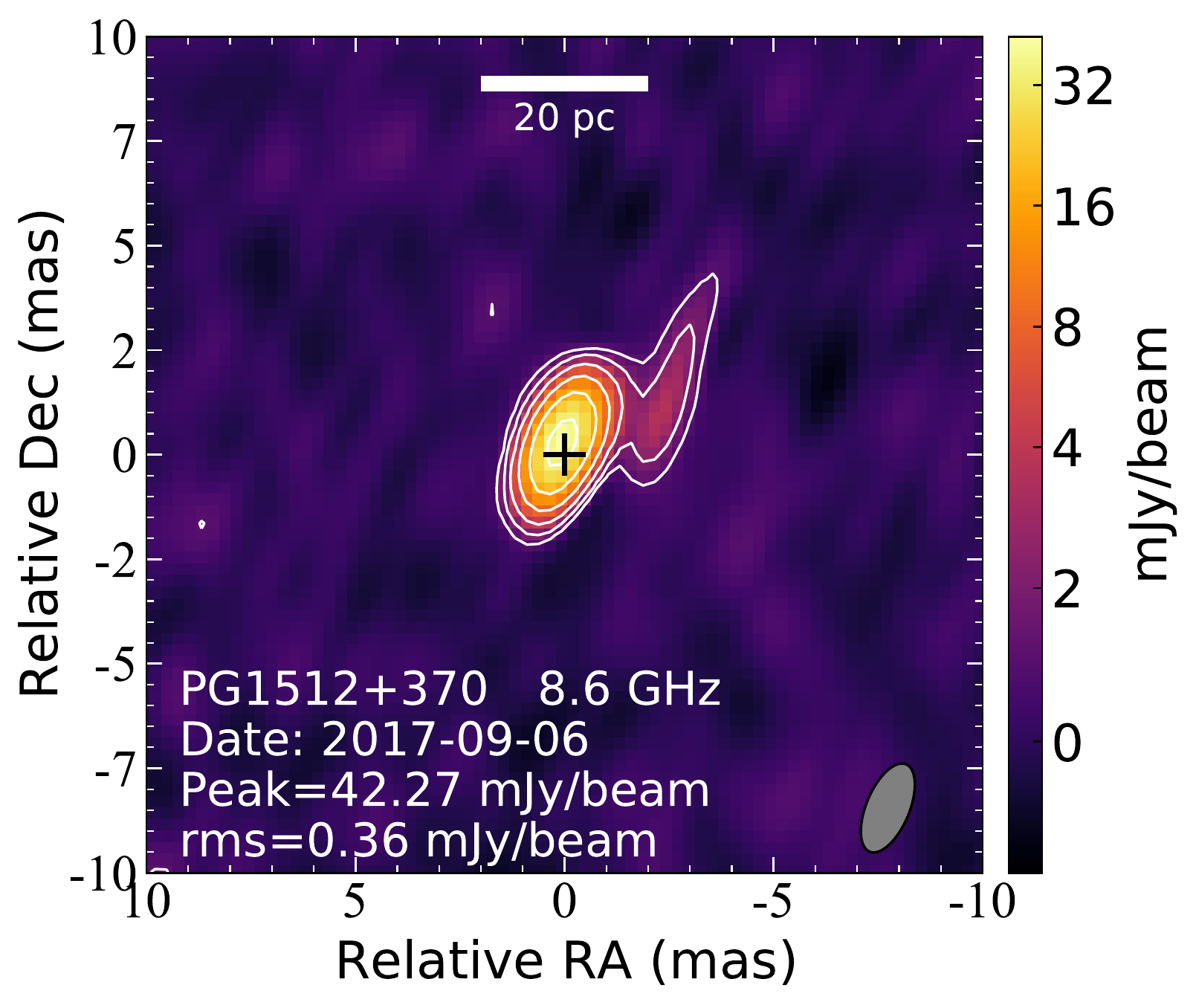}\\
  \includegraphics[width=0.24\textwidth]{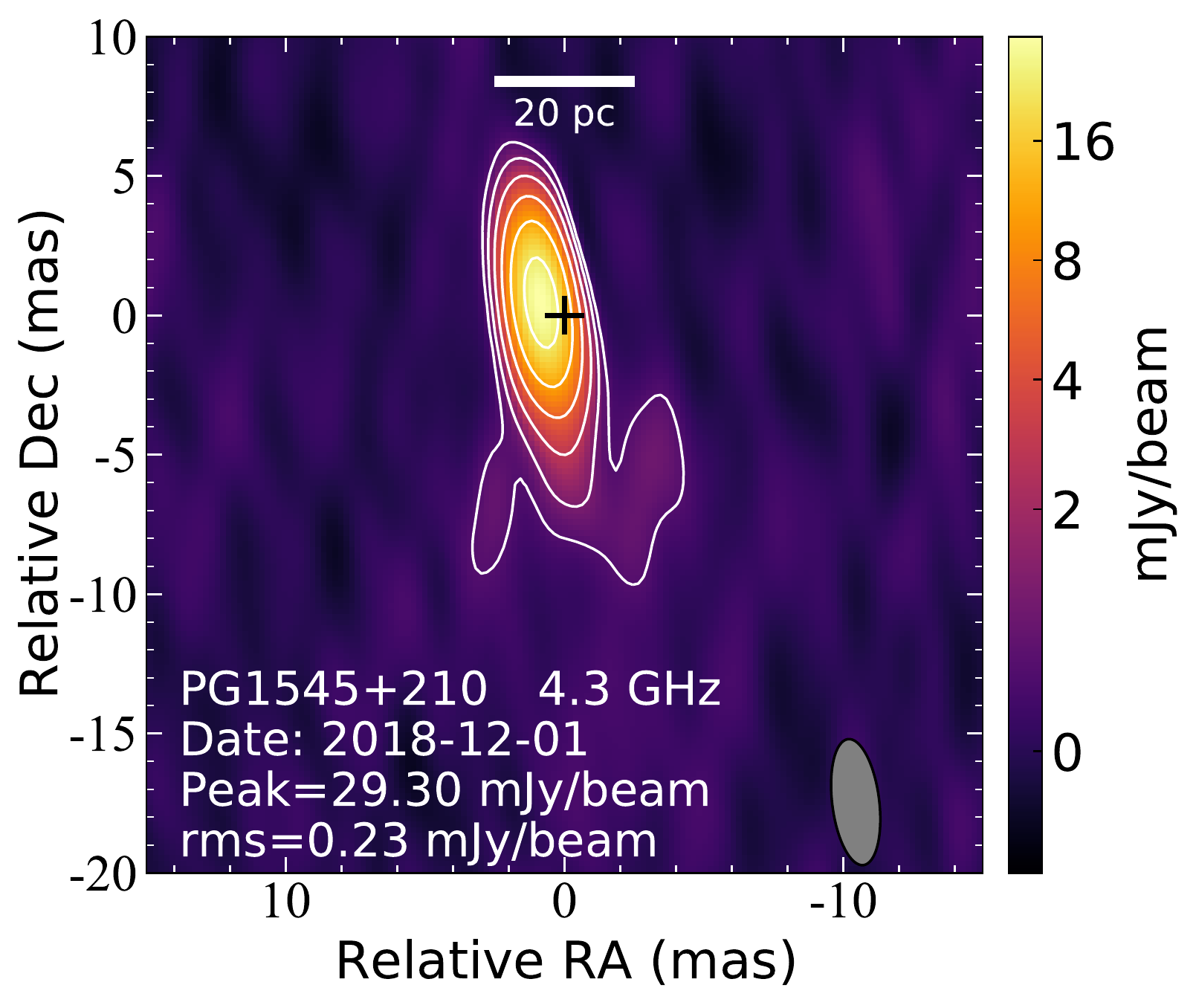}
  \includegraphics[width=0.24\textwidth]{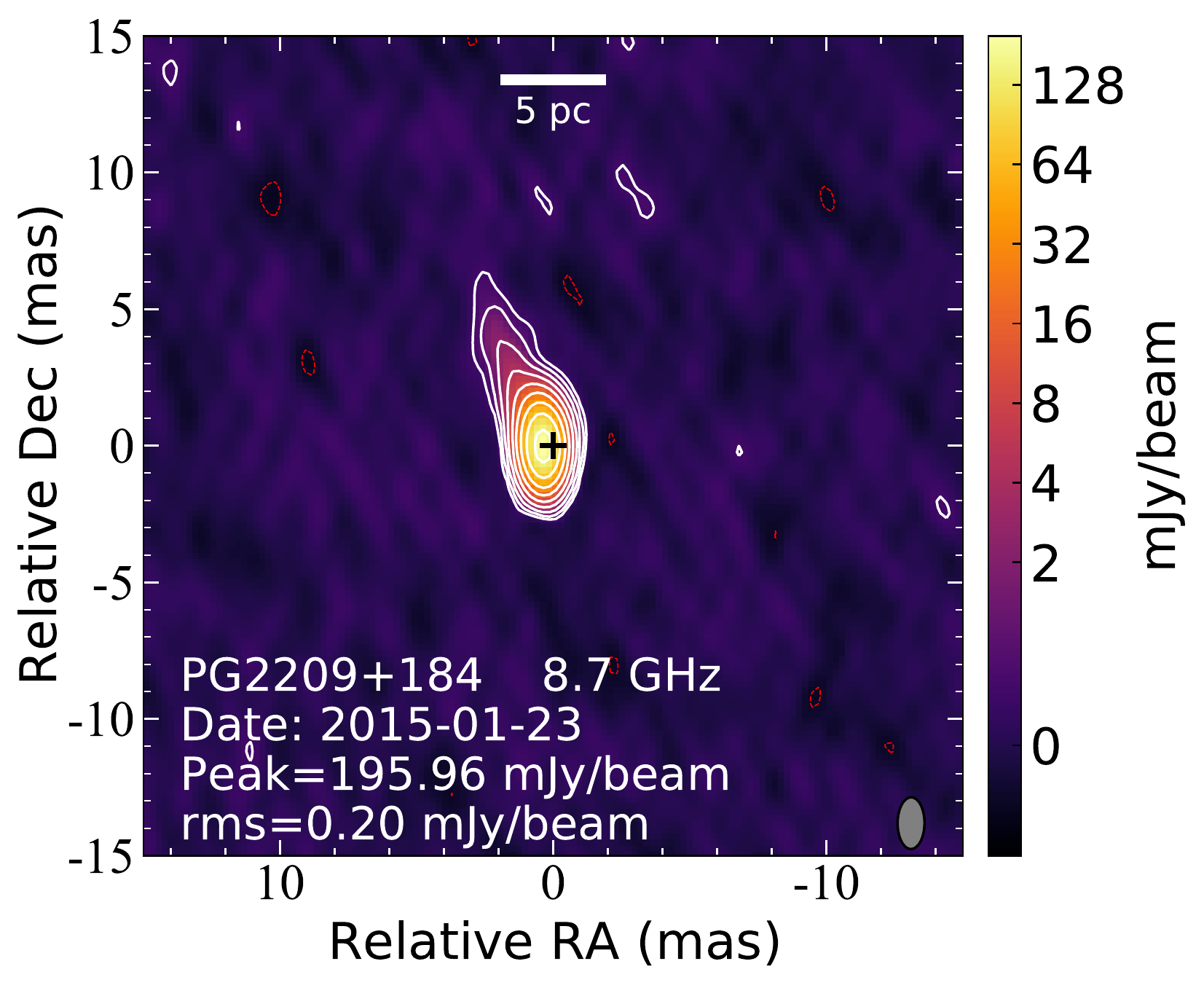}
  \includegraphics[width=0.24\textwidth]{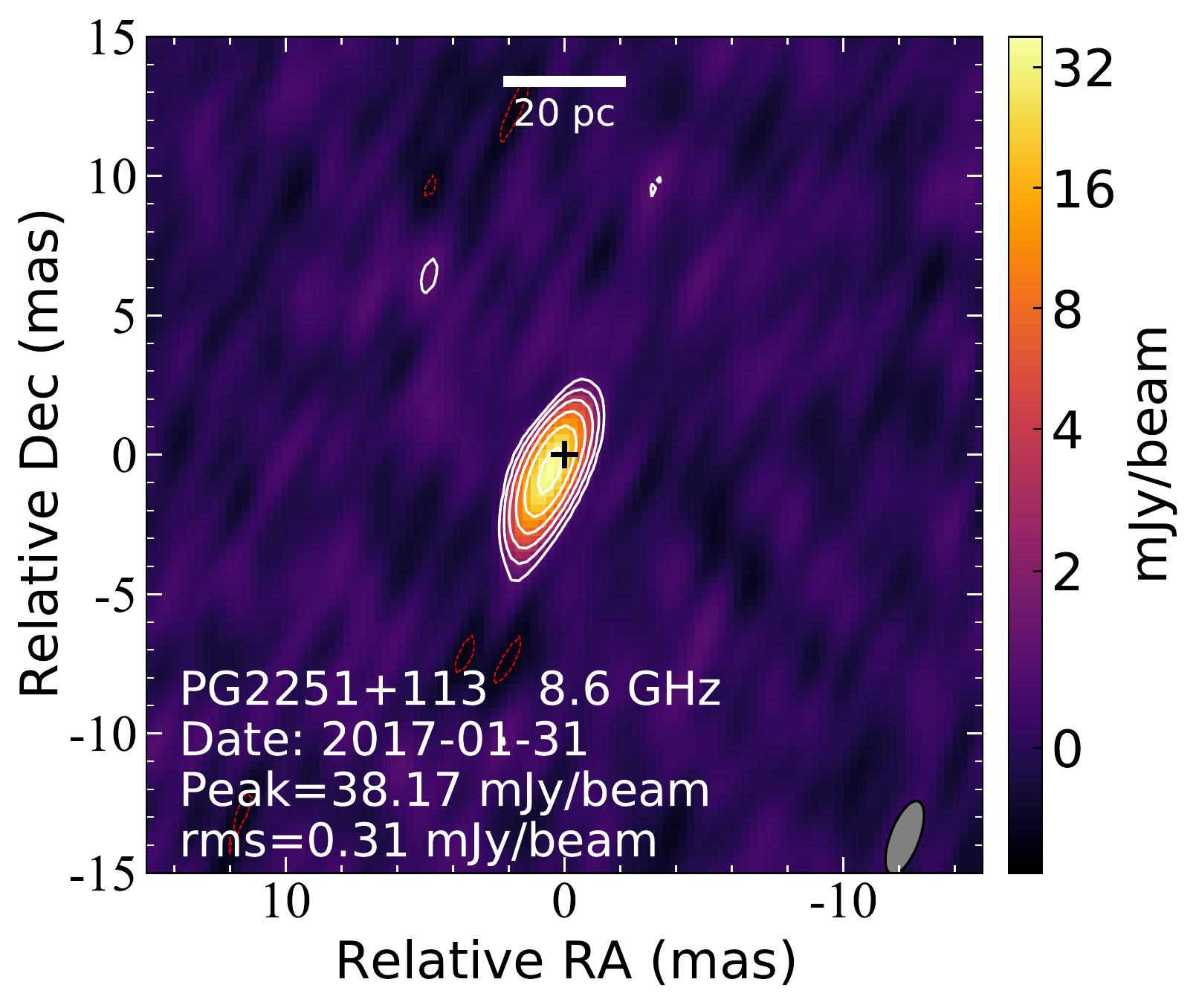}
  \includegraphics[width=0.24\textwidth]{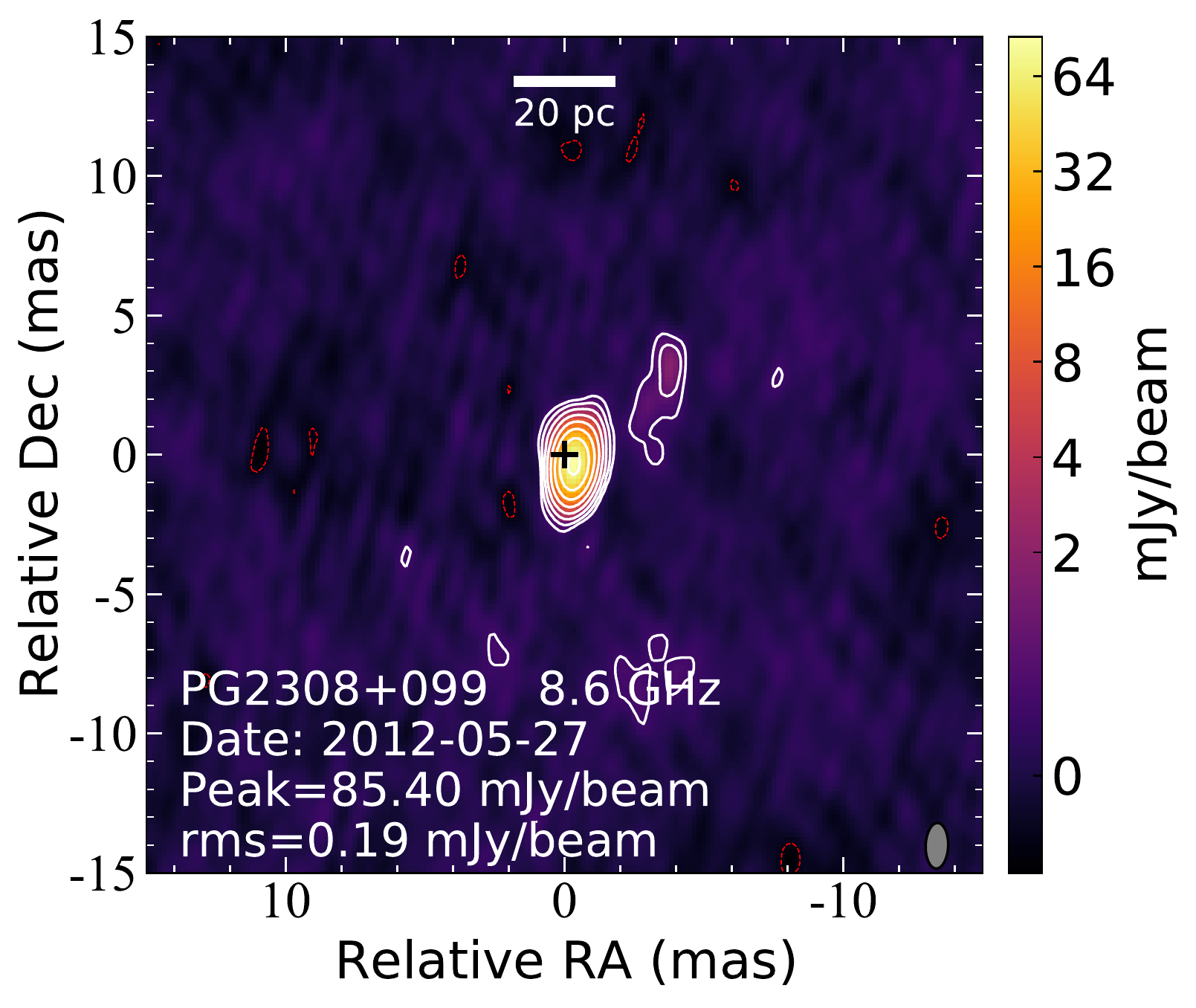}
    \end{tabular}
  \centering
  \caption{VLBI images of 12 $z < 0.5$ radio-loud PG quasars not included in our sample. The images are obtained from astrogeo VLBI archive. Image parameters are labelled in each panel and also referred to  Table \ref{tab:image}}.
  \label{fig:RLAGN}
\end{figure*}

The VLBA images of four RLQs from our sample are presented in Figure~\ref{fig:RLAGN-4}. Figure~\ref{fig:RLAGN} shows the VLBA images of 12 RLQs collected from the VLBI data archive. We chose the archival data closest to the time of our VLBA observations and compared the selected image with those from other epochs of this source to ensure image fidelity. 
Nevertheless, we can obtain fairly high-quality images by using self-calibration and hybrid-mapping technique \citep{1984ARA&A..22...97P} due to the high brightness and strong signal-to-noise ratio of these RLQs.

All four RLQs in Figure~\ref{fig:RLAGN-4} show a morphology consisting of
a compact core and a one-sided jet. The radio core contributes the majority of the total flux density. The position of the optical nucleus obtained from the \textit{Gaia} survey \citep{2022yCat.1355....0G} is very close to that of the radio core. 

\section{Radio emission from RQQs}

There are two main sources of radio emission from RQQs: 1. star formation activity, and 2. AGN-associated activity, including jets, accretion disk winds, or magnetized coronas. The main morphological difference between the two is the degree of compactness: star formation can occur  over galactic scales; in contrast, AGN-related radio emission is usually concentrated to much smaller scales located in the inner galactic region. The resolution of VLA D-array at 5 GHz is $\sim18\arcsec$, generally covering the whole galaxy. The VLA-A resolution at 5 GHz is 0.5\arcsec, corresponding to physical sizes of about 0.9 kpc at $z=0.1$ and 3 kpc at $z = 0.5$, respectively. So the ratio of $S^{\rm VLA}_{A}$ to $S_{\rm D}^{\rm VLA}$ can be used as a rough  indicator of how compact a radio source
is on the kpc scale. Here we use $f_{\rm c} = S^{\rm VLA}_{A}/S^{\rm VLA}_{D}$, the ratio of the 5 GHz flux densities obtained from the VLA A-array to VLA D-array, the same definition  as that adopted in \citet{1989AJ.....98.1195K,1994AJ....108.1163K}. The VLA data are also obtained from these papers.

\begin{figure}
\centering
  \begin{tabular}{c}
  \includegraphics[width=0.45\textwidth]{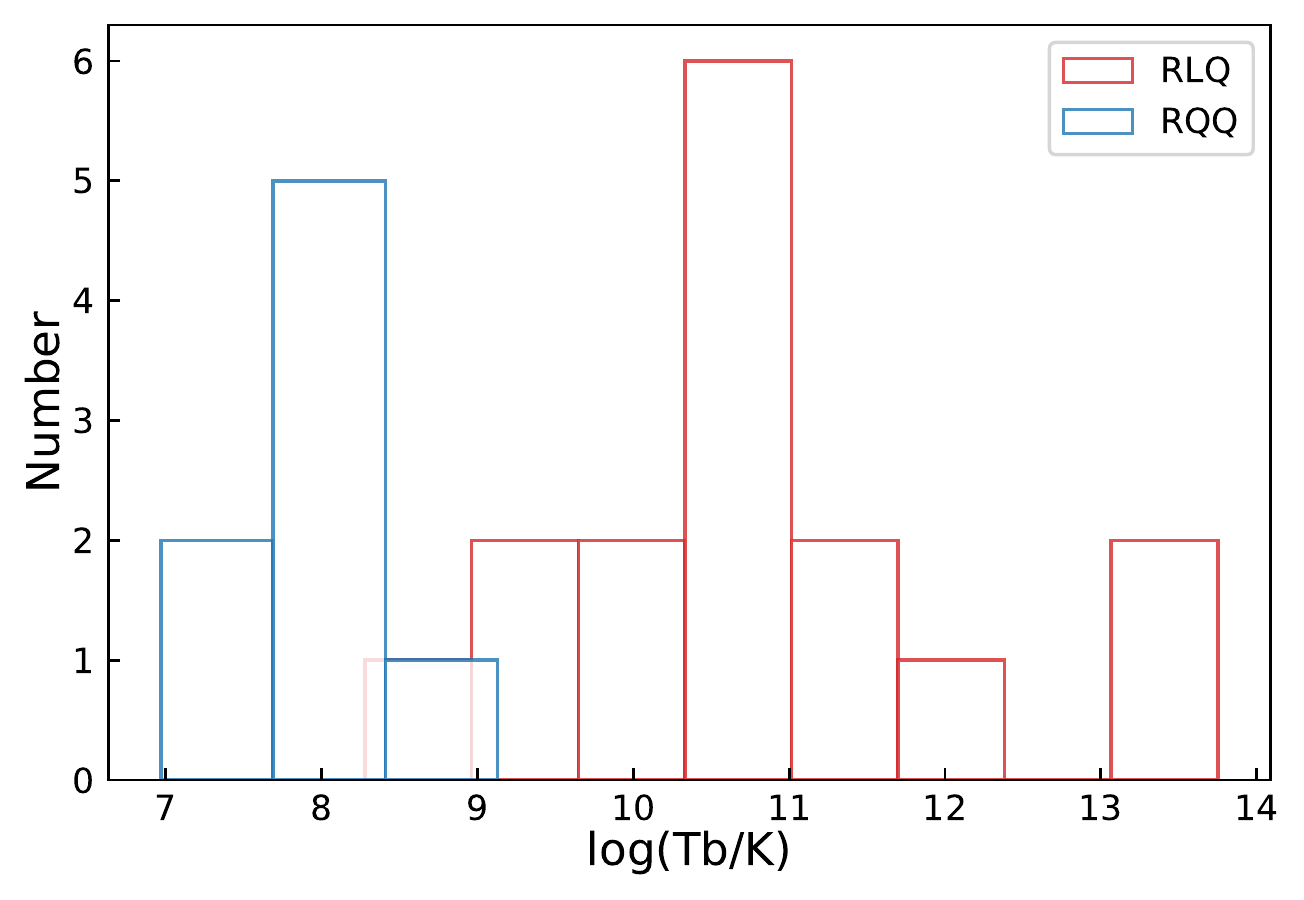}
  \end{tabular}
  \centering
  \caption{The distribution of brightness temperatures of core components (Column 6 in Table \ref{tab:radio}) for 24 PG quasars, including 8 RQQs and 16 RLQs.}
   \label{fig:Tb_hist}
\end{figure}

Figure \ref{fig:Tb_hist} shows the histogram of the core brightness temperature of RQQs and RLQs. It clearly displays a bimodal distribution. The peak of $T_{\rm b}$ of RQQs is located around $10^{8}$~K, which far exceeds the brightness temperature for star formation \citep{1992ARA&A..30..575C}, 
ruling out the possibility of a thermal origin of their radio emission.
The peak of RLQs is close to $10^{11}$K, and the brightness temperatures of the few brightest RLQs exceed the energy-equipartition brightness temperature limit \citep{1969ApJ...155L..71K,1994ApJ...426...51R}, exhibiting highly relativistic beaming effects. 
Although both $T_{\rm b}$ and $R$ show a bimodal distribution, the $R$ parameter is susceptible to contamination by the host galaxy emission in optical (more severe in nearby low-luminosity AGN) \citep[e.g.][]{2001ApJ...555..650H}, leading to ambiguous and controversial physical interpretation of the conventional $R$ parameter.  $T_{\rm b}$ is strongly correlated with the presence or absence of an AGN core, therefore the bimodal distribution of $T_{\rm b}$ clearly distinguishes between RQQs and RLQs.

\begin{figure*}
    \centering
    \includegraphics[width=0.9\textwidth]{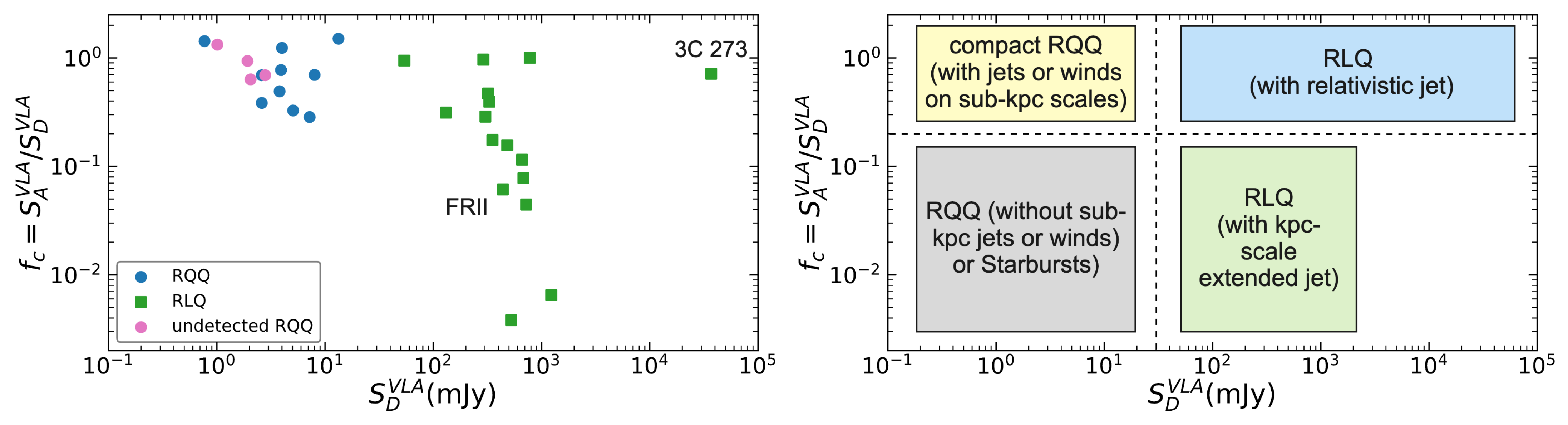}
    \caption{Flux density ratio $f_{\rm c} = S_{\rm A}^{\rm VLA}/S_{\rm D}^{\rm VLA}$ versus $S_{\rm D}^{\rm VLA}$, where $S_{\rm A}^{\rm VLA}$ and $S_{\rm D}^{\rm VLA}$ are flux densities measured by the VLA at A-rray and D-array configurations, respectively, which are obtained from \citet{1989AJ.....98.1195K,1994AJ....108.1163K}. The four undetected RQQs marked with solid pink circles have been detected by VLA but not by VLBA. Error bars are not shown because the error of $f_{\rm c}$ is tiny.}
    \label{fig:fc}
\end{figure*}

Figure \ref{fig:fc} shows $f_{\rm c}$ 
versus $S^{\rm VLA}_{D}$. Two of our 16 RQQs (PG 0923+129 and PG 1211+143) were not detected in the VLA A-array observations. Thus we only include data points from 14 RQQs. Data points of 16 RLQs with $z < 0.5$ 
are also included for comparison.

As can be seen in the left panel of Figure~\ref{fig:fc} , the total flux densities of RQQs and RLQs in this sample have a clear dividing line at about 30 mJy. 
This dividing flux density corresponds to a radio luminosity of $L_{\rm 5GHz} = 1.4\times 10^{42}$ erg s$^{-1}$ (or a radio power of $2.8 \times 10^{25}$ W~Hz$^{-1}$) at $z = 0.5$, which could be an upper limit of radio luminosity for radio-quiet quasars. 
This division line can vary due to differences in  observing frequency, observation configurations, redshift, and sample selection; for example, the division line here is slightly different from other studies in the literature. However, they all manifest that RLQs and RQQs have different radio emission properties. 
A source that exceeds this radio luminosity threshold must have developed a powerful jet.
This radio power boundary also coincides with the division  between FRII and FRI galaxies (see Figure 1 of \citealt{2012ApJ...760...77A}). 
In the luminosity-size diagram of radio galaxies, RQQs occupy considerable parameter space and represent a radio source population that has not yet been studied in depth \citep{2020NewAR..8801539H}.

The $f_{\rm c}$ of RQQs is distributed in the range of 0.28--1.50, indicating that a significant fraction or even most of the radio emission comes from the inner 1 kpc region. The resolution of 5 GHz VLA-A is not  sufficient to distinguish whether RQQs are nascent and/or short-lived jets \citep[e.g. ][]{2020ApJ...905...74N} or nuclear starbursts \citep[e.g. ][]{1992ARA&A..30..575C}. Three RQQs (PG 1216+069, PG 1351+640 and PG 2304+042) have $f_{\rm c}$ above 1.0 due to their variability (see also discussion in Section \ref{sec:individual}), and their radio structures must be very compact ($<$pc scales). 

The $f_{\rm c}$ of RLQs is distributed over a wide range of 0.004 -- 1, which is related to the various types of jet structures. Eight RLQs have $f_{\rm c} < 0.2$, they are: PG 1004+130, PG 1048$-$090, PG 1100+772, PG 1103$-$006, PG 1512+370, PG 1545+210, PG 1704+608 and PG 2251+113.
The kpc-scale morphology of all these galaxies is classified as  Fanaroff-Riley type II  (FRII, \citealt{1974MNRAS.167P..31F}) radio galaxies, which have a weak radio core but prominent lobes. 
RLQs with $f_{\rm c} > 0.2$ show a clear core dominance in the VLBA images, and most of them are classified as flat-spectrum radio quasars. The high $f_{\rm c}$ of these RLQs indicate that they have an intrinsic nature of compact emission. As the PG sample is optically selected quasars, all belong to Type 1 objects, which according to the AGN unified model have jets with a viewing angle of no more than 40\degr\ to the line of sight, while blazars with highly relativistic beamed jets have a viewing angle of no more than 10\degr. This implies that the probability of observing a blazar in a quasar sample is about 4.4\%. In our sample of 16 RLQs, there is only one obvious blazar, the brightest quasar in this sample, PG 1226+023 (3C 273), corresponding to a fraction of 6.3\%  which is very close to the probability analysed above. 3C~273 is located in the top-right corner of Figure~\ref{fig:fc}, a region occupied by highly relativistic jetted quasars. Adding more blazar samples helps to determine the distribution range of this blazar region.



The right panel of Figure \ref{fig:fc} shows a sketch of the inferences obtained from the $f_{\rm c}$--$S^{\rm VLA}_{D}$ relationship. It shows four zones representing RQQs with compact kpc-scale radio structure (yellow coloured), 
RLQs with relativistic
jets (blue), RLQs with extended kpc-scale
jets/lobes (green), and RQQs without well-defined sub-kpc-scale jets (gray) are shown separately.
We should note that the data points in this paper do not yet cover all the parameter space. For example, the region of $f_{\rm c}<0.2$ and $S^{\rm VLA}_{D} <30$~mJy may be occupied by RQQs with disrupted jets or outflows on sub-kpc scales or extended starbursts; the further to the lower left, the weaker the AGN activity. Sources in this region  lack bright compact radio sources and thus require high-sensitivity (tens of $\mu$Jy beam$^{-1}$) mapping on sub-arcsec scales, e.g., e-MERLIN \citep{2018MNRAS.476.3478B,2021MNRAS.500.4749B}. On the other hand, the zone to the right of the green-coloured region, i.e., $f_{\rm c}<0.2$ and $S_{\rm D}^{\rm VLA} >1$~Jy,  should be dominated by giant extended radio galaxies. Since the radio emission is dominated by the optically thin synchrotron emission arising from radio lobes, it is more likely that these sources will be detected in low-frequency radio surveys. For example, a sample of $>1$ Jy radio galaxies  detected \citep{2020PASA...37...18W} in the GaLactic and Extragalactic All-sky Murchison Widefield Array (GLEAM) survey \citep{2017MNRAS.464.1146H} are expected to be located in this regime.

\begin{figure*}
    \centering
    \includegraphics[width=0.9\textwidth]{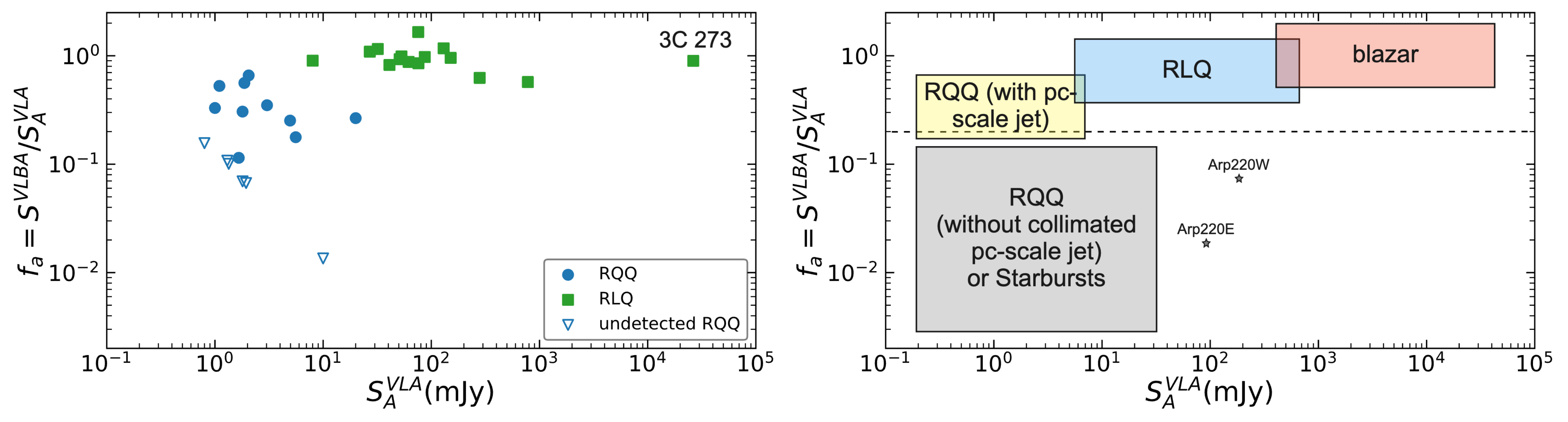}
    \caption{Flux density ratio $f_{\rm a} = S^{\rm VLBA}/S_{\rm A}^{\rm VLA}$ versus $S_{\rm A}^{\rm VLA}$, where $S_{\rm A}^{\rm VLA}$ and $S^{\rm VLBA}$ are flux densities measured by the VLA A-rray and the VLBA. The VLA data are obtained from \citet{1989AJ.....98.1195K,1994AJ....108.1163K}. The VLBA data are from the present paper and {\it astrogeo} VLBI archive. For those without 5 GHz VLBI data in the archive, we use their 2.3 and 8.4 GHz VLBI data to interpolate to estimate 5 GHz flux densities. PG 2251+113 is not presented in the image because its $f_{\rm a}$ value of 22.64 is too large. The triangle symbols mark the six RQQs which are not detected by VLBA. }
    \label{fig:fa}
\end{figure*}

Figure \ref{fig:fa} shows the correlation between $f_{\rm a}$ ($=S^{\rm VLBA}/S^{\rm VLA}_{A}$) and $S^{\rm VLA}_{\rm A}$. There are two RQQs (PG 0923+129 and PG1211+143) without $S_{\rm A}^{\rm VLA}$, so we used $S_{\rm D}^{\rm VLA}$ instead. 
As we know from Figure \ref{fig:fc}, the radio flux density ratio $f_{\rm c} = S^{\rm VLA}_{\rm A}/S^{\rm VLA}_{D}$ of RQQs has a median value of $f_{\rm c}  \sim 0.7$, so substituting $S^{\rm VLA}_{\rm A}$ with $S^{\rm VLA}_{\rm D}$ in these two sources should not change our interpretation significantly. 
The "tongue" that extends downward from RLQs with highly relativistic jets in Figure \ref{fig:fc} disappears in Figure \ref{fig:fa} because the extended lobes of FRIIs are resolved in the VLA A-array images, leaving only the central radio cores, that results in the data points shifting upward along the $f_{\rm a}$ axis. 
RLQs have typically a compact core or a core+one-sided jet on parsec scales (Figures~\ref{fig:RLAGN-4} and \ref{fig:RLAGN}).
Figure~\ref{fig:fa} shows a dividing line around $f_{\rm a}=0.2$ between VLBA-detected and VLBA-non-detected RQQs, which display different radio properties. At $f_{\rm a}>0.2$, RQQs preferentially have compact radio cores (see Figure \ref{fig:RQAGN}). In the RQQs with $f_{\rm a}<0.2$, the radio core is either not prominent (e.g. PG 1612+261), or the radio emission primarily comes from extended disk winds but not from jets (e.g. PG 0157+001), which are heavily resolved in the VLBA images. 
The radio morphology of the less compact (lower $f_{\rm a}$) RQQs lacks clear indications of collimated jets and usually shows disrupted jets or knotty structures probably resulting from the collision of extended jets/outflows with the interstellar medium (an example of such a disrupted jet is seen in PG 0050+124 (Mrk~1502) in Figure \ref{fig:RQAGN}) or AGN winds \citep[e.g., PDS~456 in ][]{2021MNRAS.500.2620Y}. Therefore, $f_{\rm a} = 0.2$ can be considered as a dividing line for the presence of compact nuclear radio structure in RQQs. This dividing line might depend on $S_{\rm A}^{\rm VLA}$, i.e., it is not necessarily a horizontal line but possibly an inclined line; it is also related to the observing frequency and redshift of the selected sample. The compact nuclear radio source may be associated with the jet or the corona. Radio emission from RQQs with $f_{\rm a}<0.2$ is either in the form of accretion disk wind with a large opening angle, or starbursts in the central region of the host galaxy.

Another dividing line appears around $f_{\rm a}=0.55$, which distinguishes the compact RQQs from RLQs. 
At $f_{a}>0.55$, all the quasars in our sample have high brightness temperatures (2--4 orders of magnitude higher than those of RQQs) and bright radio cores (with radio monochromatic luminosities $\gtrsim 10^{24}$ W Hz$^{-1}$).  More than half of RLQs in the sample show compact and collimated jets, and the remaining unresolved radio cores would hopefully be resolved in higher resolution images.
While for RQQs with $f_{\rm a}<0.55$, they lack powerful jets,
which are somewhat similar to the weak and frustrated CSOs \citep{2012ApJ...760...77A}; these sources are difficult to develop large-scale ($>$ a few kpc) jets, and the choked jets impose strong radio feedback to the host galaxies.
Similar to the $f_{\rm a}=0.2$ division, the $f_{\rm a}=0.55$ dividing line may also be affected by selection effect (e.g., the flux density limit $S^{\rm VLA}_{A}$, redshift, observing frequency). More data are required to further verify the shape of the dividing line, especially whether RLQs extend into the $f_{\rm a}<0.55$ regime.

 Although the data points used in this paper are not numerous enough to cover all parameter space, it does reveal three distinctly different zones in the $f_{\rm a}-S^{\rm VLA}_{\rm A}$ diagram. 

Subsequent work, including more data, will provide more stringent constraints on the scopes of these zones. 

In the region of $f_{\rm a}<0.2$, $S_{\rm A}^{\rm VLA}<30$~mJy, it should be occupied by 
starburst galaxies and RQQs with disrupted jets and disk winds on parsec scales. 
When $S_{\rm A}^{\rm VLA}<0.1$~mJy, we believe that both jets and starbursts are no longer dominant, and that normal galaxies with relatively lower star formation rates, spiral galaxies, and dwarf galaxies are expected to be found in this regime.

The bottom-right corner of Figure~\ref{fig:fc} is 
occupied by large-sized extended radio sources (FRIIs). However, the extended lobes of these galaxies are resolved in VLA A-array images, with only the FRII cores remaining. So $f_{\rm a}$ of these large-scale RLQs are shifted upward into the blue-coloured region in Figure~\ref{fig:fa}, leaving a void in this region of Figure \ref{fig:fa}. It is possible that sources in this region are dominated by extremely diffuse radio emission and  lack AGN activity and jets. For example, we have marked in Figure \ref{fig:fa} the positions of two nuclei of the well-known ultraluminuous infrared galaxy Arp220, where the VLBA flux densities are the sum of the observed supernove and supernova remnants in two nuclear regions \citep{1995ApJ...446..602B,2006ApJ...647..185L,2007ApJ...659..314P}. There has been  rich evidence showing that its radio emission originates from starburst activity, and no clear evidence of AGN in the two merging galaxies has been found so far.
On the other hand, few (if any) RQQs have been found in the region where $f_{\rm a}<0.2$, $S_{\rm A}^{\rm VLA}>30$~mJy. 

The position of a radio source in Figure~\ref{fig:fa} can be changed with the emission state. When radio flares occur, RLQs located in the blue-coloured region will move upward to the right. Similarly, some RQQs located in the transition zone between the compact RQQs (yellow zone) and RLQs  (blue region) also move upward to the right in flare stages, when they behave like a blazar, such as III Zw 2 \citep[][]{2001ASPC..224..265F,2005A&A...435..497B} and Mrk~231 \citep{2013ApJ...776L..21R,2021MNRAS.504.3823W}.
RQQs without powerful jets located in the grey-coloured region, under peculiar conditions such as tidal disruption events (TDEs), may have a dramatic increase in $f_{\rm a}$ and flux density and enter into the jetted RQQ region (yellow region) or even the RLQ region (blue region).

\section{Individual Detected Source Properties}
\label{sec:individual}

\subsection{PG 0003+199 (Mrk~335)}

PG 0003+199 (Mrk~335) is a low-luminosity Narrow Line Seyfert 1 galaxy at $z=0.025785$ \citep{1999ApJS..121..287H}. 
The black hole mass of Mrk~335 is $\sim10^7 M_\odot$ \citep{2004ApJ...613..682P,2014ApJ...782...45D}, putting this AGN in a super-Eddington accretion state \citep{2013PhRvL.110h1301W}. The X-ray flux of Mrk~335 have exhibited strong  variability \citep{2008ApJ...681..982G,2012ApJS..199...28G,2018MNRAS.478.2557G,2020MNRAS.499.1266T,2020A&A...643L...7K}. 
Observations at radio wavelengths, especially high-resolution imaging before and after large X-ray flares, can help to reveal the jet structure in the heart of the galaxy and the correlation between the jet and the X-ray variability.

The recent 43-GHz VLA image of Mrk~335 \citep{2022MNRAS.510.1043B}  shows a compact component on a scale of $<50$ pc with a total flux density of $\sim$0.6 mJy. 
The spectral index is $\nu_{\rm 5GHz}^{\rm 8.5GHz} = -0.86$ \citep{2019MNRAS.482.5513L}
and $\nu_{\rm 1.4GHz}^{\rm 45GHz} = -0.77$ \citep{2022MNRAS.510.1043B}, suggesting that the radio emission is dominated by optically-thin synchrotron radiation and is concentrated in the nuclear region of this galaxy \footnote{The spectral index $\alpha$ is defined as $S_\nu \propto \nu^\alpha$}.  The 1.5-GHz VLBA image obtained by \citet{2021MNRAS.508.1305Y} reveals an elongated radio structure of $\sim$40 mas ($\sim$20 pc in projection) along the north-south direction.
Our 5-GHz VLBA image (Figure \ref{fig:RQAGN}) reveals a two-sided jet, with the main structure distributed continuously along the northeast--southwest direction and a largest extent of about 30 mas ($\sim$16 pc). The jet in the 5 GHz image corresponds to the central part of the 1.5-GHz jet between N1 and S2 components in Figure 1c of \citet{2021MNRAS.508.1305Y}. The compact component corresponding to the \textit{Gaia} optical nucleus has a size smaller than 2.1 mas (1.07 pc) and a brightness temperature of $\sim10^7$ K. 

Comparison of the 5-GHz flux density between the pc scale and the kpc scale shows that 47\% of the emission comes from the compact jet components. This percentage can be regarded as a lower limit since VLBI resolves part of the extended jet. Therefore, we conclude that the radio emission from Mrk~335 is dominated by the jet. Future studies may focus on the variation of the radio jet and its correlation with the change of the accretion state - whether, similar to XRB, the jet is suppressed in the super-Eddington accretion state, and is enhanced in the X-ray low state.

\subsection{PG 0050+124 (Mrk 1502)}
PG 0050+124 (I Zwicky 001, Mrk 1502, UGC 00545) is one of the closest \citep[$z=0.06$,][]{1999ApJS..121..287H,2009ApJS..184..398H} narrow-line Seyfert 1 galaxy. Its host galaxy is an almost face-on spiral with a fainter companion to the west and two arm-like tidal tails formed by the galaxy merger \citep{1999A&A...349..735Z}. 
 PG 0050+124 is intriguing for its extreme variability: its X-ray flux 
has varied by more than an order of magnitude over the past decade \citep{2012ApJS..199...28G,2018MNRAS.478.2557G}; it is one of the most variable RQQ from optical-to-UV bands with change of brightness by a factor of $\sim$2 \citep{1998ApJ...501...82P,2012ApJS..199...28G}, but no significant variability found in radio bands.

In the 5 and 8 GHz VLA images, PG 0050+124 shows an unresolved morphology \citep{1994AJ....108.1163K,1995MNRAS.276.1262K}.  
The latest VLA 45-GHz image \citep{2022MNRAS.510.1043B} displays a marginal elongation from the radio core to the north, however this needs to be further confirmed with higher quality images. The radio SED shows a fairly steep spectrum: $\alpha_{\rm 1.4GHz}^{\rm 45GHz} = -0.83$ \citep{2022MNRAS.510.1043B} and $\alpha_{\rm 8.5GHz}^{\rm 5GHz} = -1.45$ \citep{2019MNRAS.482.5513L}. 
In \citet{2022arXiv220801488A}, PG 0050+124 shows a resolved structure along the east-west direction in the 1.4-GHz VLBA image, but only marginal detection of a few discrete faint clumps in the 5-GHz VLBA image. In our 5-GHz VLBA image, PG 0050+124 shows three discrete clumps within 15 mas east of the optical \textit{Gaia} nucleus position, and their peak flux densities are about 5--6 times the image noise. The brightness of the clump closest to the optical nucleus is only $4\sigma$.
Neither of these clumps could be identified as the AGN core from the current data. 

\subsection{PG 0157+001 (Mrk~1014)}

PG 0157+001 (Mrk 1014,  $z=0.163$) is one of the brightest quasars in a class of 'warm' ultraluminous infrared galaxies (ULIRG) \citep{1988ApJ...328L..35S,2001ApJ...554..803Y}, 
and is in the evolutionary phase from ULIRG to ultraviolet-excess quasars \citep{1988ApJ...335L...1S}. 
The host galaxy of PG 0157+001 has a prominent spiral-like tidal tail extending to the northeast, indicating a recent merger \citep{1984ApJ...283...64M}. 
The dense star formation is restricted within the central 2 kpc region \citep{2000AJ....120.1750C}.

The 8.4-GHz VLA image, with a resolution of 0.36\arcsec, shows a triple structure along the east-west direction, where two lobes are located about 1.1\arcsec\ on either side of the central core \citep{2006A&A...455..161L}. In the 43-GHz VLA image \citep{2022MNRAS.510.1043B}, there is a faint component at the optical nucleus position. The spectral index of the unresolved component is $-1.11\pm0.02$ between 5 and 45 GHz \citep{2022MNRAS.510.1043B}. 
No VLBI component is detected near the position of the optical nucleus in our 5-GHz VLBA image. The only VLBI clump  deviates from the optical nucleus by about 24 mas ($\sim$60 pc). More data are needed to identify the nature of this feature.

\subsection{PG 0921+525 (Mrk~110)}

PG 0921+525 (Mrk 110) is an X-ray bright NLS1 at low redshift of $z = 0.03529$ \citep{1985ApJ...294..106V}. 
\citet{2006ApJ...651L..13D} found significant X-ray variability in Mrk~110 which showed energy-dependent time lag. The time-scale of the variability constrained the size of X-ray emitting region to a few Schwarzschild radii. 
\cite{2022MNRAS.510..718P} also found significant radio variability on mas scales at 5 GHz from days to weeks time-scales in PG 0921+525.

The VLA-A images of PG 0921+525 show a complex structure \citep{1993MNRAS.263..461P,1994AJ....108.1163K,2022A&A...658A..12J}, with most of the emission coming from the southernmost ﬂat-spectrum component, and some diffuse emission in the north. The overall structure looks like a highly curved jet or ring-like structure. The extended emission is consistent with circumnuclear star formation, similar to ultra-luminuous infrared galaxies \citep{2017MNRAS.469..916B}. PG 0921+525 presents an unresolved feature in our 5-GHz VLBA image with a flux density of 0.92 mJy. The VLBI morphology and flux density are consistent with the results obtained from a monitoring campaign using VLBA from August 14, 2015 to May 19, 2016 \citep{2022MNRAS.510..718P}. During the monitoring observations, the peak flux density varied from a minimum value of 0.46 to a maximum value of 1.43 mJy.

\subsection{PG 1149$-$110}

PG 1149$-$110 (LEDA 37161) was imaged by the VLA in A-array  (resolution of 0.5\arcsec)  and exhibited double components along the east-west direction \citep{1993MNRAS.263..425M,1994AJ....108.1163K}. At higher resolution (0.44\arcsec$\times$0.33\arcsec) and sensitivity at 5 GHz, the source showed a triple structure extending 1.5\arcsec\ in the east-west direction \citep{2006A&A...455..161L}. However, at higher frequencies of 85 and 45 GHz, the VLA A-array only detected an unresolved component close to the optical nucleus, with a spectral index of $-0.65$\citep{1998MNRAS.297..366K,2022MNRAS.510.1043B}. 
In our VLBA image and the image obtained by \citet{2022arXiv220801488A},  PG 1149$-$110 shows an unresolved component close to the optical nucleus with a flux density of 0.32 mJy. It has a brightness temperature of $>10^{7.87}$ K and is identified as the core.

\subsection{PG 1216+069}

PG 1216+069 shows a compact  structure in the VLA images   \citep{1993MNRAS.263..425M,1994AJ....108.1163K}. It showed a flat or inverted spectral shape with a spectral index of $-0.08$ from the VLA D-array observations at 4.89$-$14.9 GHz \citep{1996AJ....111.1431B}. 
PG 1216+069 (J1219+0638) shows an unresolved core with a peak flux density of 1.04 mJy in our 5-GHz VLBA image observed on 2015 August 7. A similar structure but with a much higher flux density of 6.4 $\pm$ 0.3 mJy was revealed in the VLBA image at 5 GHz observed on 2000 January 21 \citep{2005ApJ...621..123U}, confirming the variability of this source previously found from VLA observations \citep{1996AJ....111.1431B}. 
The 8.4-GHz VLBA image obtained by \citet{1998MNRAS.299..165B} showed a possible jet extension to the south which however is not seen in our 5-GHz image.
The radio spectrum of the VLBI component is flat ($\alpha^{\rm 5GHz}_{\rm 1.4GHz} = 0.07\pm0.16$  or with an inverted spectral shape \citep{2005ApJ...621..123U}. The core brightness temperature measured in all these VLBA observations is larger than 10$^{8}$~K,  ruling out the star formation origin.

\subsection{PG 1351+640}


PG 1351+640 shows a bright component and a faint component located 1.5\arcsec\ to the west in the VLA D-array image at 5 GHz \citep{1994AJ....108.1163K}. 
This western component was clearly detected in the 1.4 GHz NVSS image, but not detected in the FIRST  and VLASS images, indicating that it has a steep spectrum and diffuse structure.
The bright VLA component shows an inverted spectrum (or peaked spectrum) and it is highly variable with a variation factor of 4 within 3 yr at 5 and 15 GHz \citep{1989ApJS...70..257B}. 
A single compact component was detected by VLBA on 2000 February 6 with a flux density of 6.8 mJy \citep{2005ApJ...621..123U}. The total flux density obtained from our 5-GHz VLBA observation is slighly lower than that of \citet{2005ApJ...621..123U}.
The morphology of PG 1351+640 in our 5-GHz VLBA image is very similar to a CSO showing double compact components separated by about 10 pc. However, the optical nucleus is close to the southeast component, making it more likely the core. 
PG 1351+640 has a 'core-jet' like morphology extending along the north-south direction in a higher resolution VLBA image at 8.4 GHz on 1996 June 9 \citep{1998MNRAS.299..165B}, very similar to our result.

\subsection{PG 1612+261}
PG 1612+261 (Ton 256) was identified as an 'extended' source with a one-sided jet toward the east-west in VLA images \citep{1993MNRAS.263..425M,1989AJ.....98.1195K,2006A&A...455..161L}. The brighter component (the VLA core) was resolved into double sources within 0.5\arcsec\ along with some more diffuse emission toward the west \citep{1998MNRAS.297..366K}. Overall, PG 1612+261 has a steep spectrum of $\alpha=-1.57$ between 5 and 8.5 GHz for the unresolved VLA core. 
PG 1612+261 is clearly detected in the 1.4-GHz VLBA image of \citet{2022arXiv220801488A}, showing an elongated structure in the east-west direction, but is not detected in their 5-GHz VLBA image.
In our 5-GHz VLBA image, a $6\sigma$ component is detected at the location of the optical nucleus of PG 1612+261.

\subsection{PG 1700+518} 

PG 1700+518 is a broad absorption line (BAL) quasar at a redshift of 0.292 \citep{1985ApJ...296..416W} and the first radio-quiet BAL QSO to discover radio jets. The broad Balmer line system has not only radial but also rotational motion, suggesting that the outflow originates from the rotating accretion disk wind \citep{2007Natur.450...74Y}. Radio images can give independent constraints on the wind geometry. 
European VLBI Network (EVN) 1.6-GHz image reveals a triple structure, with the central component corresponding to the optical and X-ray nucleus \citep{2012MNRAS.419L..74Y}. The other two components are separated by $<$1 kpc, along the north-south direction, and they are identified as terminal hotspots. 
Our 5-GHz VLBA, with a higher resolution than the 1.6-GHz EVN image, reveals a compact radio component, and a faint extension towards the southeast to be confirmed in future observations. The two hotspots observed in the EVN image are not detected in our VLBA image, due to their reduced flux density (assuming a steep spectrum) at 5 GHz and the lack of a sufficiently compact structure. Based on our VLBA data and EVN data from the literature, we estimate the spectral index of the VLBA component to be $\alpha^{\rm 5GHz}_{\rm 1.6GHz} = -0.36$ and the brightness temperature to be $1.8 \times 10^8$K. The flat spectrum and high brightness temperature of the compact VLBA component are consistent with it being the radio core of the AGN.

\subsection{PG 2304+042}

On kpc scales, PG 2304+042 was detected with a single compact component in the VLA D-array images and an unresolved component. This component coincides with a second elongated component in the VLA A-array images at 5 GHz \citep{1994AJ....108.1163K}. \citet{2005ApJ...618..108B} indicated extreme radio variability in PG 2304+042.  PG 2304+042 shows a compact core component with a flux density of 0.54 mJy in our VLBA image at 5 GHz.  \cite{2022arXiv220801488A} also successfully detected a flat VLBI component in PG 2304+042 ($\alpha^{5GHz}_{1.4GHz}=-0.09$).





\section{Undetected RQQs} \label{sec:undetected}

\subsection{PG 0923+129 (Mrk 705)}

PG 0923+129 (Mrk~705) is a RQ NLS1 whose host galaxy is an SA-type spiral galaxy with an outer and an inner rings \citep{2017MNRAS.471.4027B}. 
PG 0923+129 shows a two-component structure in VLA-D images at 5 GHz \citep{1994AJ....108.1163K}. The southern part corresponds to the core. In the new 5 GHz VLA A-array image, \cite{2022A&A...658A..12J} detected a central compact component surrounded by several patches with a steep spectral index of $-0.75$. 

The VLBI image of PG 0923+129 in \citet{2013ApJ...765...69D} shows a compact core as well as a possible eastward jet extending to $\sim$45 mas.  The brightness temperature of the core is larger than 10$^{7.9}$~K, which indicates the non-thermal nature of the emission. The core of PG 0923+129 has a flux density of 1.8 mJy at 1.7 GHz, but is not detected in our 5 GHz VLBA image ($5\sigma$ upper limit is 0.135 mJy). If the compact core is indeed present, but bleow our VLBA detection capability, 
a spectral index of $-2.4$ is inferred for the core using the $5\sigma$ upper limit. However this steep spectral index is clearly impractical,  which is much steeper than that of a conventional optically thin jet. We notice that in \citet{2013ApJ...765...69D} the radio core was not detect in the initial CLEAN image, yet was detected after using self-calibration as they claimed. The use of self-calibration for very faint sources with insufficient signal-to-noise ratio may cause false detections as well as excessive fluxes.

\subsection{PG 1116+215}

 The VLA-D image of PG 1116+215 showed a triple structure with a faint central component and two bright components distributed on each side extending along north-south; the VLA-A image only detected the brightest southern component \citep{1994AJ....108.1163K}. The overall structure looks like a core-jet structure, with the core lying at the southernmost end of the jet.
PG 1116+215 (Ton 1388) was detected with an upper limit flux density less than 0.3 mJy at 8.4-GHz by VLBA  \citep{1998MNRAS.299..165B}. It is not detected in our 5-GHz VLBA image either.

\subsection{PG 1211+143}
PG 1211+143 is one of the AGNs with potentially mildly relativistic ultra-fast outflow 
\citep{2003MNRAS.345..705P,2006MNRAS.372.1275P}.
The radio structure is unresolved in the VLA-A image \citep{1994AJ....108.1163K}.
PG 1211+143 is not detected in our 5-GHz VLBA images, suggesting the lack of compact component on pc scales. The radio emission  may be in the form of extended outflows or diffuse winds distributed on the scales of tens to hundreds of parsecs. 

\subsection{PG 1448+273}

PG 1448+273 is a NLS1 galaxy \citep{2015ApJS..219...12A} showing high variability in X-rays \citep{2012A&A...542A..83P} and  ultra-fast outflows \citep{2020MNRAS.495.4769K}.
The black hole mass is $9\pm2 \times 10^{6}$ M$_{\odot}$ \citep{2006ApJ...641..689V}, and the bolometric luminosity is $\sim$10$^{45.5}$erg s$^{-1}$, suggesting a super-Eddington accretion AGN with an Eddington ratio of $L/L_{\rm Edd} \sim 3$.
In VLA images, it is unresolved \citep{1994AJ....108.1163K}.
There is no reported VLBI detection so far including the non-detection in our observation, probably due to the intrinsic suppression of jet activity in the high accretion state. 

\subsection{PG 1534+580 (Mrk 290)}

PG 1534+580 (Mrk 290) 
was unresolved in the VLA-A image at 5 GHz \citep{1994AJ....108.1163K}.
No VLBI detection of PG 1534+580 has been reported in literature. The source is not detected in our VLBA image.

\subsection{PG 2130+099}

PG 2130+099 (Mrk 1513) showed a triple FRI-type structure in the  VLA-A images, with a bright core straddled by two weaker steep-spectrum lobes and with an overall extent of 2.5\arcsec\ \citep{1993MNRAS.263..425M,1994AJ....108.1163K,1998MNRAS.297..366K}. The overall source has a steep spectral index of $-2.32$ between 1.4 and 8.4 GHz \citep{1998MNRAS.297..366K,2006A&A...455..161L}.
The VLA core is not detected in our VLBA image and in \cite{2022arXiv220801488A}. 

\section{Conclusion}

We have observed 20 low-redshift ($z < 0.5$) Palomar-Green (PG) quasars at 5 GHz using the VLBA and obtained their parsec-scale radio images.  We also collected archive VLBA and VLA data for 12 radio-loud quasars with $z<0.5$ for a comparative study with RQQs. By doing so, we have a complete sample of 32 PG  quasars with $z<0.5$ and  $S_{\rm D}^{\rm VLA}>1$ mJy. 

A single core or a 'core + one-sided jet' structure was found in all radio-loud PG  quasars.
However, the radio morphology of RQQs is more complicated because the radio emission comes from a combination of star formation and supermassive black hole-driven jets.
Compact radio components are detected in 10 of the 16 RQQs, with a detection rate of 62.5 per cent. The high detection rate on pc scales suggests that compact radio emission associated with AGN  is prevalent among the flux density-limited (the brightest) RQQs.
The six sources not detected in 5-GHz VLBA images could be due to their optically thin jets becoming too weak at 5 GHz or their intrinsic lack of compact emission components.
A compact VLBA component is detected near the \textit{Gaia} optical nucleus in eight RQQs, probably the radio core or jet knot close to the central engine. For the compact components discovered in the VLBA images, the available single-frequency images are insufficient to  discern their physical nature. Further multi-frequency VLBA observations are needed to obtain the radio spectral indices of the compact components to determine whether the parsec-scale radio emission is from jet or corona. 

The RQQs and RLQs in our sample have a division around $S_{\rm D}^{\rm VLA}=30$~mJy. 
At $z=0.5$ this flux density approximately corresponds to a radio luminosity of $\sim10^{42}$ erg s$^{-1}$. This seems to imply that quasars below this radio power have not developed powerful relativistic jets.
By comparing $f_{\rm c}=S_{\rm A}^{\rm VLA}/S_{\rm D}^{\rm VLA}$, we found that all RQQs have $f_{\rm c}>0.2$, implying that more than 20\% of the 5-GHz flux density of RQQs comes from within the central kpc scale. 
On the kpc scale, RQQs generally show compact and unresolved core, or FRI-like structures (the core is dominated). Very few RQQs are of FRII type. 
The radio emission from RLQs is clearly dominated by the jet. When the jet inclination angle is small, RLQs show compact cores or 'core + one-sided jet' structures with  relativistic beaming effect in the jet; when the jet inclination angle is large, RLQs display FRII-type morphology and have small $f_{\rm c}$.

In the $f_{\rm a}$--$S_{\rm A}^{\rm VLA}$ plot, where $f_{\rm a} = S^{\rm VLBA}/S_{\rm A}^{\rm VLA}$, there seems a division line at $f_{\rm a}=0.2$ that separates the RQQ sample into two classes: compact and non-compact RQQs. These two classes differ significantly in morphology, compactness, and total flux densities. Quasars with  $f_{\rm a}>0.2$ all have compact radio cores, or core + well-defined jets in the VLBI images. 
A comparison between compact RQQs and RLQs reveals significant differences in the radio morphology and compactness of the two classes as well. On pc scales, RLQs show a one-sided core-jet structure or unresolved core, 
while compact RQQs show diverse morphologies: unresolved core, compact core-jet, two-sided jet, and knotted jet. Moreover, the fraction of pc-scale jets in the total flux density of RLQs (i.e. the compactness factor $f_{\rm a}> 0.55$) is systematically higher than that of compact RQQs ($f_{\rm a} = 0.2 - 0.55$), indicating that the extended jets or relic jets in RQQs on scales of tens to hundreds of parsecs contribute to the VLA-A flux density but are resolved out in VLBA images.

The brightness temperatures of the VLBA components of the RQQs and RLQs show a bimodal distribution. 
The core brightness temperatures of the RLQs are in the range of $10^9$ K -- $10^{13}$ K, while the core brightness temperatures of the RQQs are less than $10^{9.13}$ K.
In the $f_{\rm a}<0.2$ regime, quasars, on the other hand, tend to lack well-defined jets, and their radio emission on VLBI scales is either from disrupted jets, or accretion disk winds, or a clumps of supernovae or supernova remnants. These weak jets are not powerful enough to break through the confinement from the host galaxy and grow into extended radio sources of tens of kpc.

In conclusion, the study in this paper is based on a 
sample of nearby ($z<0.5$) PG  quasars with total flux densities above 1 mJy, and does not have other selection bias. Although this sample does not cover all the parameter space, we still obtain some interesting conclusions, which are very helpful for studying the radio emission mechanism of RQQs, and the intrinsic connection between RQQs and RLQs.
The brightest RQQs on arcsec scales exhibit properties that are intermediate between  jet-dominated and starburst-dominated AGNs. The radio emission consists of a combination of galactic scale star formation and AGN-related activity in the nuclear region. Future expansion of the RQQ sample to lower flux densities is needed to determine whether radio AGN is present in weak RQQs with total flux densities below 1 mJy. This will also  help to determine whether there is a clear dichotomy between radio loud and quiet quasars, and whether the central engines (black hole -- accretion disk systems) of these two classes are intrinsically different.

\section*{Acknowledgements}
This work is supported by the SKA pre-research funding granted by the National Key R\&D Programme of China (grant number 2018YFA0404603).  
X-P. Cheng is supported by the Brain Pool Programme through the National Research Foundation of Korea (NRF) funded by the Ministry of Science and ICT (2019H1D3A1A01102564).
LCH was supported by the National Science Foundation of China (11721303, 11991052, 12011540375) and the China Manned Space Project (CMS-CSST-2021-A04, CMS-CSST-2021-A06).
WAB acknowledges the support from the National Natural Science Foundation of China under grant No.11433008 and the Chinese Academy of Sciences President's International Fellowship Initiative under grants No. 2021VMA0008 and 2022VMA0019. 
The VLBA observations were sponsored by Shanghai Astronomical Observatory through an MoU with the NRAO. 
The National Radio Astronomy Observatory is a facility of the National Science Foundation operated under cooperative agreement by Associated Universities, Inc. Scientific results from data presented in this publication are derived from the following VLBA project codes: BA114.
The VLBI data processing made use of the compute resource of the China SKA Regional Centre prototype, funded by the Ministry of Science and Technology of China and the Chinese Academy of Sciences. 

\section*{DATA AVAILABILITY}
The correlated data used in this article are available in the VLBA data archive (\url{https://archive.nrao.edu/archive/archiveproject.jsp}). The calibrated visibility data underlying this article can be requested from the corresponding author.




\bibliographystyle{mnras}
\bibliography{ref} 





\bsp	
\label{lastpage}

\end{document}